\documentclass{jfm}

\usepackage{graphicx}
\usepackage{dcolumn}
\usepackage{epstopdf, epsfig, amsmath, bm}
\usepackage{etoolbox}

\usepackage[utf8]{inputenc}
\usepackage[T1]{fontenc}

\usepackage{newtxtext}
\usepackage{newtxmath}
\usepackage{natbib}
\usepackage{hyperref}
\usepackage{comment}
\hypersetup{
    colorlinks = true,
    urlcolor   = blue,
    citecolor  = black,
}

\newcommand{\RomanNumeralCaps}[1]
\linenumbers

\newcommand{\be}{\begin{equation}}
	\newcommand{\ee}{\end{equation}}
\newcommand{\bes}{\begin{equation*}}
	\newcommand{\ees}{\end{equation*}}
\newcommand{\bea}{\begin{eqnarray}}
	\newcommand{\eea}{\end{eqnarray}}
\newcommand{\bi}{\begin {itemize}}
\newcommand{\ei}{\end {itemize}}
\newcommand{\benm}{\begin{enumerate}}
	\newcommand{\eenm}{\end{enumerate}}
\newcommand{\bmn}{\begin{minipage}}
	\newcommand{\emn}{\end{minipage}}
\newcommand{\bfig}{\begin{figure}}
	\newcommand{\efig}{\end{figure}}
\newcommand{\ig}{\includegraphics}
\newcommand{\lnw}{\linewidth}

\newcommand{\bcls}{\begin{columns}}
	\newcommand{\ecls}{\end{columns}}
\newcommand{\bcl}{\begin{column}}
	\newcommand{\ecl}{\end{column}}

\newcommand{\bmat}{\begin{matrix}}
	\newcommand{\emat}{\end{matrix}}
\newcommand{\bpmat}{\begin{pmatrix}}
	\newcommand{\epmat}{\end{pmatrix}}
\newcommand{\bvmat}{\begin{vmatrix}}
	\newcommand{\evmat}{\end{vmatrix}}
\newcommand{\bbmat}{\begin{bmatrix}}
	\newcommand{\ebmat}{\end{bmatrix}}

\newcommand{\ptl}{\partial}

\newcommand{\lla}{\left\langle}
\newcommand{\rra}{\right\rangle}
\newcommand{\lal}{\langle}
\newcommand{\ral}{\rangle}

\newcommand{\ep}{\epsilon}
\newcommand{\al}{\alpha}

\newcommand{\lm}{\lambda}
\newcommand{\om}{\omega}

\renewcommand{\k}{{\bm k}}

\renewcommand{\v}{{\bm v}}

\newcommand{\e}{{\bm e}}
\newcommand{\f}{{\bm f}}

\newcommand{\x}{{\bm x}}

\newcommand{\bo}{{\bm \omega}}
\newcommand{\bom}{{\bm \omega}}

\newcommand{\bk}{{\bm k}}

\newcommand{\bu}{{\bm u}}
\newcommand{\bv}{{\bm v}}
\newcommand{\bx}{{\bm x}}

\newcommand{\DD}{{\mathcal D}}
\newcommand{\EE}{{\mathcal E}}

\newcommand{\NN}{{\mathcal N}}

\newcommand{\PP}{{\mathcal P}}

\graphicspath{{./data/}}

\title{The spectral dynamics and spatial structures of 
the conditional Lyapunov vector in slave Kolmogorov flow}

\author{Jian Li\aff{1}, Wenwen Si\aff{1}, Yi
Li\aff{2}\corresp{\email{yili@sheffield.ac.uk}}
} 
\affiliation{
  \aff{1}School of Naval Architecture and Maritime, Zhejiang Ocean
University, Zhoushan, 316022, China
  \aff{2}School of Mathematical and Physical Sciences, University of
  Sheffield, Sheffield, S3 7RH, UK 
}

\begin{document}
\maketitle

\begin{abstract}
We conduct direct numerical simulations to investigate the
synchronization of Kolmogorov flows in a periodic box, with a focus on the mechanisms
underlying the asymptotic evolution of infinitesimal velocity
  perturbations, also known as conditional leading Lyapunov vectors.
  This study advances
previous work with a spectral analysis of the perturbation, 
which clarifies the behaviours of the production and dissipation
  spectra at different coupling wavenumbers. We show that, in
  simulations with moderate Reynolds numbers, 
  the conditional leading Lyapunov exponent can be smaller than a lower
  bound proposed previously based on a viscous estimate.   
A quantitative analysis of
the self-similar evolution of the perturbation energy spectrum is
  presented, 
extending the existing qualitative discussion. 
The prerequisites for obtaining self-similar solutions are
  established,
  which include an interesting
relationship between the integral length scale of the perturbation
velocity and the local Lyapunov exponent. 
By examining the governing equation for
  the dissipation rate of the
velocity perturbation, 
we reveal the previously neglected roles of the strain
rate and vorticity perturbations
and uncover their unique geometrical characteristics.  
\end{abstract}

\section{Introduction \label{sect:intro}}

Chaos synchronisation concerns the 
process by which a characteristic of one chaotic system (the
master system) is transmitted to another (the slave system) through
specific coupling mechanisms, thus enabling the slave system to
emulate or replicate the essential properties of the master system. 
The phenomenon \citep{Boccalettietal02}
was initially investigated in the study of
coupled oscillators \citep{FujisakaYamada83}, and has been found
in diverse fields such as communication 
technology, electrical power systems, and biomedical sciences. 
Recently, the topic has garnered 
attention in turbulence research. 

To synchronise turbulent flows, commonly 
two coupling methods are employed: master-slave 
coupling \citep{Lalescuetal13} and nudging coupling
\citep{Leonietal20}. The coupling can be broadly categorised as
unidirectional and bi-directional. In unidirectional coupling,
one flow is influenced by the other, but it does not exert any
influence in return. In bi-directional coupling, the two flows
will influence each other.  Master-slave coupling is unidirectional, 
where part of
the slave system is directly replaced by the corresponding part of
the master system, driving the former towards a complete replica of
the latter. 
In nudging coupling, a nudging
term is introduced, which either drives one system towards the 
other (if the coupling is unidirectional) or enables mutual 
convergence (if the coupling is bi-directional). 
To the best of our knowledge, 
though bi-directional coupling has been used in chaos
synchronisation experiments in other fields \citep{Boccalettietal02}, 
only unidirectional coupling has been investigated 
in turbulent synchronisation. 
Several questions are at the centre
of the research into the synchronisation of turbulent flows. 
The first one is on the threshold
for synchronisation, which usually is in the form of a threshold
coupling strength for nudging coupling or in the form of
a threshold coupling scale for master-slave coupling.
The threshold measures the amount of 
data required to be imparted from the master to the slave to
achieve synchronisation. 
The mathematical literature on this question dates back several decades.
\citet{NikolaidisIoannou22} highlighted these efforts in a way that we find most
accessible. More recently, \citet{OlsonTiti03} and \citet{Henshawetal03} both  
derived analytical bounds for the threshold 
(although for slightly different systems), and observed that the
synchronisation  
could be achieved with much less data in numerical experiments.
\citet{Yoshidaetal05} 
established the criterion $k_c\eta \approx 0.2$ for isotropic 
turbulence with master-slave coupling, where $k_c$ is the
threshold wavenumber and $\eta$ is the Kolmogorov length scale.
\citet{Lalescuetal13} hypothesized that $k_c\eta$ might 
depend on small scale intermittency. 
\citet{Leonietal18} found that large-scale columnar vortices can
enhance the  
synchronisation between rotating turbulence, although recently
\citet{Lietal2024} showed that the forcing terms and the rotational
rates may have larger impacts on the threshold $k_c$.
Another central question is
the relationship between the threshold and the characteristics of the
flows. \citet{Yoshidaetal05} related the threshold to the ratio of the
enstrophy contained in the master modes. \citet{Leonietal20} remarked
that the threshold seemed to coincide with the end of the inertial
range.
\citet{NikolaidisIoannou22} investigated the synchronisation between
two Couette flows by coupling selected streamwise
modes. They showed that synchronisation took place when the conditional leading
Lyapunov exponent (LLE) \citep{Boccalettietal02} was negative, and the threshold
was reached when the conditional
LLE was zero. They also corroborated the observation that the
threshold corresponds to where the inertial range ends. 
\citet{Inubushietal2023} analysed the conditional LLEs of isotropic
turbulence at higher
Reynolds numbers and showed that the threshold depended on the
Reynolds number mildly. Note that the conditional LLE are referred to as
transverse Lyapunov exponent in \citet{Inubushietal2023}.  \citet{WangZaki22}
investigated the synchronisation
between two channel flows by coupling in physical space. They
documented the size and location of the 
coupling regions (i.e., the threshold) required for synchronisation,
and examined their relationship with the time and length
scales of the flow. They also employed Lyapunov exponents to quantify the
decay rate of
synchronisation errors. \citet{Wang2022} looked into non-continuous
coupling and showed that the gap
between episodes of coupling can be increased by one to two
orders of magnitude. This investigation shed lights on the coupling
threshold from another perspective. 
The third central question is on, broadly characterised, imperfect
synchronisation.  
\citet{BuzzicottiLeoni20} and \citet{Lietal22} examined the
synchronisation between large 
eddy simulation and direct numerical simulations (DNS). While the
former applied synchronisation as a way to optimise
subgrid-scale
stress models, the latter focused on the
threshold and synchronisation errors, and they reported
that under certain conditions, the 
standard Smagorinsky model exhibited the smallest synchronisation
error. The impacts of noise in the data were also investigated by
\citet{Lietal22}, an issue that was touched upon in   
\citet{Wang2022} in the context of
channel flows. \citet{VelaMartin21} considered partial synchronisation
of isotropic turbulence coupled below threshold. They argued that
synchronisation is better in strong vortices.  
\citet{Wang2023} fine-tuned the coupling to maximise
synchronisation when only partial synchronisation is achievable.

Related to the question about the threshold for synchronisation mentioned above, an interesting
observation was made by \citet{Lietal2024}, which states that the
energy spectrum of the velocity perturbation of the slave system has a
peak near the threshold wavenumber $k_c$. Same
observation is made in \citet{Lietal24a} for the synchronisation
between large eddy simulations coupled 
via a DNS master. This
observation suggests that there is a non-trivial link
between the velocity perturbation of the slave system, also
known as the leading Lyapunov vector (LLV) \citep{Nikitin2018}, and the
synchronisability of turbulent flows. The aim of present research is to further 
look into the properties of the LLV  
and hopefully shed lights on this
relationship. 
Several previous investigations have cast their eyes on 
the LLV, though not in the context of turbulence
synchronisation. 
\citet{Nikitin2008} looked into properties of the LLV in a
channel flow, in particular its relationship with
the near wall structures of the base flow. The
growth of the LLV was shown to depend crucially on flow inhomogeneity
and the span-wise velocity. 
\citet{Geetal23} analysed the production of the
velocity perturbation in isotropic
turbulence and found, among others, that the energy
spectrum of the perturbation is self-similar 
over a period of time (see also \citet{KatsunoriAriki19}). 
However, one main difference sets current investigation
apart from previous work, which is our focus on the effects of
coupling. From the perspective of
turbulence synchronisation, it is crucial to understand the effects of
coupling, especially its effects
on the spectral dynamics of the velocity perturbations. 
These effects were not covered in previous research. 
We present a systematic investigation of the coupling effects on
the production and dissipation of the LLVs, in both the Fourier space
and the physical space. On top of that, we revisit the self-similar
evolution of the LLVs, which puts previous qualitative discussion on a
firmer footing and leads to new insights.  
The analysis of the
dissipation of the LLVs employs the
transport equation of the dissipation rate for the velocity
perturbation, which allows us to reveal and examine mechanisms
that have been overlooked before.  

In the next section, we present the equations governing various
properties of the velocity perturbations. The numerical methods and
the flow parameters are given in Section \ref{sect:param}. The results
are discussed in Section \ref{sect:rd}. Conclusions are summarised in
Section \ref{sect:con}. 

\section{Governing equations\label{sect:eq}}

We consider the Kolmogorov flow in a triply periodic box
$B=[0,2\pi]^3$ as in our previous work \citep{Lietal22, Lietal24a}. Some relevant
equations and definitions have been given therein, but they are
repeated here for completeness. 
The flow is governed by the incompressible Navier-Stokes 
equations (NSE), which reads 
\begin{equation} \label{eq:nse}
\ptl_t \bu + (\bu \cdot \nabla) \bu = - \nabla p + \bar{\nu} \nabla^2 \bu + \f,
\end{equation}
and the continuity equation
\begin{equation} 
\label{eq:cont}
\nabla \cdot \bu = 0, 
\end{equation}
where $\bu \equiv (u_1, u_2, u_3)$ is the velocity field, $p$ is the 
pressure (divided by the constant density), $\bar{\nu}$ is the viscosity, 
and $\f$ is the forcing term defined by 
\begin{equation}
\f \equiv (a_f\cos k_f x_2,0,0) 
\end{equation}
with $a_f = 0.15$ and $k_f =1$. As is customary in the literature on turbulence, 
it is assumed that the parameters have been
non-dimensionalised with arbitrary length and velocity scales, although for
notational simplicity we do not replace $\bar{\nu}$ with the reciprocal of the
corresponding Reynolds number.  We consider the synchronisation between two
flows governed by
Eqs. (\ref{eq:nse}) and (\ref{eq:cont}), 
where one is labelled the master system $M$ 
and the other the slave systems $S$.
Let $\bu^{(M)}(\bx,t)$ be the velocity of system $M$, and its Fourier 
mode be $ \hat{\bu}^{(M)}(\k,t)$, where $\k$ represents the wavenumber 
vector. $\bu^{(S)}$ and $\hat{\bu}^{(S)}$ are defined similarly for system $S$. 
The two systems are simulated concurrently.
System $S$ is driven by system $M$ via one-way master-slave coupling \citep{Yoshidaetal05}. 
Specifically, at every time step, we replace the Fourier modes 
of $\hat{\bu}^{(S)}(\k,t)$ with $\vert \k \vert \le k_m$ 
by their matching counterparts from $\hat{\bu}^{(M)}(\k,t )$, where
$k_m$ is termed the coupling wavenumber. This coupling modifies system
$S$ by enforcing
\begin{equation} \label{eq:abm}
\hat{\bu}^{(S)}(\k,t) = \hat{\bu}^{(M)}(\k,t), 
\end{equation} 
for $\vert \k \vert \le k_m$ at all time $t$, but system $M$ is not
affected by system $S$. The Fourier modes of the
two systems with $\vert \k \vert \le k_m$ are called 
the \emph{master modes}, while those in system $S$ with $\vert \k
\vert > k_m$ are called the \emph{slave modes}.

When $k_m$ is sufficiently large, system $S$ will be synchronised to
system $M$ exactly as $t\to \infty$, as was shown in isotropic turbulence
\citep{Yoshidaetal05} 
and rotating turbulence in a periodic box \citep{Lietal2024}. 
The smallest $k_m$ for which synchronisation occurs defines 
the threshold wavenumber, denoted as $k_c$. 

The synchronisation process has been 
analysed using the LLEs, the conditional LLEs and the LLVs of the
flows previously. Synchronisation takes place when the conditional LLE is
negative, as having been shown for channel flows \citep{NikolaidisIoannou22,
WangZaki22}, isotropic turbulence \citep{Inubushietal2023}, and rotating
turbulence \citep{Lietal2024}. The conditional LLEs are defined in such a way that they measure
the mean growth rate of the perturbation applied
specifically to the slave modes. 
Let $\bu$ be the velocity of a slave system, 
and $\bu^\delta$ be an infinitesimal perturbation to the 
\emph{slave modes} of $\bu$, where $\bu$ is  also referred to as the base flow
in the analysis of $\bu^\delta$. 
Since the \emph{master modes} are not perturbed, we have, by definition,
\begin{equation} \label{eq:cle0}
\hat{\bu}^\delta(\k,t) = 0 \quad \text{for}\quad \vert \k \vert \le k_m. 
\end{equation}
The perturbation $\bu^\delta$ is governed by the linearised NSE 
\begin{equation} \label{eq:udel1}
\ptl_t \bu^\delta + (\bu \cdot \nabla) \bu^\delta + (\bu^\delta \cdot \nabla) \bu = - \nabla 
p^\delta + \bar{\nu} \nabla^2 \bu^\delta + \bm f^\delta,
\end{equation}
and the continuity equation
\begin{equation} \label{eq:udel2}
\nabla \cdot \bu^\delta = 0,
\end{equation}
where $p^\delta$ and $\bm f^\delta$ are the pressure perturbation and the
forcing perturbation, respectively.
Although $\bm f^\delta$ is included for completeness, in practice it
is zero as $\bm f$ is a constant, 
and will be dropped from now on.
The conditional LLE corresponding to coupling wavenumber $k_m$, denoted by $\lambda(k_m)$, is defined as
\citep{Boccalettietal02, NikolaidisIoannou22, Inubushietal2023} 
\begin{equation} \label{eq:lyadef}
\lambda(k_m) = \overline{\lim_{t\to \infty}}
\frac{1}{t}\log \frac{\Vert \bu^\delta(\x, t+t_0)\Vert}{\Vert \bu^\delta (\x, t_0)\Vert},
\end{equation}
where $\bu^\delta (\x,t_0)$ is the perturbation at an arbitrary initial time $t_0$, and 
$\Vert \cdot \Vert$ denotes the $L^2$ norm, defined for an arbitrary vector field $\bm w$ as
\begin{equation}
\Vert \bm w \Vert \equiv  \lal \bm w \cdot \bm w \ral_v^{1/2},
\end{equation}
with $\lal ~ \ral_v$ denoting spatial average, i.e., 
\begin{equation}
\lal \bm w \cdot \bm w \ral_v = \frac{1}{(2\pi)^3}\int_{[0,2\pi]^3} \bm w \cdot \bm w dV.
\end{equation}
Though mathematically $\lambda(k_m)$ depends on the initial perturbation $\bu^\delta(\x,t_0)$, in practice a randomly chosen $\bu^\delta(\x,t_0)$ will almost surely lead to the same $\lambda(k_m)$. Therefore, we assume $\lambda(k_m)$ to be independent of $\bu^\delta(\x,t_0)$ in what follows.  
The velocity perturbation $\bu^\delta$, as $t\to \infty$, 
will also be called the conditional LLV where appropriate, 
extending the terminology of (unconditional) LLV 
used in \citet{Nikitin2018}. The conditional LLVs have not
received as much attention as the conditional LLEs.   
We will focus on the conditional and unconditional LLVs and their
relationship with the conditional LLEs in this study.
To analyse the conditional LLVs, it is useful to explore some of the consequences of 
the linearised NSE. An immediate result of Eqs.
(\ref{eq:udel1}) and (\ref{eq:udel2}) 
is the
equation for the energy of $\bu^\delta$, defined as 
\begin{equation}
K_\delta (t) \equiv \frac{1}{2}\Vert \bu^\delta \Vert^2,
\end{equation}
It is not difficult to show that
\begin{equation} \label{eq:udelnorm}
\frac{d K_\delta}{dt} = \lal \PP \ral_v - \lal \DD
\ral_v, 
\end{equation}
where 
\begin{equation} \label{eq:pdf}
\PP \equiv - u_i^\delta u_j^\delta s_{ij},\qquad 
\DD \equiv \bar{\nu} \ptl_j u_i^\delta \ptl_j u_i^\delta, ~~
\end{equation}
are the instantaneous production and the viscous dissipation density,
respectively, and 
$s_{ij} = (\ptl_j u_i + \ptl_i u_j)/2$ is the rate of strain tensor for 
the base turbulent flow. Eq. (\ref{eq:pdf}) shows that the
velocity perturbation is produced by the straining effects of the
base flow whereas it is destroyed by viscous dissipation
associated with its gradient. 
Eq. (\ref{eq:udelnorm}) has been used 
previously (see, e.g., \citet{Lietal2024, NikolaidisIoannou22, WangZaki22, 
Inubushietal2023, Geetal23}). It
provides an alternative way to calculate the conditional LLEs.
We introduce the normalised  velocity perturbation $v_i = u^\delta_i/\Vert \bu^\delta\Vert$, and 
let
\begin{equation}
\PP_e = - v_i v_j s_{ij}, \qquad \DD_e = \bar{\nu} \ptl_j v_i \ptl_j v_i.
\end{equation} 
We then obtain from Eq. (\ref{eq:udelnorm})
\begin{equation}
\frac{d \ln \Vert \bu^\delta \Vert}{dt} = \lal \PP_e \ral_v - \lal \DD_e
\ral_v, 
\end{equation}
which, upon integrating over time, leads to 
\begin{equation} \label{eq:lyadef2}
\lambda (k_m) =\overline{\lim_{t\to\infty}}\frac{1}{t}
\int_{t_0}^{t+t_0} (\lal \PP_e\ral_v - \lal \DD_e\ral_v) dt.
\end{equation}
Using $\lal ~ \ral$ to denote the combination of 
spatial and temporal averages, we obtain
\begin{equation} \label{eq:lyadef3}
\lambda (k_m) =\lal \PP_e\ral - \lal \DD_e\ral.
\end{equation}
The rate of change $d \ln \Vert \bu^\delta\Vert/dt$ is called the local
conditional LLE, which is denoted as $\gamma(k_m,t)$, i.e.,
\begin{equation}
\gamma(k_m,t) \equiv \frac{d  \ln \Vert \bu^\delta \Vert}{dt} = \lal \PP_e \ral_v -
\lal \DD_e \ral_v. 
\end{equation}
Obviously, the conditional LLE is the long time average of $\gamma$, i.e., $\lm(k_m) = \lal
\gamma(k_m,t)\ral$. 

The expressions for $\PP_e$ and $\DD_e$ show that the LLEs
(conditional or unconditional) crucially depend on 
the spatial structures of $\bu^\delta$
and its correlation with the base flow, which can be understood from
the equation for $v_i$. The equation for $v_i$ reads
\begin{equation} \label{eq:vi}
	\ptl_t v_i + u_j \ptl_j v_i = - v_j \ptl_j u_i - 
 \ptl_i p_e + \bar{\nu} \nabla^2 v_i - \gamma v_i ,
\end{equation}
where $p_e \equiv p^\delta/\Vert \bu^\delta\Vert$.
The transport equation for the kinetic energy 
$e\equiv v_i v_i/2$ follows readily:
\begin{equation} \label{eq:e}
    \ptl_t e + u_j \ptl_j e = \PP_e - \ptl_i (p_e v_i) - \DD_e 
    + \bar{\nu} \nabla^2 e - 2 \gamma e.
\end{equation}
Note that, from the definition of $v_i$, one can show that $\lal e \ral_v = 1/2$.

The small scale spatial structure of 
$\bv$ can be studied using its gradient
$B_{ij}  \equiv \ptl_j v_i$. The expression for the dissipation
rate $\DD_e$ shows that $B_{ij}$ is a
crucial quantity. 
The equation for $B_{ij} $ is:
\begin{equation} \label{eq:bij}
\ptl_t B_{ij} + u_\ell \ptl_\ell B_{ij} 
= - B_{i\ell}  A_{\ell j} - A_{i\ell}B_{\ell j} - \ptl^2_{ij} p_e
+ \nabla \cdot \left( \bar{\nu} \nabla B_{ij}
+ \bv A_{ij}\right) - \gamma B_{ij},
\end{equation} 
with $A_{ij} \equiv \ptl_j u_i$ being the velocity gradient of the base flow
$\bu$. The first two terms on the right hand side of the equation represent the
interaction between the gradients $B_{ij}$ and $A_{ij}$,
which is a key mechanism by
which $B_{ij}$ is amplified. The other terms on the right hand
side of Eq. (\ref{eq:bij})
are the pressure
Hessian \citep{Meneveau11}, the transport term, and 
the damping due to normalisation in $\bv$. 
Introducing
the strain rate tensor 
$s_{ij}^v$ and the vorticity $\omega_i^v$ of $\bv$, where $s_{ij}^v = (\ptl_j
v_i + \ptl_i v_j)/2$ and $\omega_i^v = \varepsilon_{ijk} \ptl_j v_k$ with
$\varepsilon_{ijk}$ being the Levi-Civita symbol, we obtain
\begin{align} 
\ptl_t  \DD_e  + u_\ell \ptl_\ell 
\DD_e = & \NN_e - 2\gamma \DD_e + \bar{\nu} \nabla^2  \DD_e 
- 2\bar{\nu} s^v_{ij} \ptl^2_{ij} p_e \notag \\ 
- & 2\bar{\nu}^2 \nabla B_{ij} \cdot \nabla B_{ij} 
 + 2\bar{\nu} B_{ij}  \nabla \cdot (\bv A_{ij}),\label{eq:bb}
\end{align} 
where $\NN_e$ represents the interaction terms:
\begin{equation} 
\NN_e = \underbrace{-4 \bar{\nu} s_{ij}^v s_{jk} s_{ki}^v}_{\NN_{es}} +
\underbrace{\bar{\nu} s_{ij} \om^v_i \om^v_j}_{\NN_{eo}},
\end{equation} 
with 
$\NN_{eo}$ representing the vortex stretching effect and  
$\NN_{es}$ the interactions between the base flow and
perturbation strain rate
tensors. 
The equation for the mean dissipation $\lal \DD_e \ral$ is 
\begin{equation} \label{eq:mBB} 
2 \lal \gamma
\DD_e \ral  = \lal \NN_e \ral - 2\bar{\nu} \lal s^v_{ij} \ptl^2_{ij} p_e \ral
  - 2\bar{\nu}^2 \lal \nabla B_{ij} \cdot \nabla B_{ij} \ral + 2\bar{\nu} \lal
B_{ij} \nabla \cdot (\bv A_{ij})\ral .
\end{equation}
The periodic boundary condition has been applied to arrive at 
the above equation. Eq.
(\ref{eq:mBB}) provides a breakdown of the contributions to the
mean dissipation. To keep the scope of current research manageable, we 
will focus on the interaction term $\NN_e$ in what follows. 

One can gain insights into the spectral properties of $\bu^\delta$ 
from the energy spectrum of $\bu^\delta$, $E_\delta(k,t)$, 
defined as
\begin{equation}
E_\delta (k,t) = \frac{1}{2}\sum_{k -\frac{1}{2}\le \vert \k \vert \le k+\frac{1}{2}} 
\hat{\bu}^\delta(\k,t) \cdot \hat{\bu}^{\delta *}(\k,t),
\end{equation} 
where $\hat{\bu}^\delta(\k,t)$ is the Fourier mode for $\bu^\delta$ with
wavenumber $\k$ and we have used $\hat{~~}$ to represent the Fourier transform 
and $^*$ to represent complex conjugate. Similarly, one can
consider the spectrum of $\bv$, defined as
\begin{equation} \label{eq:Ekv_def}
E_v (k,t) = \frac{1}{2}\sum_{k -\frac{1}{2}\le \vert \k \vert \le k+\frac{1}{2}} 
\hat{\bv}(\k,t) \cdot \hat{\bv}^{ *}(\k,t) = \frac{1}{2K_\delta(t )} E_\delta(k,t).
\end{equation} 
Taking the Fourier transform of Eq. (\ref{eq:vi}), 
we find, after some simple algebraic manipulations,
the equation for $E_v(k,t)$, which reads
\begin{equation} \label{eq:Ekv}
\ptl_t E_v(k,t) = \PP_v(k,t) - \DD_v(k,t) - 2 \gamma(t) E_v(k,t),
\end{equation} 
where 
\begin{equation} \label{eq:pvkdvk}
\PP_v(k,t) = \sum_{k-\frac{1}{2} \le \vert \bk\vert \le k+\frac{1}{2}} \hat{P}_v(\bk,t), ~~
\DD_v(k,t) = 2 \bar{\nu} k^2  E_v (k,t),
\end{equation}
with $\hat{P}_v(\bk,t)$ given by
\begin{align} 
\hat{P}_v(\bk,t) = k_n \left(\delta_{i\ell} - \frac{k_i k_\ell}{k^2}\right) 
  \Re &\left[ \imath \sum_{\bm{q}} \hat{v}_\ell ^{*} (\bm{q},t)  
\hat{u}_n^* (\bk - \bm{q},t) \hat{v}_i (\bk,t)\right.  \notag \\   
  + & \phantom{[} \left. \imath \sum_{\bm{q}} \hat{v}_n ^{*} (\bm{q},t)  
  \hat{u}_\ell^*(\bk - \bm{q},t) \hat{v}_i(\bk,t) \right] ,
\label{eq:pvkh}
\end{align}
where $\delta_{i\ell}$ is the Kronecker delta tensor, $\imath$ is the imaginary unit, 
and $\Re$ indicates the real part. The summation $\sum_{\bm q}$ is taken over
all Fourier modes with wavenumber $\bm q$. 
Eq. (\ref{eq:Ekv}), together with Eqs. (\ref{eq:pvkdvk}) and
(\ref{eq:pvkh}), forms the basis of the spectral analysis of the
LLV
$\bv$. It is easy to see that
\be
 \lal \PP_e \ral_v = \int_0^\infty \PP_v(k,t) d k , \quad
\lal \DD_e \ral_v = \int_0^\infty \DD_v(k,t) d k, 
\ee
and 
\be 
\quad \lal e
\ral_v = \int_0^\infty E_v(k,t) d k = \frac{1}{2}.
\ee 

\section{Parameters and numerics \label{sect:param}}
        
\begin{table} 
\begin{center}
\def~{\hphantom{0}}
		\begin{tabular}{cccccccccc}
			Case & $N$ & $Re_\lambda$ & 
   $u_{\rm rms}$ & $\epsilon$ & $\eta$ & $\bar{\nu}$ & $\lambda_a$ & $\tau$ & $k_\text{max}\eta$ \\
			\hline
			R1 & 128 & 75 & 0.63 & 0.072 & 0.042 & 0.0060 & 0.71& 0.30 & 1.79\\ 
			R2 & 192 & 90 & 0.65 & 0.074 & 0.033 & 0.0044 & 0.61& 0.24 & 2.11\\ 
			R3 & 256 & 112 & 0.66 & 0.077 & 0.024 & 0.0030 & 0.51& 0.20 & 2.05\\ 
			R4 & 384 & 147 & 0.66 & 0.076 & 0.016 & 0.0017 & 0.38& 0.15 & 2.05\\
      F1 & 256 & 75 & 0.64 & 0.072 & 0.042 & 0.0060 & 0.71 & 0.29 &  3.50
		\end{tabular}
	\caption{\label{tab:cases} Parameters for the simulations.
		$N^3$: the number of grid points.
		$u_{\rm rms}$: root-mean-square velocity. 
		$\bar{\nu}$: viscosity. 
		$\epsilon$: mean energy dissipation rate of the base flow.
		$\lambda_a$: Taylor length scale.
		$Re_\lambda =  u_{\rm rms}\lambda_a/\bar{\nu}$: the Taylor-Reynolds number. 
		$\eta =  (\bar{\nu}^3/\epsilon)^{1/4}$: Kolmogorov length scale.  
		$\tau= (\bar{\nu}/\epsilon)^{1/2}$: Kolmogorov time scale. $k_\text{max}$: the maximum effective wavenumber. 
	}
\end{center}	
\end{table}

The NSE and the continuity equation 
are solved in the Fourier space numerically with the pseudo-spectral
method. The two-thirds rule \citep{Pope00} is used to de-aliase the  
advection term so that the maximum effective wavenumber is $k_\text{max} = N/3$, 
where $N^3$ is the number of grid points.
Time stepping uses an explicit second order Euler scheme. 
The viscous diffusion term is treated with an integration factor. More
details about the numerical methods can be found in 
\citet{Lietal2024}.
The step-size $\delta t$ is chosen in such a way that the 
Courant-Friedrichs-Lewy number $ u_{\rm rms} \delta t/\delta x$ is
less than $0.1$, where $\delta x  = 2\pi/N$ is the grid
size, and $u_{\rm rms}$ is the root-mean-square velocity defined by 
\begin{equation}
u_{\rm rms} = \left(\frac{2}{3} \int_0^\infty \lal E\ral (k) dk
  \right)^{1/2} 
\end{equation} 
where $\lal E \ral(k)$ is the average energy spectrum of the DNS base
flow, defined as
\begin{equation} 
\lal E\ral (k) = \frac{1}{2} \sum_{k-\frac{1}{2} \le \vert \k \vert \le k+\frac{1}{2}}
\lal \hat{\bu}^*(\k, t)
\cdot \hat{\bu}(\k,t)\ral . 
\end{equation}
The mean energy dissipation rate $\ep$ and the Taylor micro-scale
$\lambda_a$
are defined in the usual way. That is,
\begin{equation}
\epsilon = \int_0^\infty 2\bar{\nu} k^2 \lal E\ral (k)dk, \qquad
  \lambda_a = \left(\frac{15\bar{\nu} u_{\rm rms}^2}{\ep}\right)^{1/2}.
\end{equation}
Further parameters can be calculated from the above key
quantities, including the Kolmogorov length scale  $\eta\equiv
(\bar{\nu}^3/\epsilon)^{1/4}$ and the Kolmogorov time scale 
		$\tau\equiv (\bar{\nu}/\epsilon)^{1/2}$.
The values of these parameters 
are summarised in Table \ref{tab:cases}.
Mainly four groups of DNS with different Reynolds numbers are 
conducted and analysed, which
are called groups R1, R2, R3, and R4, respectively. 
Each group includes several simulations with different 
coupling wavenumber $k_m$. 
Where necessary, we append `K$b$' to 
differentiate such simulations, where $b$ is the value of $k_m$.
For example, R$3$K$11$ refers to the case with $k_m = 11$ and
$Re_\lambda = 112$. The resolution of the simulations is indicated 
by the value of $k_\text{max} \eta$. Its values are 
above the recommended minimum value $1.5$ \citep{Pope00} in all cases, as 
one can see in Table \ref{tab:cases}. 
Furthermore, an additional DNS with 
$k_\text{max} \eta \approx 3.50$ is conducted  
to verify the conditional LLEs results calculated from group 
R1 (c.f. Fig. \ref{fig:energy_decay}). 
This simulation is labelled F1 in Table \ref{tab:cases}, 
where the parameters of the simulation are also recorded.

The computation of the conditional LLEs follows the algorithm
as outlined in, e.g., \citet{BoffettaMusacchio17}, with specific
implementation given in our previous work \citep{Lietal2024}.
The detail is thus not repeated here. One key aspect of the
algorithm is that $\bu^\delta$ is approximated by the difference
between two DNS velocity fields evolving from very close initial
conditions, and the difference is re-scaled
periodically to keep it small so that it can be treated as
infinitesimal at all times. 

The computation of $\hat{P}_v (\k, t)$ uses
the following alternative expression for the quantity:
\[\hat{P}_v(\bm k,t) = k_n \left(\delta_{i\ell} - k_i k_\ell/k^2\right) \Re \left\{\imath
\hat{N}^*_{\ell n}(\bm k, t) \hat{v}_i(\bm k, t)\right\}\]
with 
\[\hat{N}_{\ell n}(\bm k,t) \equiv \widehat{v_\ell u_n}(\bm k,t) + 
\widehat{v_n u_\ell}(\bm k, t), 
\]
where $\hat{N}_{\ell n}$ is calculated using the pseudo-spectral method. 
Some data points are also calculated using 
Eq. (\ref{eq:pvkh}) as a way to cross check the results. As only 
negligible differences are found, these data have been omitted for clarity. 

\section{Results and discussion \label{sect:rd}}

\subsection{Basic statistics}
\bfig
\centering
\ig[width=0.55\lnw]{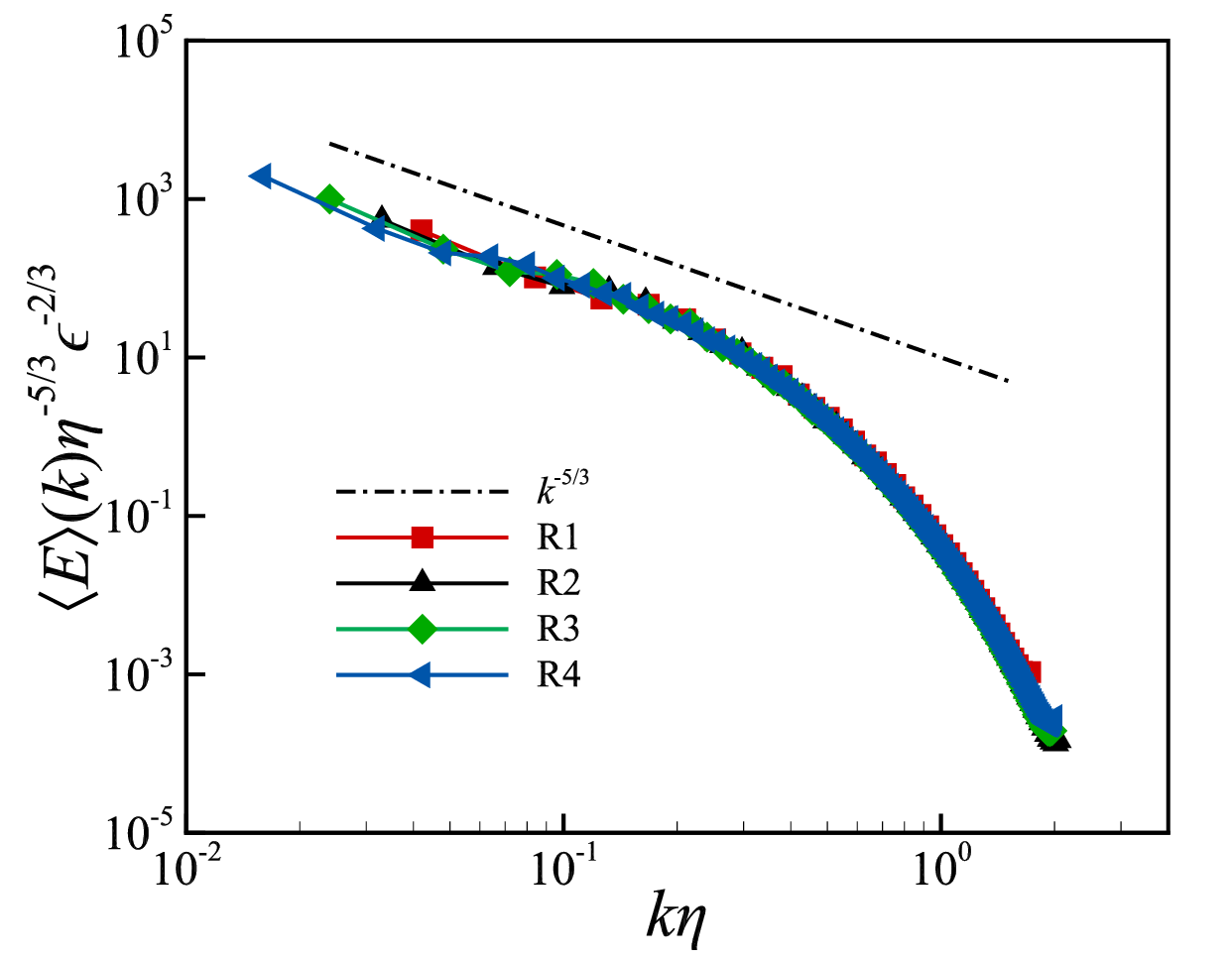}
\caption{\label{fig:Ek_keta} Mean energy spectra $\lal E
\ral(k)$ of the base flow. Dash-dotted line: the 
$k^{-5/3}$ power law.} 
\efig 

\bfig
\centering
\ig[width=0.48\lnw]{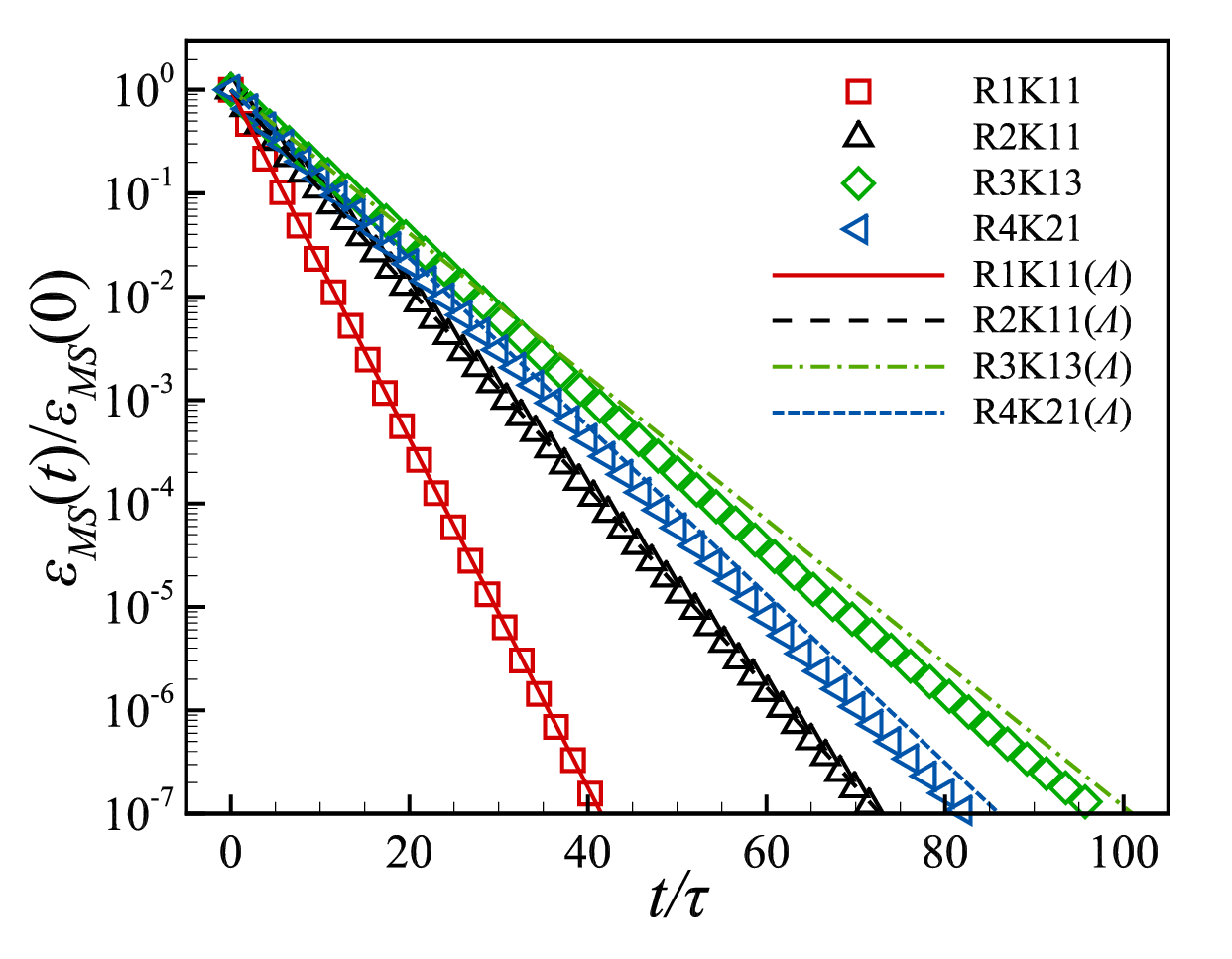} 
\ig[width=0.48\lnw]{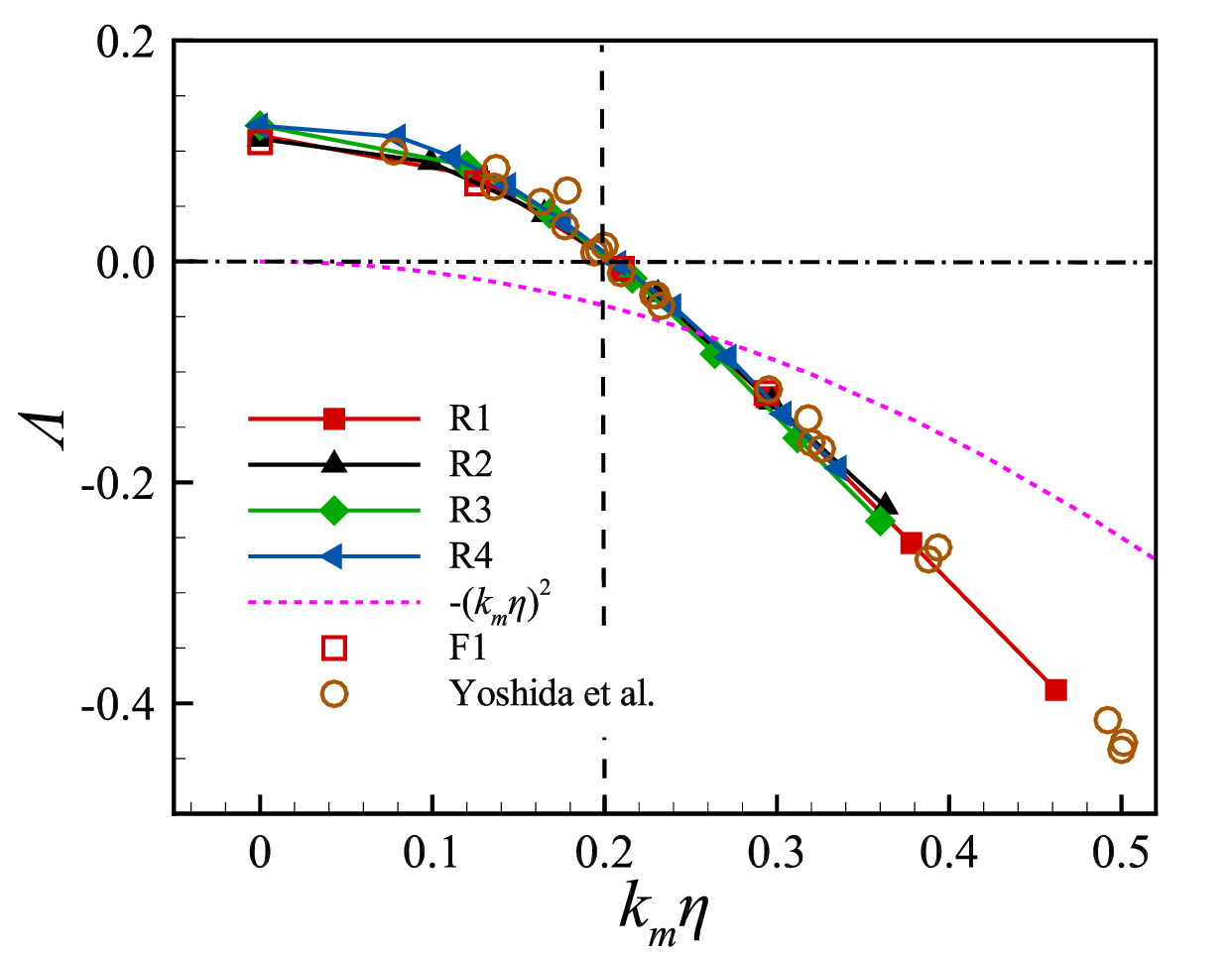}
\caption{\label{fig:energy_decay} Left: Comparison between the decay rates
of the synchronisation error $\EE_{MS}(t)$ and the conditional LLEs. 
Right: Normalised conditional LLEs 
$\Lambda \equiv \lambda \tau$, with the short-dashed line
showing $-(k_m \eta)^2$ and the vertical line marking
$k_m\eta=0.2$. Only four data points are shown for the case F1. The empty circles 
are the decay rate data from \citet{Yoshidaetal05} (see the main text
for more detail).  }
\efig

\bfig
\centering
\ig[width=0.55\lnw]{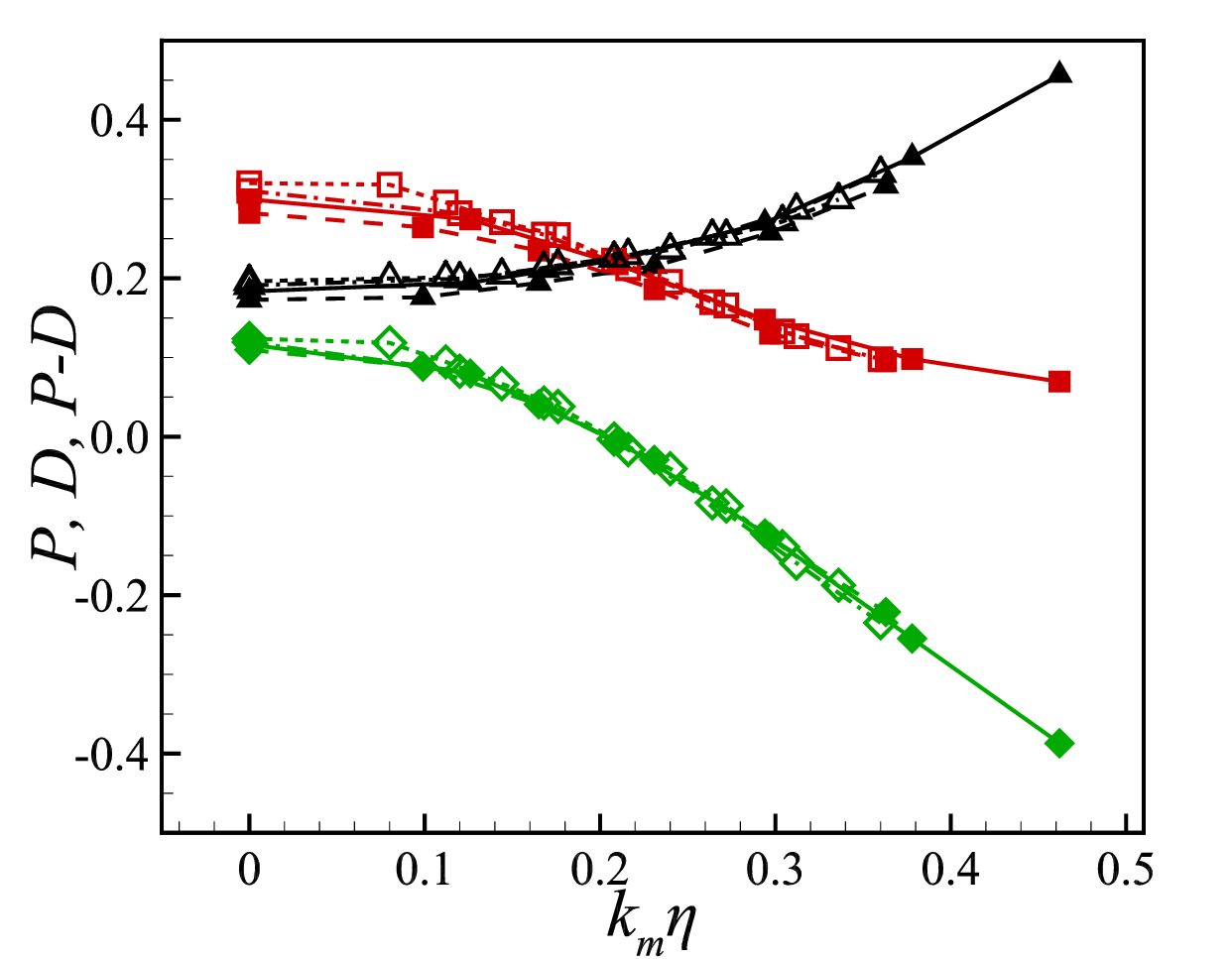}
\caption{\label{fig:P_D_P-D_kmeta} The production $P$,  
dissipation $D$ and $P - D$ as functions of $k_m\eta$.
$P$: lines with both solid and empty squares; $D$: lines with both solid 
and empty triangles; $P-D$: lines with both solid and empty diamonds.
Solid lines: R1; long dashed lines: R2; dash-dotted lines: R3;
short-dashed lines: R4.  }
\efig

We start with a few results that characterise the basic properties of
the flows and the synchronisation process. For reference, 
the average energy spectra are documented in Fig. \ref{fig:Ek_keta},
which is compared with the $k^{-5/3}$ scaling law.
To monitor the synchronisation between $\bu^{(M)}$ and $\bu^{(S)}$, we
use the synchronisation error
\begin{equation} \label{eq:error}
\EE_{MS}(t) = \Vert \bu^{(M)}-\bu^{(S)}\Vert,
\end{equation}
which will decay exponentially 
when the two flows synchronise, and the
rate of decay of $\EE_{MS}$ is related to the conditional LLE $\lambda(k_m)$, as having been shown in 
\citet{Henshawetal03, Yoshidaetal05, Inubushietal2023, Lietal2024}.
The left panel of Fig. \ref{fig:energy_decay} compares the
results for $\EE_{MS}(t)$ 
(symbols) with $\exp(\Lambda t/\tau)$ (lines), 
where 
\begin{equation} \label{eq:Lam}
\Lambda (k_m) \equiv \lambda(k_m) \tau
\end{equation} 
is the conditional LLE non-dimensionalised with $\tau$.
Some small discrepancies can be seen between the two quantities, 
which we attribute to statistical uncertainty. 
This result is consistent with previous findings
\citep{NikolaidisIoannou22,Lietal2024}, which shows
that the non-dimensional decay rate of the synchronisation error
can be given by
$\Lambda$. 

The right panel of Fig. \ref{fig:energy_decay} shows 
$\Lambda$ as functions of $k_m\eta$. The conditional LLEs are known to
depend on the Reynolds number weakly (see, e.g.,
\citet{Inubushietal2023}). Here the curves for different $Re_\lambda$
show some differences at small $k_m\eta$, but they collapse on each
other for larger $k_m \eta$ (for, e.g., $k_m\eta \ge 0.15$)
because $Re_\lambda$ varies
only mildly. The conditional LLEs decrease as $k_m \eta$
increases, but the variation appears to be small for small $k_m\eta$. 
The threshold wavenumber $k_c$, for which $\Lambda = 0$, is found 
to be $k_c \eta \approx 0.2$, which is also the value obtained in
\citet{Yoshidaetal05} for isotropic turbulence. The dashed line
without symbols is the function
$-(k_m\eta)^2$, which is an
estimate of the non-dimensional conditional LLE when the
evolution of the velocity perturbation is
determined solely by viscous diffusion (see, e.g.,
\citet{Inubushietal2023}). We will discuss this estimate below
together with Fig. \ref{fig:P_D_P-D_kmeta}. 

The right panel of 
Fig. \ref{fig:energy_decay} also 
includes data from two other sources to cross check the results from groups R1-R4. 
The conditional LLEs from 
case F1 at four different $k_m \eta$ are plotted with empty squares. 
They display only 
negligible differences with those from R1, which shows that 
the results are essentially grid independent. 
The empty circles are 
calculated from the decay rates $\tilde{\alpha}$ obtained 
in \citet{Yoshidaetal05}, plotted in Fig. 3 therein. 
More specifically, the empty circles 
are the values of $-\tilde{\alpha}/2$. Per \citet{Yoshidaetal05},
 $\tilde{\alpha}$ is defined by 
$\Vert \bu^{(M)} - \bu^{(S)}\Vert^2 \sim \exp(- \tilde{\alpha} t/\tau)$. 
Since $\Vert \bu^{(M)} - \bu^{(S)}\Vert \sim \exp(\Lambda t/\tau)$ as is shown 
in the left panel of Fig. \ref{fig:energy_decay}, we 
expect $\Lambda = - \tilde{\alpha}/2$. This relation is verified by the data, 
as the empty circles fall closely on the 
curves for $\Lambda$. Note that 
the empty circles correspond to 
a wide range of Reynolds numbers. 
Also, $\tilde{\alpha}$ is obtained by measuring 
the decay rate of $\Vert \bu^{(M)} - \bu^{(S)}\Vert^2$, 
which is a procedure that is very different from how $\Lambda$ is calculated
\citep{Yoshidaetal05}. 
Thus,
the agreement between $\Lambda$ and $-\tilde{\alpha}/2$ is a strong 
validation of our results. 

Further insights on
$\Lambda$ can be explored according to 
Eq. (\ref{eq:lyadef3}). We use  
\begin{equation}
P \equiv   \tau\lla \PP_e \rra , \qquad D \equiv
  \tau\lla \DD_e \rra
\end{equation}
to denote the non-dimensional mean production and mean dissipation,
respectively. It follows from Eq. (\ref{eq:lyadef3}) that  
\begin{equation} \label{eq:lampd}
\Lambda =  P - D. 
\end{equation}
Note that a pre-requisite for Eq.
(\ref{eq:lampd}) is that the external forcing has no impact on the
evolution of the velocity perturbation (c.f. Eq. (\ref{eq:udel1}) and
the comments following Eq. (\ref{eq:udel2})). Otherwise, the
equation would contain a forcing term. 

The values of $P$, $D$, and their difference $P-D$ are shown 
in Fig. \ref{fig:P_D_P-D_kmeta} for different coupling wavenumber
$k_m$. As dictated by Eq. (\ref{eq:lampd}), the curves for $P-D$
agree precisely with those of the conditional LLEs 
$\Lambda$ shown in Fig. \ref{fig:energy_decay} (the right panel thereof). 
The main observation about Fig. \ref{fig:P_D_P-D_kmeta} is that 
$P$ decreases, whilst $D$ increases, 
with increasing $k_m$. The two curves intersect 
at the threshold wavenumber $k_c \eta$. Both seem to contribute 
roughly equally to the change in $\Lambda$ as $k_m$ varies. The
results are slightly 
different for different $Re_\lambda$. 

We now return to the discussion of the estimate $-(k_m\eta)^2$
for $\Lambda(k_m)$ shown in Fig. \ref{fig:energy_decay}. It was found
in \citet{Inubushietal2023} 
that $\Lambda(k_m)$ is always larger than $-(k_m\eta)^2$, and
approaches $-(k_m \eta)^2$ from above (see Fig. 3 therein). These
trends suggest that $P$ becomes negligible for large $k_m\eta$, and
$D$ approaches the pure viscous estimate as $k_m\eta$ increases.
Our results, on the other hand, appear
to display different behaviours, as we can see from Figs.
\ref{fig:energy_decay} and \ref{fig:P_D_P-D_kmeta}. Firstly, Fig.
\ref{fig:energy_decay} shows that $-(k_m\eta)^2$ can be larger than
$\Lambda(k_m)$ in our simulations. Secondly, Fig.
\ref{fig:P_D_P-D_kmeta} shows that,
though $P$ decreases as $k_m\eta$
increases, it is still significantly larger than zero for the largest
$k_m \eta$, even when $\Lambda(k_m )$ is already smaller than $-(k_m
\eta)^2$. In sum, the dissipation contribution in our
simulations is significantly higher than what is implied by the estimate
$-(k_m\eta)^2$. The cause for the difference is unclear. An
explanation possibly, though unlikely, lies in 
the difference in the Reynolds numbers. We will discuss this further in Section
\ref{sect:spec} when we discuss the spectral dynamics of the
LLVs.

\bfig
\centering 
\ig[width=0.48\lnw]{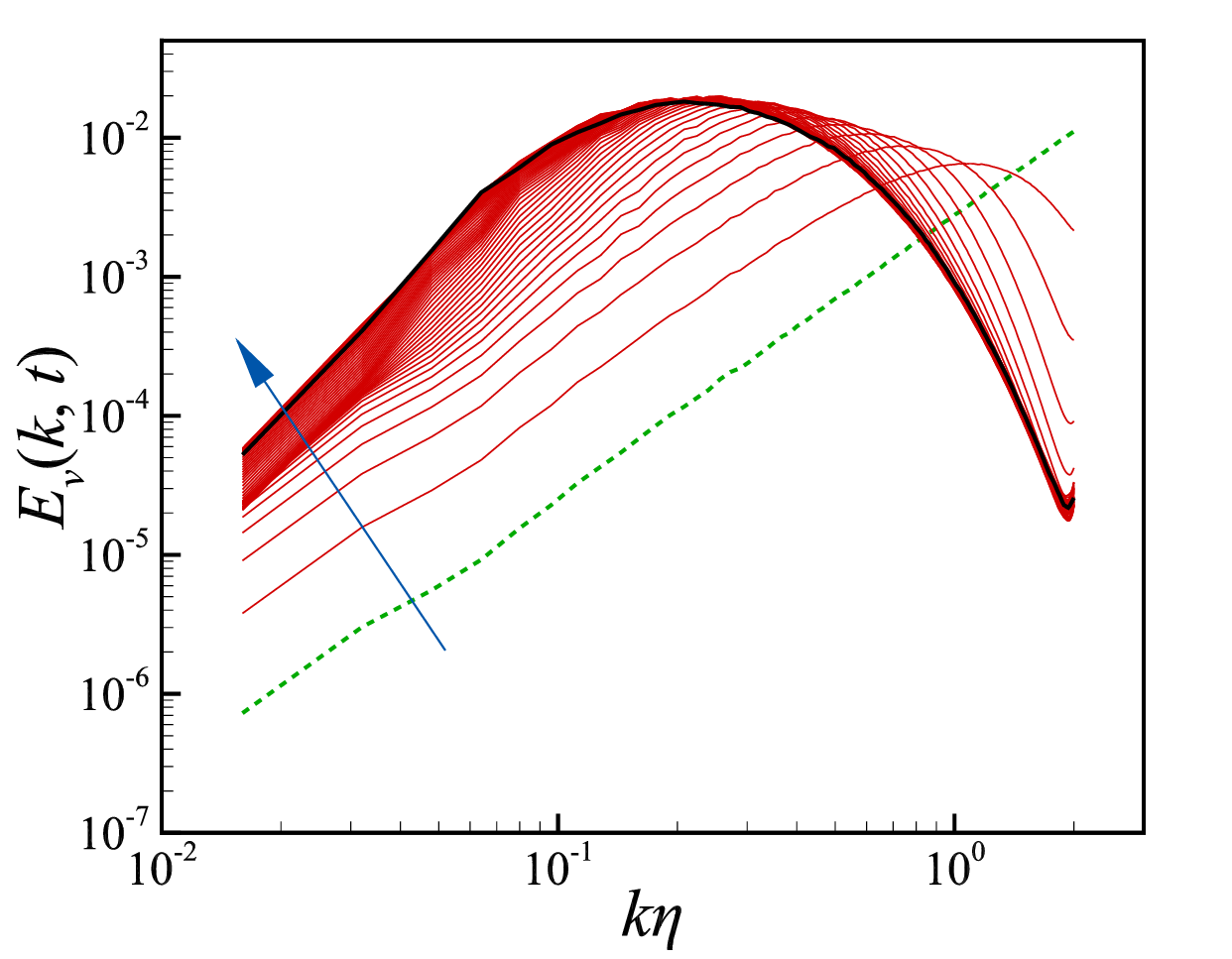} 
\ig[width=0.48\lnw]{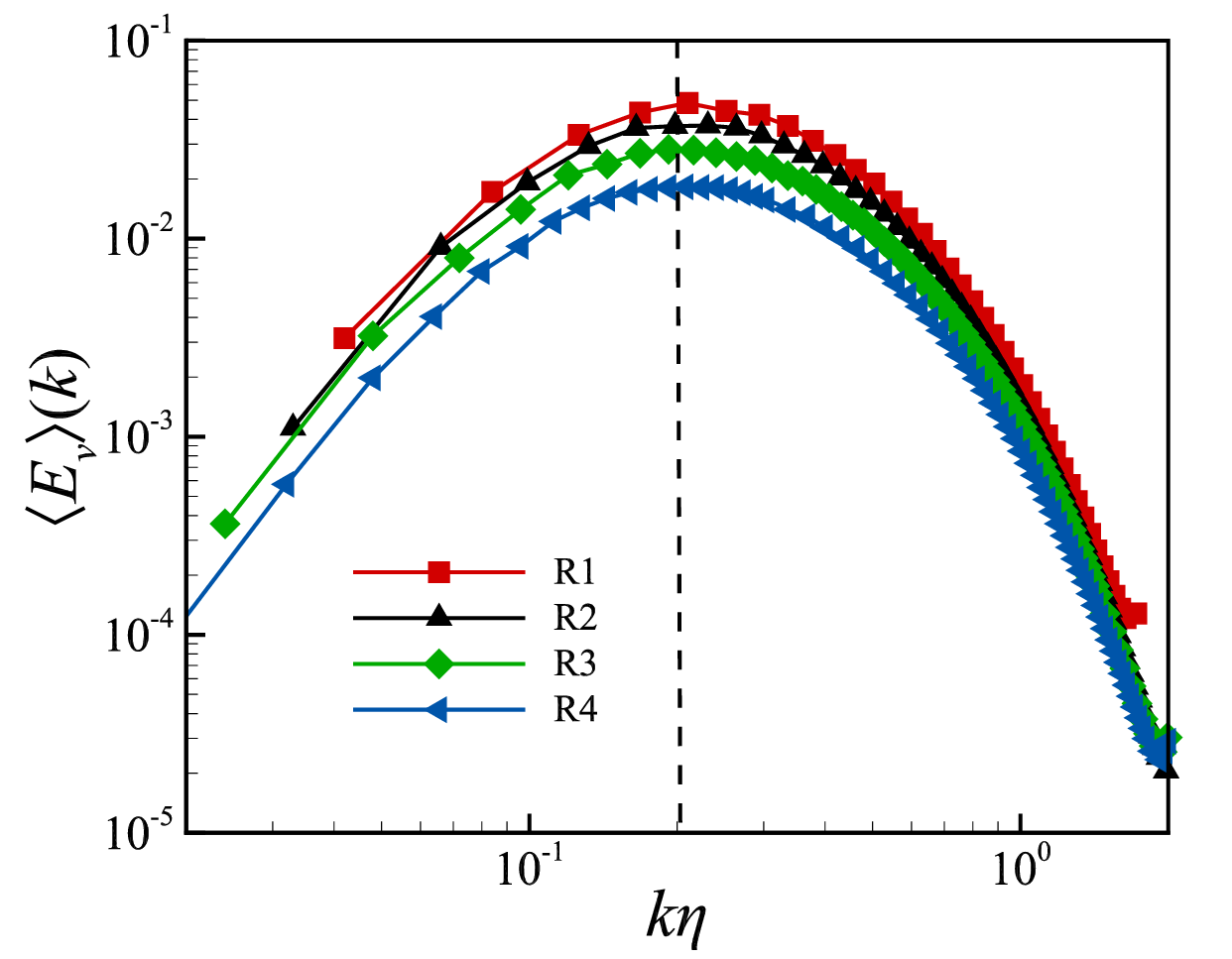}
\caption{\label{fig:Ekv0} Left: Short time evolution of
$E_v(k,t)$ for $t$ between $0$ and approximately $ 20 \tau$, with the dashed line showing
$E_v(k,t=0)$ and the black thick line showing $E_v(k,t)$ for $t
\approx 20\tau$. The arrow indicates the 
increasing direction of time. Right:
average
energy spectra $\lal E_v\ral$ for the unconditional LLV, with the
vertical dashed line marking the value $k\eta=0.2$.}
\efig

Finding the threshold wavenumber $k_c$ via the conditional
LLEs requires considerable computational cost, as it entails
calculating $\lambda(k_m)$ for many $k_m$. On this issue, 
\citet{Lietal2024} made an interesting observation 
in the context of rotating turbulence. They found  
that the average energy spectrum of the unconditional LLV peaks at
a wavenumber which appears to be close to, or the same as $k_c$.
The observation is reproduced in
\citet{Lietal24a} for the synchronisation of large eddy
simulations of periodic turbulence. Elementary results for the spectra of
the LLV in the present study, i.e., $E_v(k,t)$, 
are shown in Fig. \ref{fig:Ekv0}.  In the left panel 
of Fig. \ref{fig:Ekv0}, the early evolution of $E_v(k,t)$ is plotted.
We initialise the Fourier modes of the perturbation with independent random
numbers with identical probability distributions. As a consequence, $E_v(k,t=0) \sim k^2$
for large $k$, as the number of modes in the spherical shell with radius $k$ is
proportional to the area $4\pi k^2$ of the shell. $E_v(k,0)$ is shown with the
green dashed line. The spectrum at time $t$, $E_v(k,t)$, exhibits a 
period of transient evolution, as
depicted by the think red lines, which show that the peak of the spectrum moves
towards lower wavenumbers. At $t \approx 20 \tau$, the spectrum converges
towards a distribution shown with the thick black line, which then  fluctuates
over time. 
The long time average $\lal E_v \ral$ 
is shown in the right panel of Fig.
\ref{fig:energy_decay}. Clearly, the peaks of the spectra
are found at $k\eta \approx 0.2$, reproducing previous findings. 
The flow here is very different from the rotating
turbulence investigated in \citet{Lietal2024}. For example,
the base flow energy
spectrum $\lal E \ral (k)$ 
follows the $k^{-2}$ or $k^{-3}$ power laws in \citet{Lietal2024}, whereas
here it follows the canonical $k^{-5/3}$ scaling.
Therefore, it is non-trivial
for the same relationship to hold in both cases. 
As a step towards understanding the origin of this relationship, 
we look into the spectral dynamics of the
LLVs and the conditional LLVs in what follows.

\subsection{Production and dissipation: spectral analyses \label{sect:spec}}

To understand the dependence of $P$ and $D$ (hence $\Lambda$) 
on $k_m$ better, 
we look into the production spectrum $\PP_v(k,t)$ and the 
dissipation spectrum $\DD_v(k,t)$. We first consider their
non-dimensional ensemble averages 
\begin{equation} \label{eq:pvdv}
P_v(k) \equiv \tau \lal \PP_v (k,t)\ral, \qquad 
  D_v(k) \equiv
\tau\lal 
  \DD_v(k,t) \ral . 
\end{equation}
The expression for $D_v(k)$ can also be written as 
\be
  D_v(k) =
  2 \bar{\nu} k^2 \tau \lal E_v \ral(k) = 2 (k\eta)^2
  \lal E_v \ral(k). 
\ee
The behaviours of $P_v$ and $D_v$ are related by Eq.
(\ref{eq:Ekv}).  
Taking the average of Eq. (\ref{eq:Ekv}), and noting $\ptl_t
\lal E_v \ral = 0$, we find
\begin{equation} \label{eq:Ekv2}
P_v(k) - D_v(k) = 2\tau\lal \gamma E_v\ral.  
\end{equation}
Our data show that $\gamma$ is essentially uncorrelated to  $E_v$
(figure omitted). Therefore $\lal \gamma E_v\ral \approx \lal \gamma
\ral \lal E_v \ral = \lambda \lal E_v \ral $, and we obtain 
\begin{equation} \label{eq:pdev2}
  P_v (k) = D_v(k) + 2 \Lambda \lal E_v \ral = 2 \left[ (k\eta)^2
  + \Lambda (k_m) \right] \lal E_v \ral.
\end{equation}
The equation delineates the spectral balance of the energetics of the
velocity perturbation.  Integrating Eq. (\ref{eq:pdev2}) over $k$, we obtain 
\be \label{eq:speclam}
\Lambda(k_m) = \int_0^\infty P_v(k) dk - \int_0^\infty 2
(k\eta)^2 \lal E_v \ral (k) dk, 
\ee 
which makes clear that Eq. (\ref{eq:speclam}) is the spectral version of Eq.
(\ref{eq:lampd}). 

\bfig 
\centering
\ig[width=0.48\lnw]{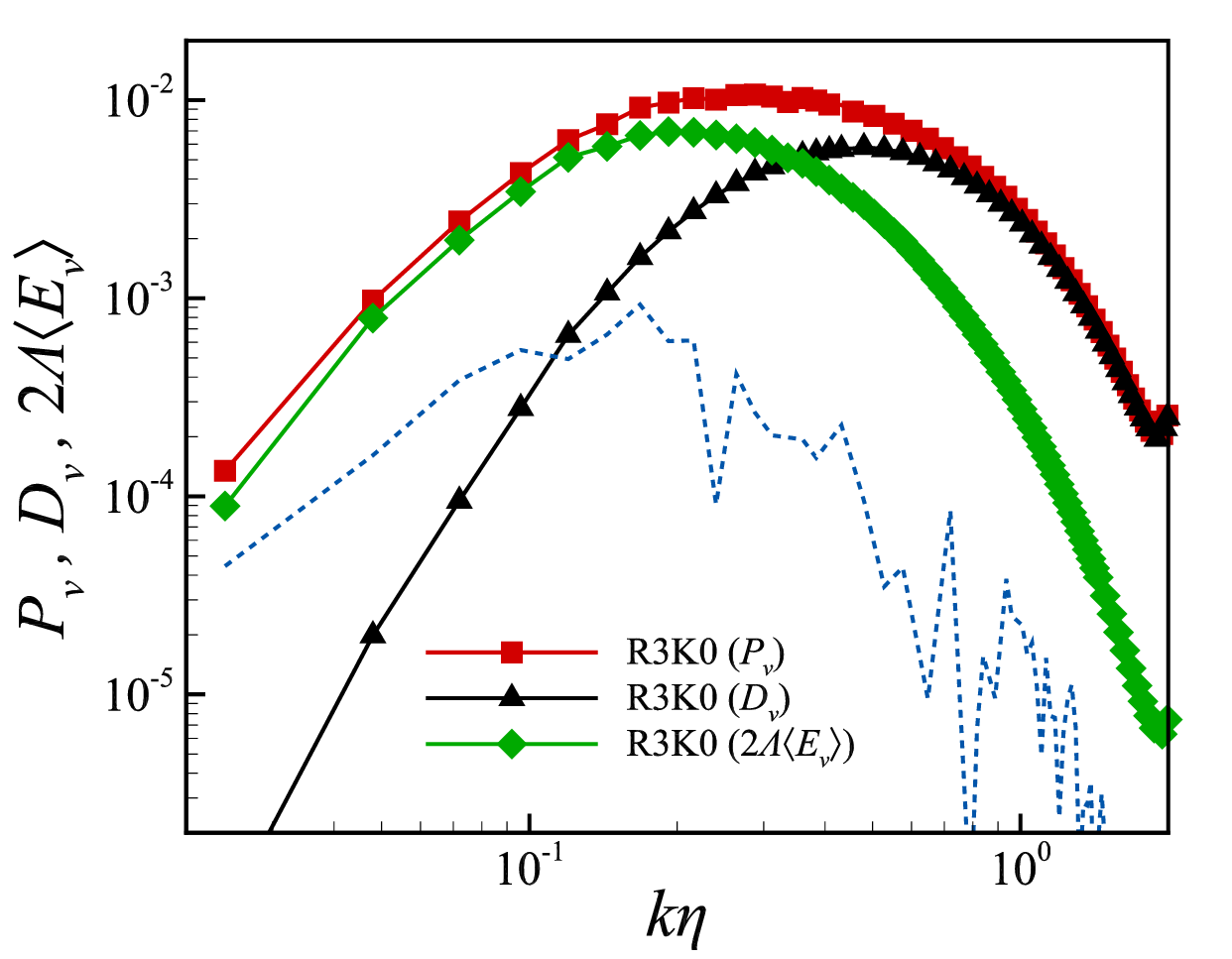} 
\ig[width=0.48\lnw]{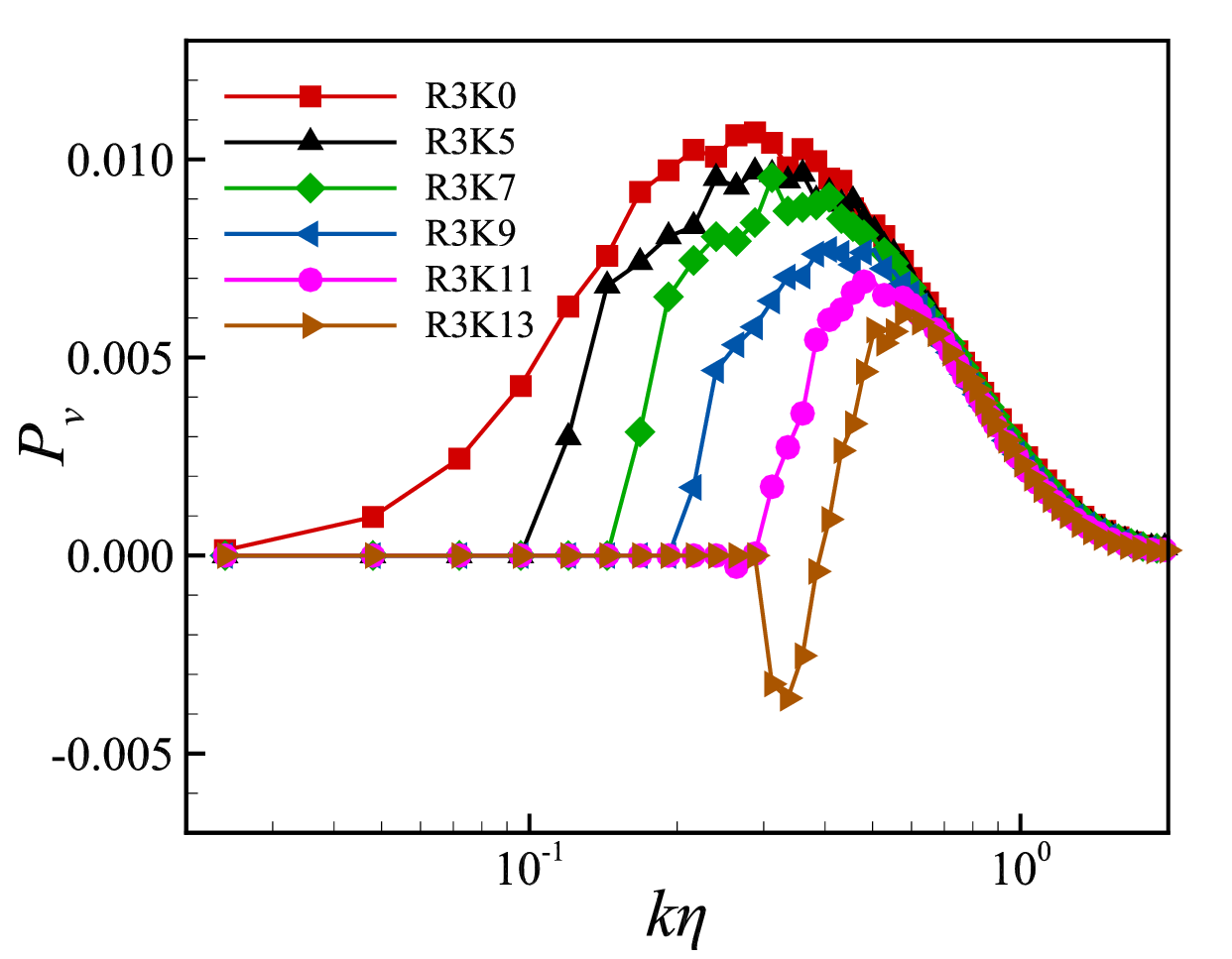} 
\ig[width=0.48\lnw]{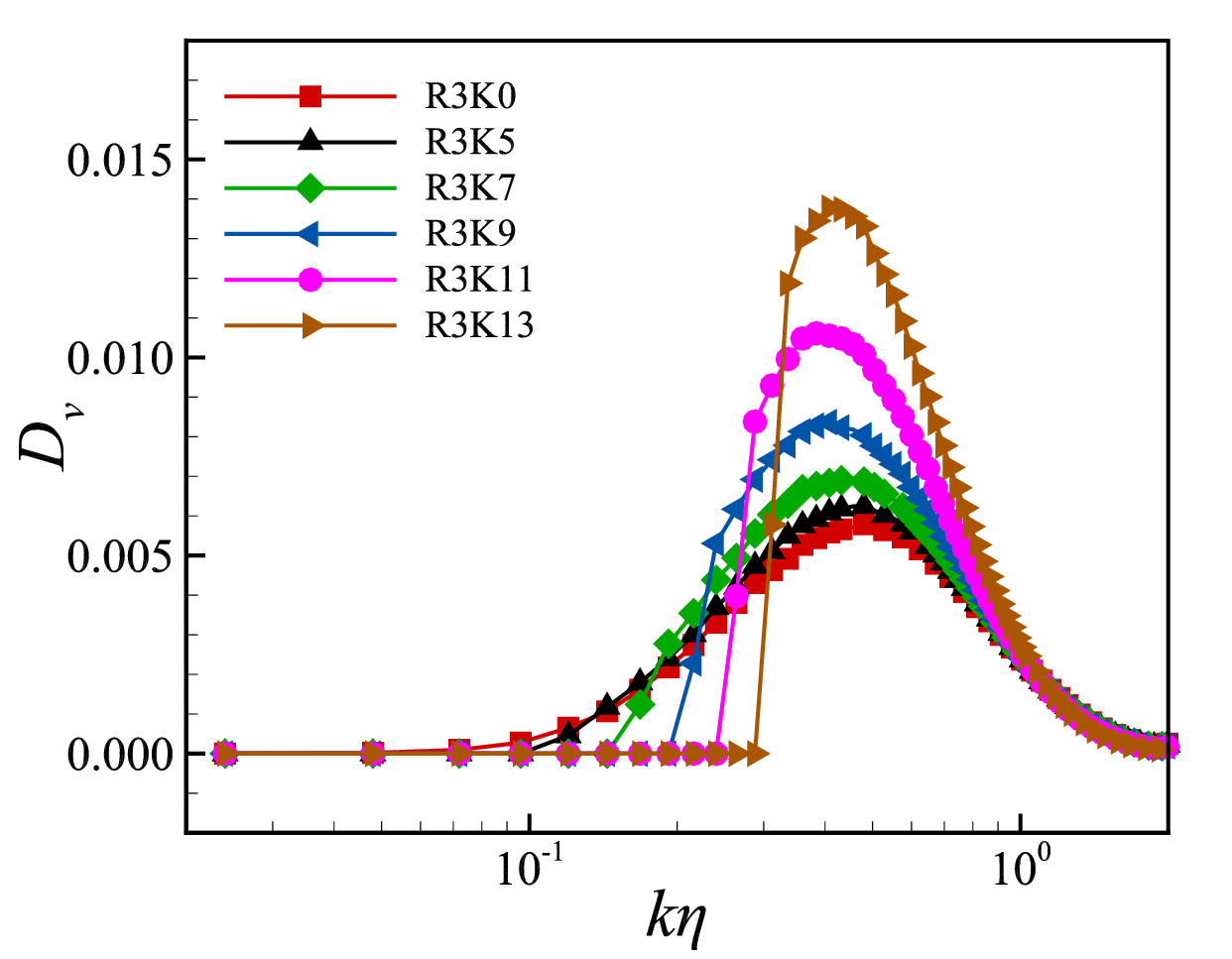}
\ig[width=0.48\lnw]{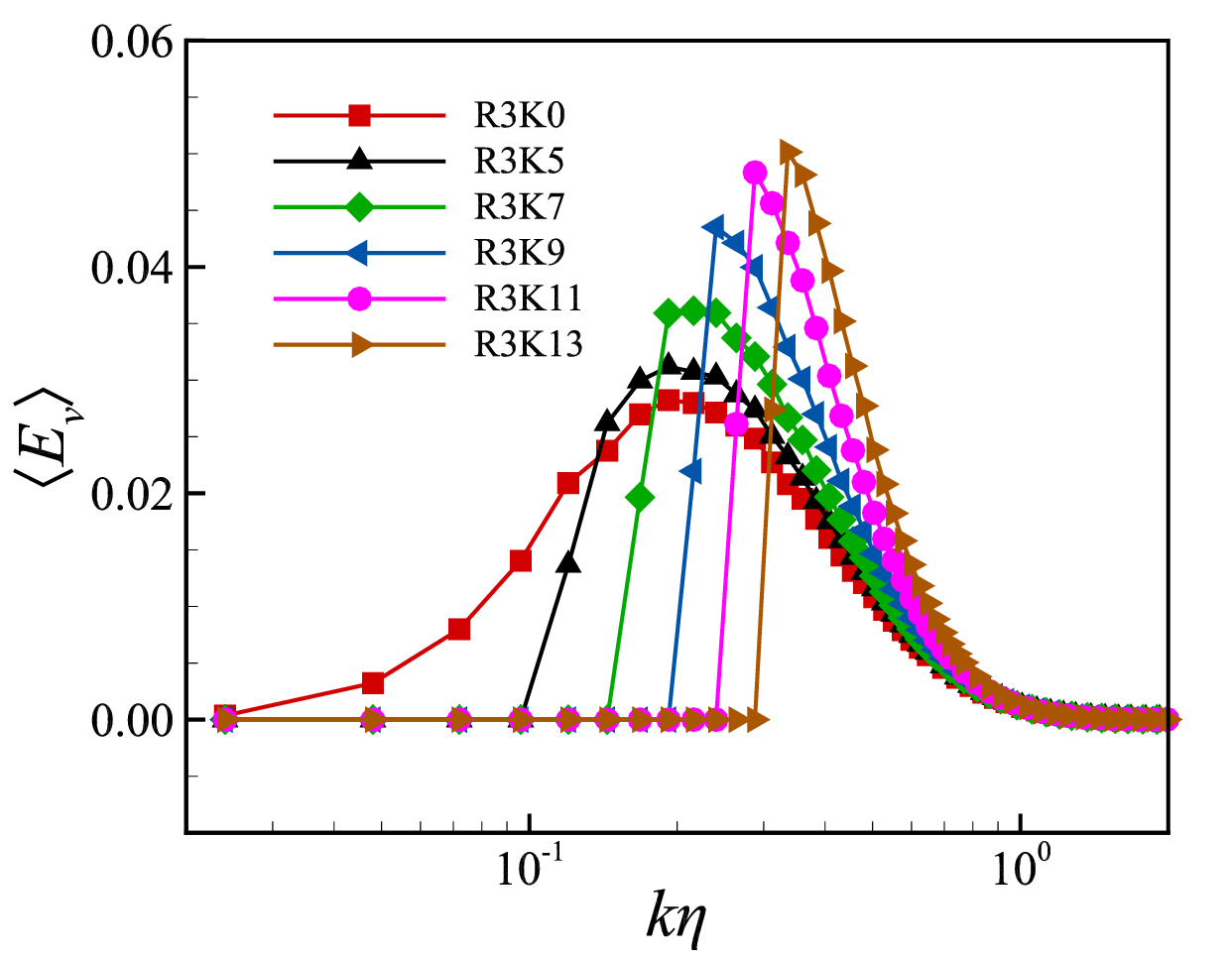}
\caption{\label{fig:PDEv_k_R3} Top-left: production spectrum
$P_v$, dissipation spectrum $D_v$ and $2\Lambda \lal E_v \ral$ as a
function of $k\eta$ for $k_m=0$ with the dashed line showing the
residual 
$P_v - D_v - 2\Lambda \lal E_v \ral$. Top-right: $P_v$.
Bottom-left: $ D_v $. Bottom-right: $\lal E_v \ral$. The non-dimensionalised coupling wavenumbers $k_m\eta$ are $0$, $0.12$, $0.17$, $0.22$, $0.26$, and $0.31$.  For group
R3. }
\efig  
\bfig
\centering
\ig[width=0.48\lnw]{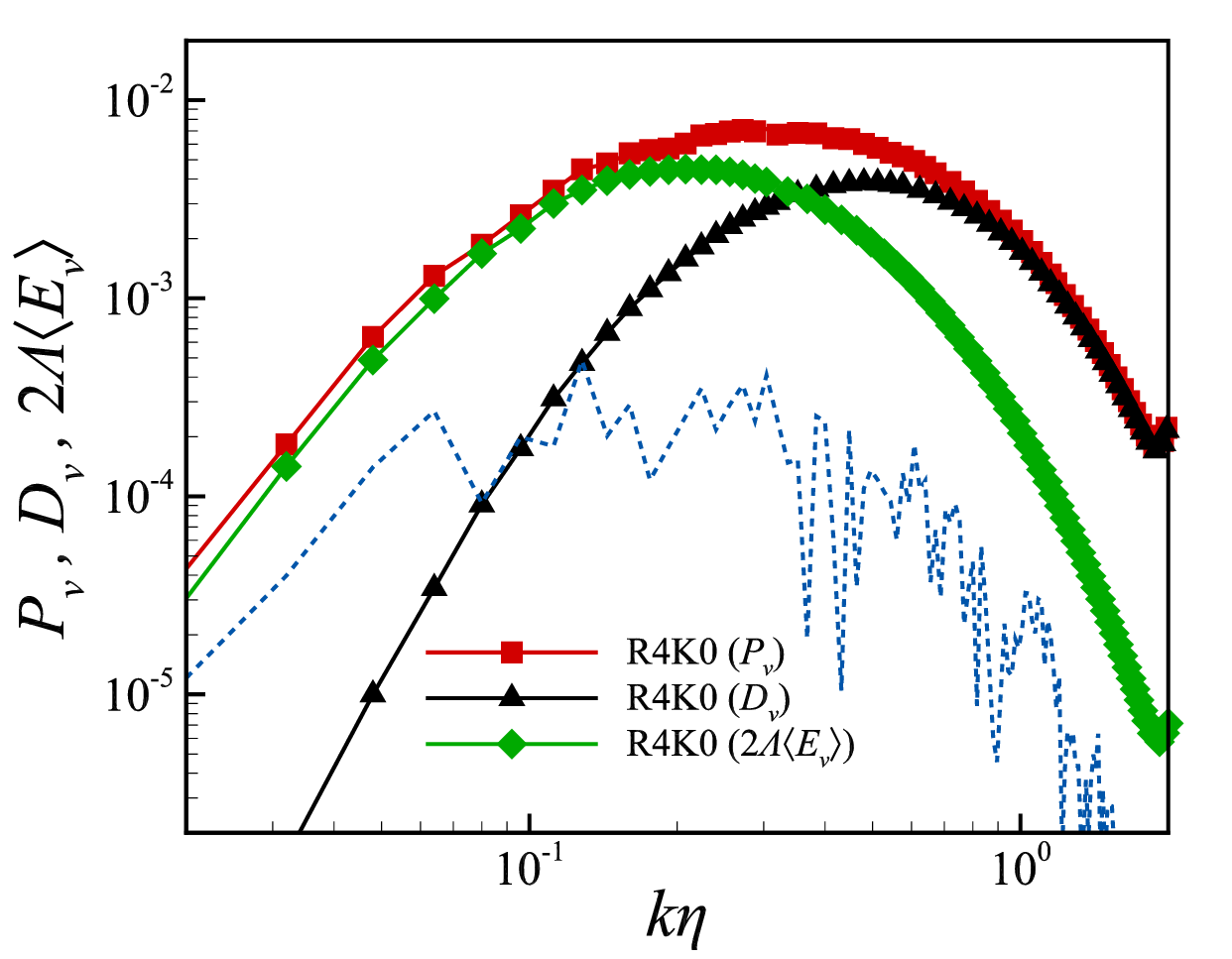} 
\ig[width=0.48\lnw]{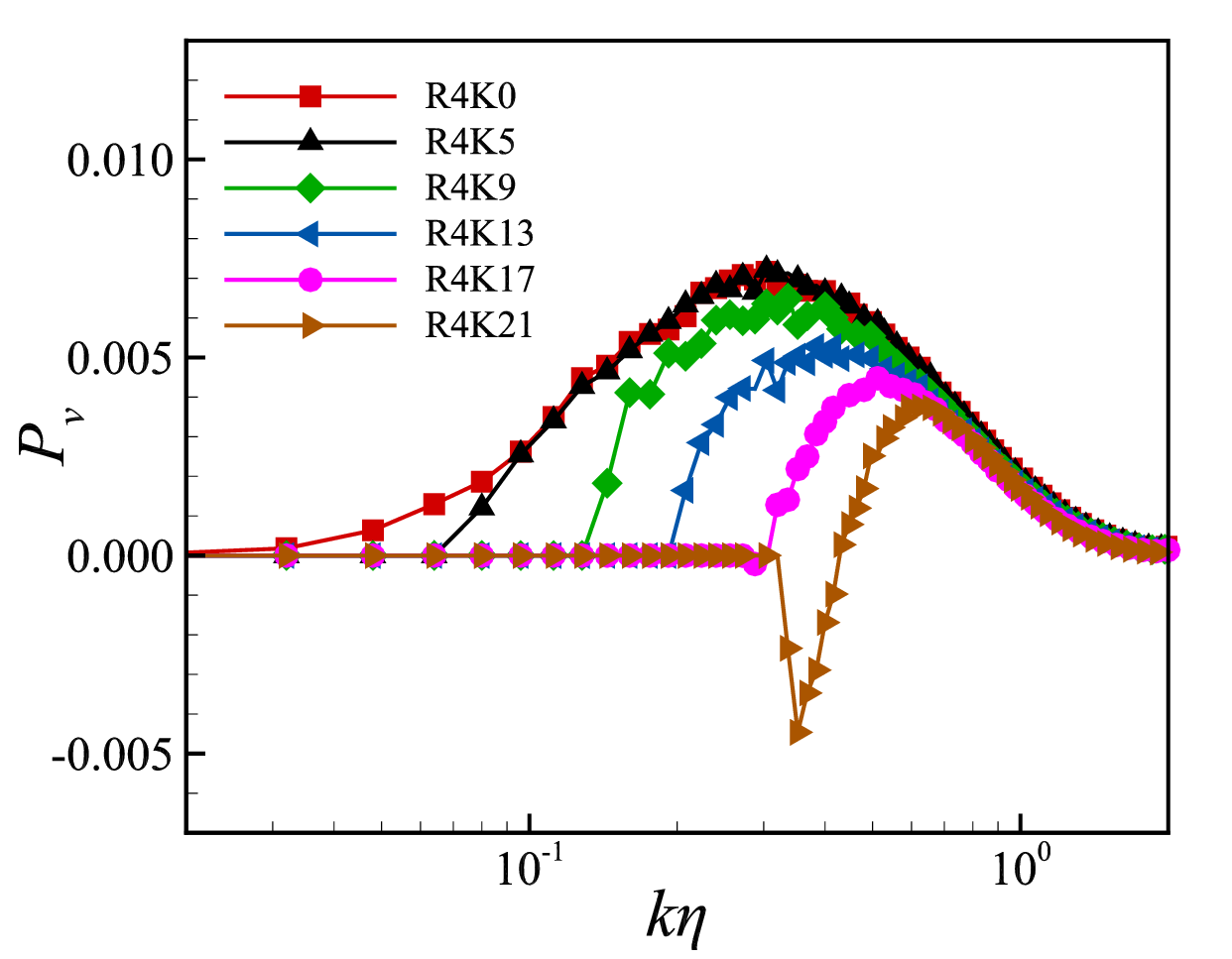}
\ig[width=0.48\lnw]{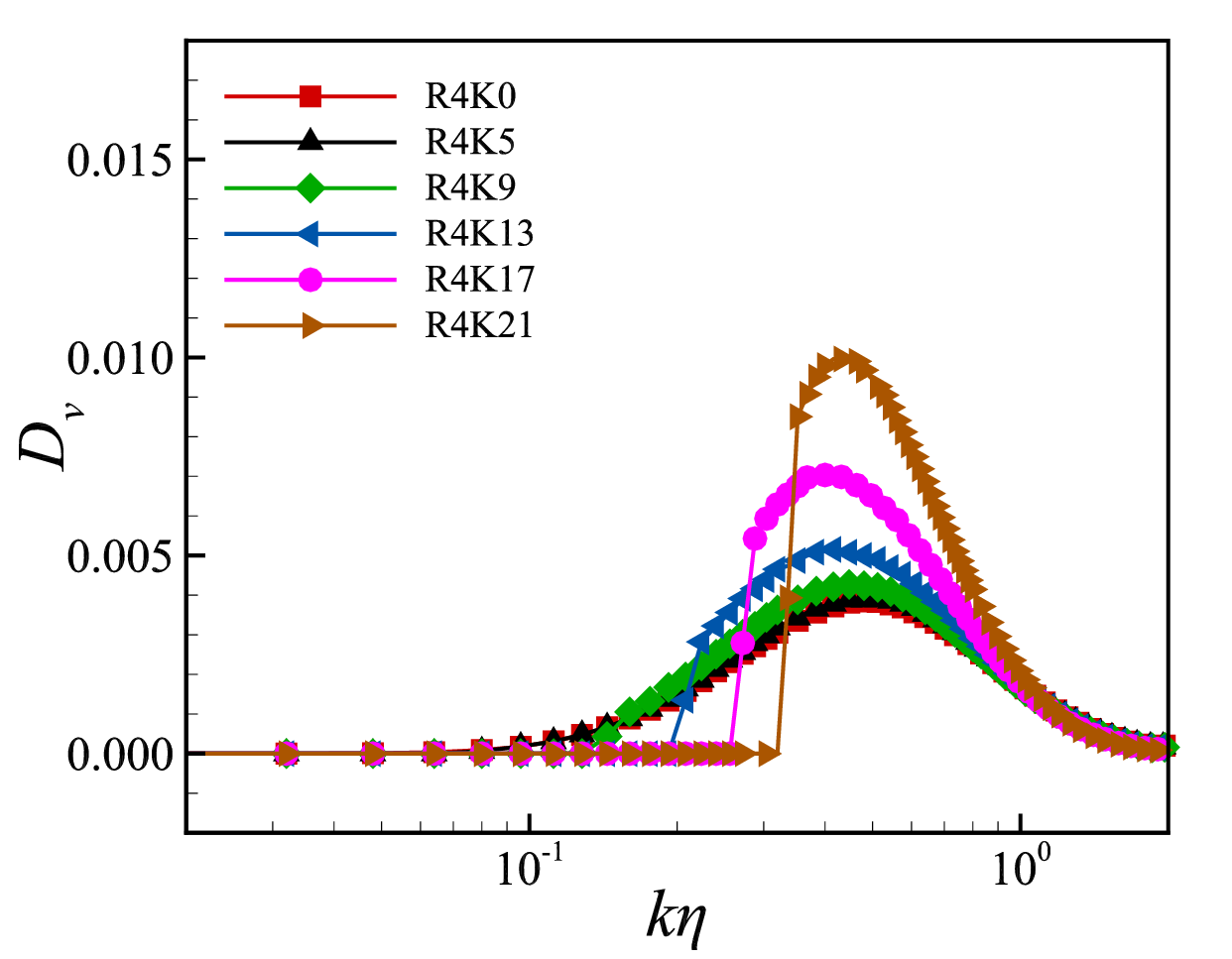}
\ig[width=0.48\lnw]{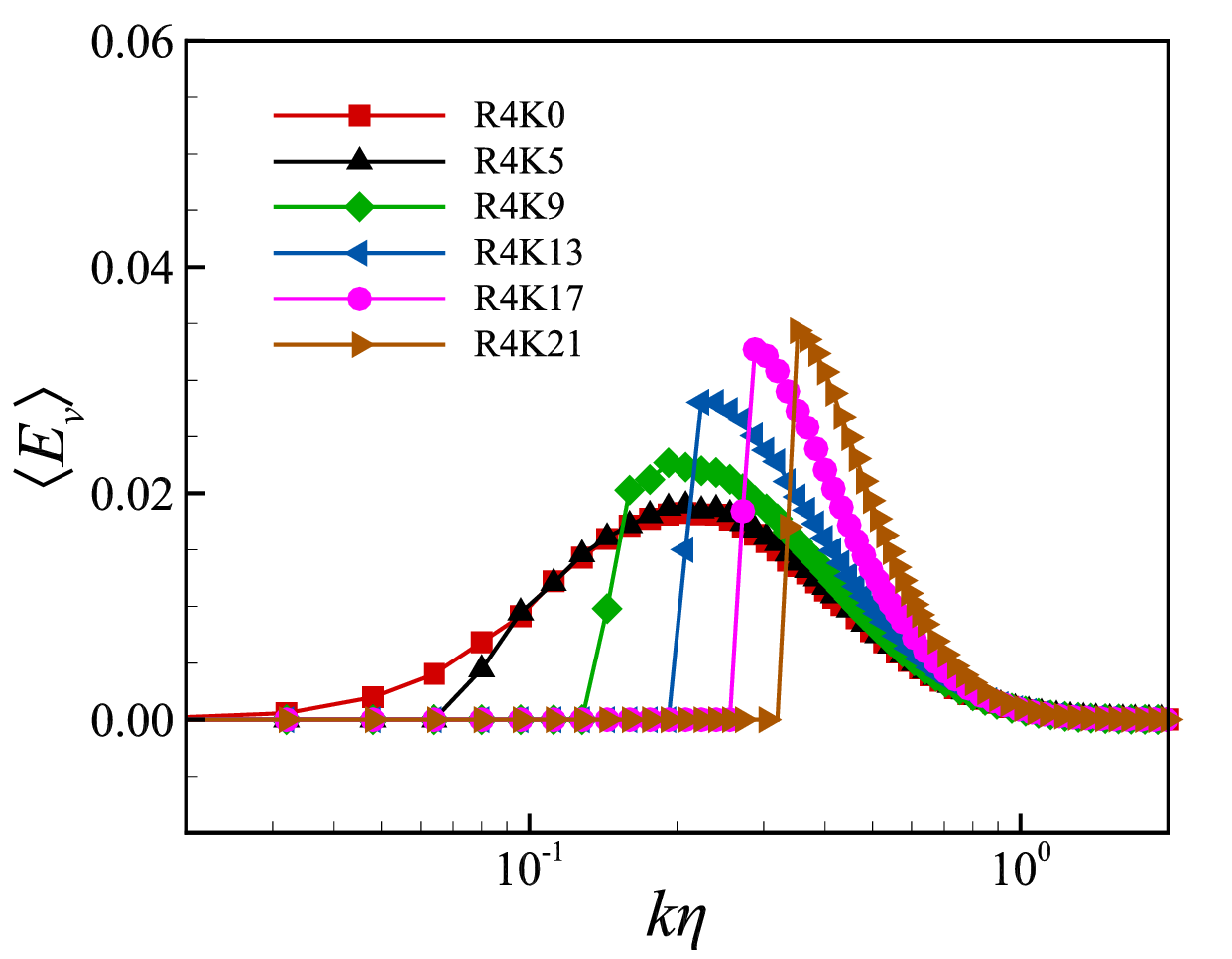}
\caption{\label{fig:PDEv_k_R4} Same as Fig. \ref{fig:PDEv_k_R3}
but for group R4. The non-dimensionalised coupling wavenumbers $k_m\eta$ are $0$, $0.08$, 
$0.14$, $0.21$, $0.27$, and $0.34$.}
\efig  

Fig. \ref{fig:PDEv_k_R3} shows the spectra
$P_v$, $D_v$, and $\lal E_v \ral$ with 
different $k_m$ for the cases in group R3. The same
results for group R4 are shown in Fig. \ref{fig:PDEv_k_R4} as 
corroboration. The two figures depict same behaviours. Therefore we
will only discuss Fig. \ref{fig:PDEv_k_R3} in detail.  
The top-left panel of Fig. \ref{fig:PDEv_k_R3} 
compares the three distributions
for $k_m=0$, i.e., for the cases where no coupling is imposed. 
The dashed line shows 
the residual $P_v - D_v - 2\Lambda \lal E_v \ral$, which, according to
Eq. (\ref{eq:pdev2}), should be essentially zero.
Though it is not exactly zero, the dashed line shows that it is
negligible compared with the dominant terms at all wavenumbers. Not
surprisingly, the dissipation $D_v$ peaks at
a higher wavenumber compared with $P_v$ and $2\Lambda \lal
E_v\ral$, 
as dissipation dominantly comes from small
scales. 
On the other hand, $2\Lambda \lal E_v \ral$ peaks at 
$k\eta \approx 0.2$ (same as that of 
$\lal E_v \ral$). 
The peak of $P_v$ is found at a
wavenumber in between, which shows that the strongest
production is found at wavenumbers well inside the dissipation range
of the base flow. Also, $P_v(k)$ is positive definite, implying
that the velocity perturbation is amplified by the production
mechanism at all scales.  

The production spectra 
for several different $k_m$ are shown
in the top-right panel of Fig. \ref{fig:PDEv_k_R3}. 
Because $\hat{\bv}(\k,t) = 0$ for $\vert \k\vert <
k_m$ due to the coupling between the synchronised flows, 
$P_v(k) = 0$ for $k < k_m$. 
What is noteworthy is that $P_v(k)$ is  
also reduced by the coupling for $k> k_m$, 
and it is increasingly smaller for larger $k_m$. The attenuating
effect of the coupling is
localised in the 
wavenumbers around $k_m$, with 
$P_v(k)$ at large $k$ little affected. 
To interpret this feature, we refer back to Eq.
(\ref{eq:pvkh}). Since $\hat{\v}(\bm{q},t) =
0$ when $\vert \bm{q}\vert < k_m$, the summation in Eq.
(\ref{eq:pvkh}) does not include the Fourier modes with $\vert
\bm{q} \vert < k_m$. Thus 
increasing $k_m$ means excluding more Fourier modes from the
summation, which thus likely leads to smaller $P_v(k)$, because
the top-left panel shows that the contributions from these
excluded modes are likely to be positive.   

Another observation is that, though $P_v(k)$ is
positive definite for $k > k_m$ in most cases, it assumes negative
values for some wavenumbers when $k_m$ is large enough (e.g. for
case R$3$K$13$). This observation can be understood from Eq.
(\ref{eq:pdev2}). Obviously, $P_v(k)$ would be positive definite for $k>k_m$ if
$(k\eta)^2 + \Lambda(k_m ) > 0$ for all $k> k_m$. This is clearly
satisfied if $\Lambda(k_m) >0$, which is the case when $k_m=0$ or
$k_m$ is small. If $\Lambda(k_m) <0$, on the other hand, then 
$(k\eta)^2 +  \Lambda <0$ when 
$k\eta < (-\Lambda)^{1/2}$. As a result,
$P_v(k) $ would be negative for wavenumber $k$ if $k_m\eta
< k\eta < (-\Lambda)^{1/2}$. The inequality can be satisfied by some
wavenumbers if $ - (k_m \eta)^2 > \Lambda(k_m)$, which is satisfied
when $k_m\eta$ is large enough as shown in the right panel of
Fig. \ref{fig:energy_decay}. Therefore, when $k_m$ is large
enough, the velocity perturbations at the scales just below the
coupling scale would be suppressed by the production term. 

The dissipation spectrum $D_v$ is shown in the bottom-left panel,
while the energy spectrum $\lal E_v \ral$ is shown in the bottom-right
panel. As $D_v = 2 (k\eta)^2 \lal E_v \ral$, the two parameters
are similar in many ways. The most conspicuous feature for both is
that their distributions are elevated as $k_m$ increases. This
behaviour could be a simple consequence of the normalisation condition $\lal
e \ral_v = 1/2$ which fixes the total integral of $\lal E_v \ral$. As
the support of $\lal E_v \ral$ is reduced when $k_m$ increases, its
values have to increase. The values of $D_v(k)$, as a
consequence, have to increase too. 
Nevertheless, there is another mechanism by
which $D_v(k)$ is enhanced. Eq. (\ref{eq:pdev2}) implies that
\be \label{eq:dvkpvk}
D_v(k) = P_v(k) \frac{(k\eta)^2}{(k\eta)^2 + \Lambda(k_m)}.
\ee
The above equation shows that reduced $P_v(k)$ tends to reduce $D_v(k)$. However, reduced $P_v(k)$
also tends to reduce $\Lambda(k_m)$, which
in turns enhances the factor $
(k\eta)^2/[\Lambda(k_m) + (k\eta)^2]$, thus potentially increases
$D_v(k)$. That is, there is a mechanism by which
$D_v(k)$ increases as a consequence of reduced $P_v(k)$. This
effect is stronger at lower wavenumbers as the factor $
(k\eta)^2/[\Lambda(k_m) + (k\eta)^2]$ is more sensitive to the
change in $\Lambda(k_m)$ when $k\eta$ is smaller. 
Unfortunately, it is unclear which of the above two mechanisms contributes 
more to the enhancement of $D_v(k)$.  

We now turn to a brief discussion on the viscous estimate
$-(k_m\eta)^2$ for $\Lambda(k_m)$.  
Note that $\lal E_v \ral (k) = 0$ for $k< k_m$.
Therefore, 
\begin{align}
D & = \int_{k_m}^\infty 2 (k\eta)^2 \lal E_v \ral dk = 2(k_m\eta)^2
\int_{k_m}^\infty (k/k_m)^2 \lal E_v \ral dk \notag \\
&\ge 2 (k_m\eta)^2
\int_{k_m}^\infty \lal E_v \ral dk = (k_m\eta)^2, 
\end{align}
which implies that $(k_m\eta)^2$ tends to underestimate the
dissipation $D$. Therefore, it might not be surprising that in
our simulations 
$\Lambda(k_m)$ becomes smaller than $-(k_m\eta)^2$ for some
$k_m$. The estimate $(k_m\eta)^2$ would, however, become exact
if $\lal E_v\ral$ was proportional to the Dirac delta function concentrated at
$k_m$ (with the strength being $1/2$). Our results in the bottom-right panel of Fig.
\ref{fig:PDEv_k_R3} do show a tendency for $\lal E_v \ral$ to
concentrate around $k_m$ as $k_m$ increases.

We have been able to present some semi-analytical discussions of
the results in Fig. \ref{fig:PDEv_k_R3} based on Eq.
(\ref{eq:pdev2}). Deriving the relationship between
the coupling wavenumber $k_c\eta$ and the peak wavenumber of $\lal E_v
\ral$ analytically requires, as a very first step,
finding an analytical expression for the peak wavenumber for $\lal E_v
\ral$ at $k_m=0$. Attempts at such
analyses, however, quickly run into the classical closure problem, as we
do not have an expression for $P_v(k)$ in terms of $\lal
E_v \ral$. Nevertheless, some elementary results can be obtained. 
Since the peak wavenumber is given by $d
\lal E_v \ral /d (k\eta) = 0$, we can find from
$\lal E_v \ral = P_v(k)/[2(k\eta)^2 + 2 \Lambda(0)]$ that the
peak wavenumber $k\eta$ satisfies
\be
\frac{ 2 k\eta}{ (k \eta)^2 + \Lambda(0)} = \frac{1}{P_v} \frac{ d
P_v}{d (k\eta)} = \frac{ d \ln P_v}{d (k\eta)}, 
\ee
which can be solved for $k\eta$ if we have an 
expression for $P_v(k)$. Purely as a demonstration, we let
$P_v(k) \sim a
(k\eta)^b$, assuming the expression provides a good approximation for the slope of $\ln
P_v(k)$ around the peak wavenumber for some constants $a$ and $b$. 
The peak wavenumber is then given by 
\be
k\eta = \left(\frac{b \Lambda(0)}{2-b}\right)^{1/2}.
\ee
If we let $\Lambda(0) = 0.12$ (hence ignoring $\Lambda(0)$'s
dependence on $Re_\lambda$), then $b=1/2$ would give $k\eta =
0.2$. Progress may be made by developing an EDQNM-type model
for $P_v(k)$, but this is beyond the scope of this investigation. 

\bfig
\centering
\ig[width=0.5\lnw]{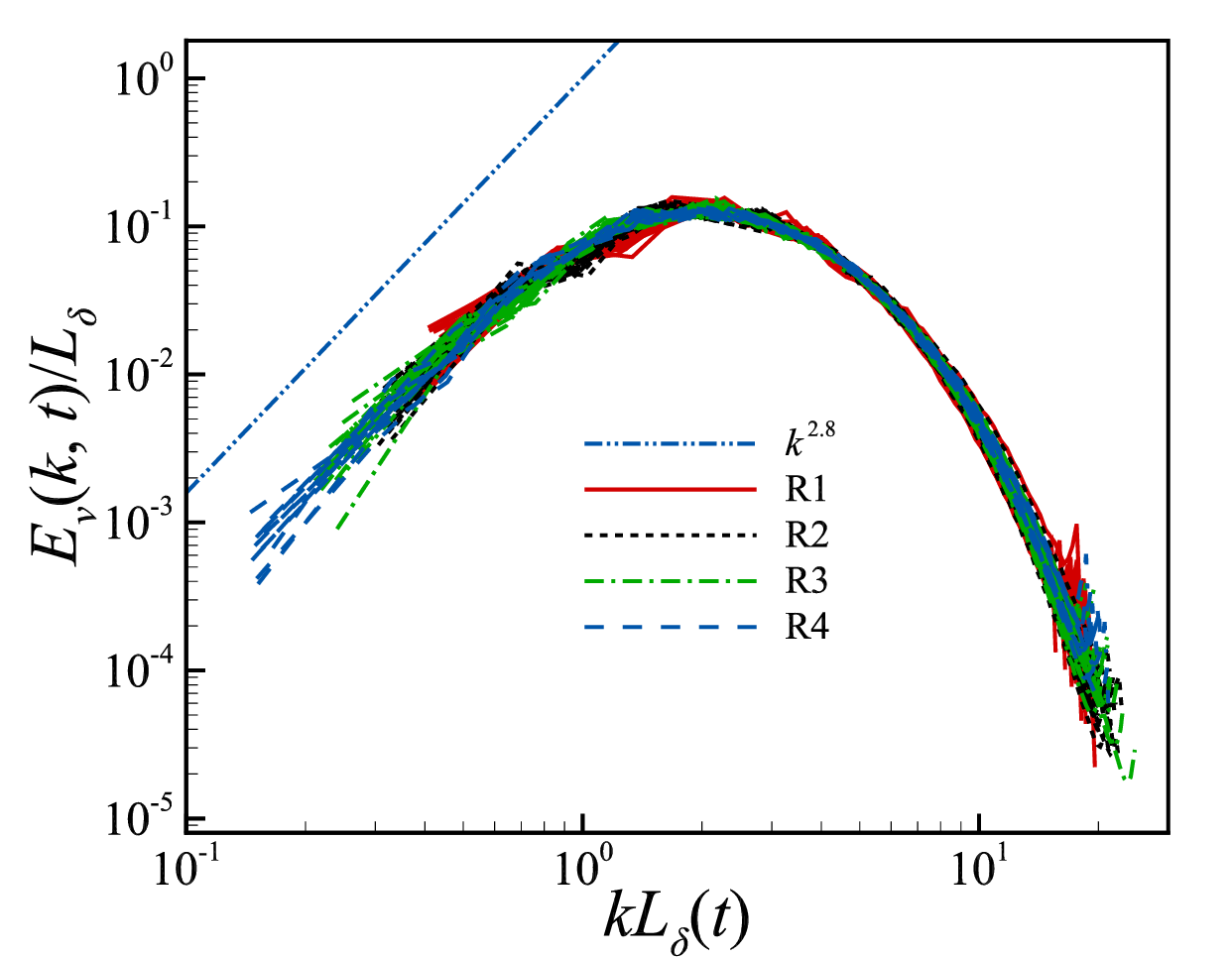}
\caption{\label{fig:g_kL} Normalised spectrum $
E_v(k,t)/L_\delta$ as a function of $k L_\delta$ at
different times. For cases with $k_m=0$ only. }
\efig

We now explore some aspects of the time evolution of the spectrum $E_v(k,t)$. 
An interesting observation is made in \citet{Geetal23,
KatsunoriAriki19}, which shows that $E_\delta
(k,t)$ evolves in a self-similar manner over a period of time. To
examine this 
phenomenon in our simulations, we follow \citet{Geetal23}, and define
an integral length scale for the velocity perturbation by 
\begin{equation} \label{eq:ldel}
L_\delta(t) = \frac{3\pi}{4K_\delta(t)} \int_0^\infty k^{-1}
  E_\delta(k,t) dk  = \frac{3\pi}{2} \int_0^\infty k^{-1} E_v(k,t)
  dk.
\end{equation}
Self-similar evolution takes place if 
$E_\delta(k,t)/K_\delta L_\delta$ is a function of $k L_\delta$ alone,
independent of time. In terms of $E_v(k,t)$, it implies that we have
\begin{equation} \label{eq:Ekv_sim}
	E_v(k,t) = L_\delta(t) g(kL_\delta),
\end{equation}
for some function $g(\cdot)$. Eq.
(\ref{eq:Ekv_sim}) implies that, $E_v / L_\delta$, when 
plotted against $k L_\delta$, should collapse on a single curve. 
Fig. \ref{fig:g_kL} plots the results obtained from our data over
a period of time spanning over $100 \tau$. The immediate
observation  
is that the curves mostly
fall on each other. The agreement is the best for wavenumbers
somewhat larger than the wavenumber where the curves peak. This
feature is also observed in \citet{Geetal23}. The
discrepancies are larger at the two ends of the spectra. 
The peak of the normalised spectra
is found approximately at $kL_\delta = 2$, the same as in
\citet{Geetal23}. 
There are attempts to deduce
analytically the slope of the spectra as $kL_\delta \to 0$
\citep{KatsunoriAriki19}, but various values have been observed in
DNS. For example, $k^4$ is found in \citet{KatsunoriAriki19}, and
slopes closer to $k^{3.3}$ are reported in \citet{Geetal23}. For
the cases in group R4, which have the largest Reynolds number in our
simulations, 
the slope appears to scale with $k^{2.8}$, as shown by the
dash-double-dotted line. 

Our observation broadly agrees with 
those in \citet{Geetal23, KatsunoriAriki19}. 
Note that the spectra shown in Fig. \ref{fig:g_kL} are calculated from
the long time limit of $\bv$. In contrast, 
\citet{Geetal23, KatsunoriAriki19} observe self-similarity in an
intermediate stage of the evolution of the velocity perturbation where
no rescaling is applied to keep the perturbation small. The
agreement between the results shows that the intermediate stage of
evolution observed in the latter appears to be the same as the long
time asymptotic state of an infinitesimal perturbation. 

\begin{figure}
	\centering 
	\includegraphics[width=0.5\lnw]{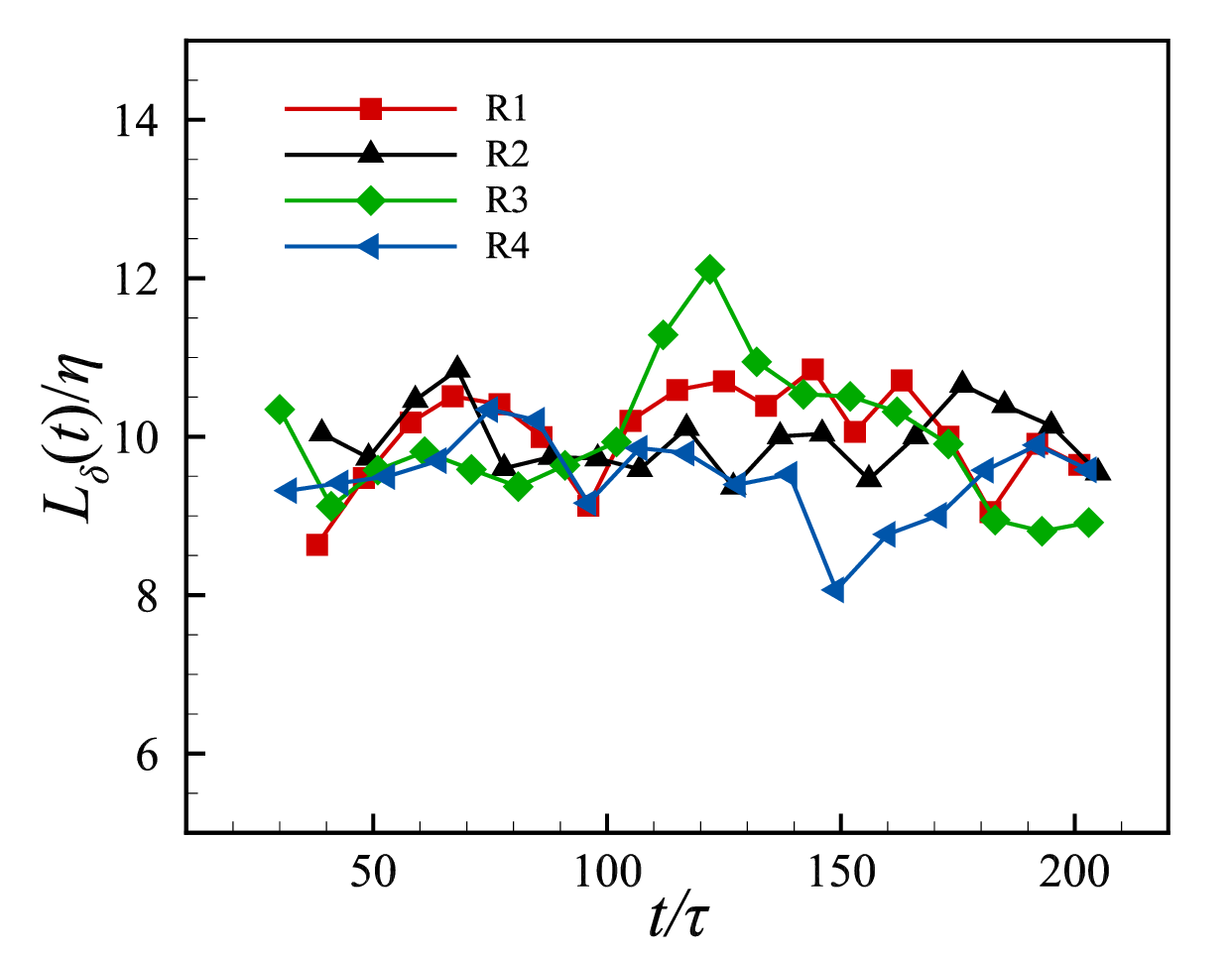} 
	\caption{\label{fig:Lt} Normalised integral length scale $L_\delta/\eta$ for the perturbation
	velocity.}
\end{figure}

\bfig
\centering
\ig[width = 0.5\lnw]{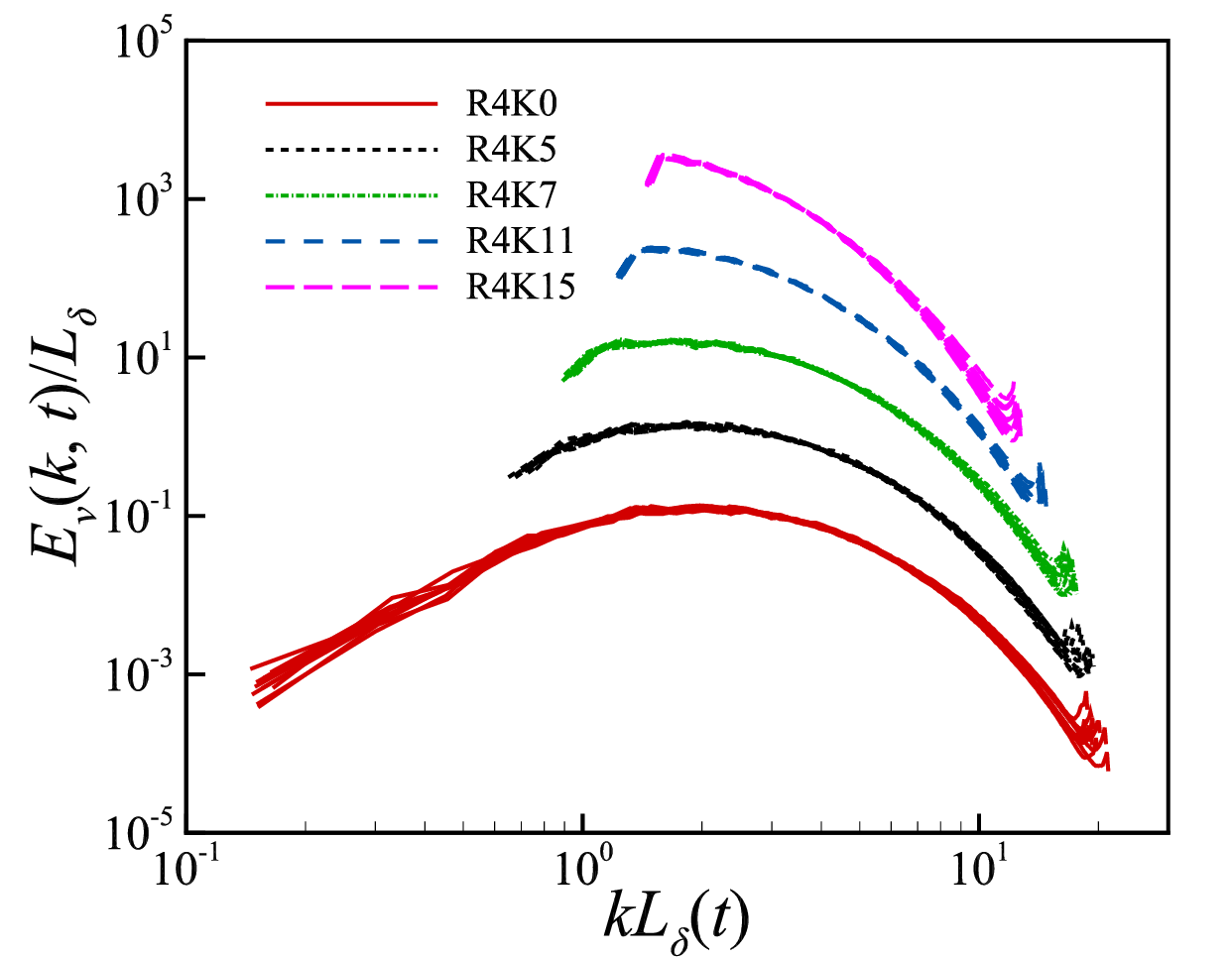}
\caption{\label{fig:err_g_R4} Normalised instantaneous energy spectrum $
E_v (k,t)/ L_\delta$ as a function of $k L_\delta$
for different $k_m$ at different times. For cases in group R4.}
\efig

It is natural to explore the relationship between the peak location of the self-similar spectrum and the 
peak location of $\lal E_v \ral$ shown in the right panel of Fig. \ref{fig:Ekv0}. This can be
inferred from Fig. \ref{fig:Lt}, which shows that the ratio $L_\delta/\eta$ fluctuates
around $10$. Therefore $kL_\delta \approx 2$ is equivalent to $k\eta
\approx 0.2$, which is the peak location of $\lal E_v \ral(k)$. This
relationship lends further support to the conjecture that the peak
wavenumber of $\lal E_v \ral$ is a physically significant parameter
for turbulence synchronisation. 
Another finding is that self-similarity is also observed
for the spectra of the conditional LLVs, as shown in Fig.
\ref{fig:err_g_R4}. For clarity, the curves for larger $k_m$ are
shift upwards by a factor of $10$ successively. Evidently, there is a
very good agreement between the curves at different times, as
required by self-similarity.

\bfig
\centering
\ig[width=0.5\lnw]{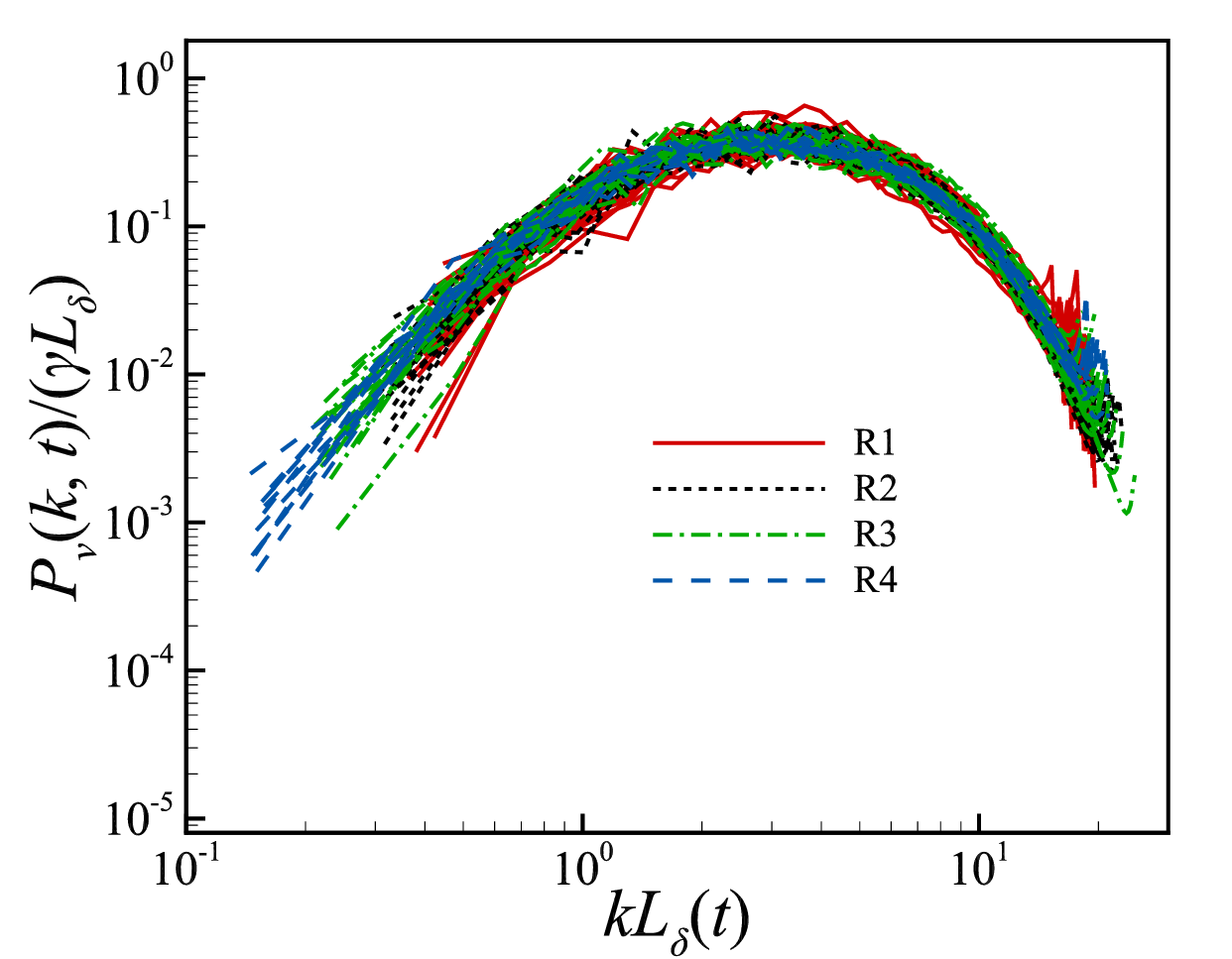}
\caption{\label{fig:P_kL} Normalised instantaneous production spectrum $
P_v(k,t)/(\gamma L_\delta)$ as a function of $k L_\delta$. For
cases with $k_m = 0$ only. }
\efig

The self-similar evolution can be examined quantitatively via the
equation for the spectrum $E_v (k,t)$, i.e.,
Eq. (\ref{eq:Ekv}). 
Substituting Eq. (\ref{eq:Ekv_sim}) into Eq. (\ref{eq:Ekv}), we obtain
\begin{equation} \label{eq:gxi}
	g(\xi) + \xi \frac{ d g}{d\xi} = \frac{P_v(k,t)}{\dot{L}_\delta}  -
  \frac{2\bar{\nu}}{L_\delta \dot{L}_\delta} \xi^2 g(\xi) -
  \frac{\gamma(t)L_\delta(t)}{\dot{L}_\delta} g(\xi),
\end{equation}
where $\xi \equiv k L_\delta$ and $\dot{L}_\delta$ is the time
derivative of $L_\delta$. Therefore, a fully self-similar solution
(over all wavenumbers) is possible only if 
\begin{equation} \label{eq:constraints_sim}
	\frac{P_v(k,t)}{\dot{L}_\delta} = h(\xi), \qquad
  \frac{2\bar{\nu}}{L_\delta \dot{L}_\delta} = \alpha,\qquad \frac{\gamma(t)
  L_\delta(t)}{\dot{L}_\delta} = \beta,
\end{equation}
where $h(\xi)$ is some function to be determined, and $\alpha$ and
$\beta$ are constants. 
Fig. \ref{fig:g_kL} shows that $E_v(k,t)$ is mainly self-similar for
intermediate wavenumbers. 
For these wavenumbers, we may drop the viscous effect, i.e., the
second term on the right hand side of 
Eq. (\ref{eq:gxi}). With that, only
the first and the third equations in Eq.
(\ref{eq:constraints_sim}) are required for there to be a self-similar solution. The third equation
establishes a relation between $\gamma(t)$ and $L_\delta(t)$.
It shows that $L_\delta$ grows exponentially if $\gamma(t)$ is a
constant. The growth rate of $L_\delta$ is given
by $\gamma/\beta$ and $L_\delta \sim \exp(\gamma t/\beta)$. This regime appears to be the one observed in
\citet{Geetal23}.  However,
Fig. \ref{fig:Lt} shows that $L_\delta(t)$ does not grow
exponentially for our data (and $\gamma(t)$ is generally not a
constant). Therefore, the self-similarity in our data
belongs to a different regime, characterised more generally by the
third equation in Eq. (\ref{eq:constraints_sim}).  

In order for the self-similar solution to exist, 
$P_v(k,t)$ must also have a self-similar form,
as shown by the first equation in Eq. (\ref{eq:constraints_sim}). 
Together with the third equation
in Eq. (\ref{eq:constraints_sim}), we obtain 
\begin{equation}
	P_v(k,t) = \dot{L}_\delta h(\xi) = \beta^{-1} \gamma(t)
  L_\delta(t) h(\xi) \sim \gamma(t) L_\delta(t) h(\xi).
\end{equation}
We plot $P_v(k,t)/(\gamma(t) L_\delta(t))$ against $k L_\delta$ in
Fig. \ref{fig:P_kL}. The agreement between 
the curves at 
different times is less satisfactory compared with that shown in Fig.
\ref{fig:g_kL}, but the curves still largely fall on each other. The
deviation from a clear
self-similarity in $P_v$ could be due to the
contamination from the two ends of $E_v(k,t)$. Note that $E_v(k,t)$ is
self-similar mainly in the mid-wavenumber range, 
which means $g(\xi)$ is
well-defined only for a finite range of values for $\xi$. 
Since $h(\xi) \sim d g/d\xi$ according to Eq. (\ref{eq:gxi}), it
is plausible that $h(\xi)$
is well defined 
over a narrower range of $\xi$. This argument suggests that
simulations covering a wider range
of wavenumbers are required to ascertain 
whether strict 
self-similarity in $P_v(k,t)$ exists or not. 

\subsection{Production and dissipation: physical space analyses}

Additional understanding of the
production and dissipation of the LLV can
be obtained with complementary analyses in the physical space.
It has been known for a while that the spatial structures of the
velocity perturbations \citep{Nikitin2008,Nikitin2018,Geetal23}
are non-trivial.
Physical space analyses are well-suited if one is interested in
the impacts of these spatial structures.

In physical space analyses, it is more meaningful to
express the 
production term in the intrinsic coordinates formed by the
eigenvectors of the 
strain rate tensor. 
Let $s^+_{ij} \equiv \tau s_{ij}$ be the non-dimensional strain rate
tensor,  
so that we can write $P = -\lal v_i v_j s^+_{ij} \ral$. 
We let $\lambda_\al\ge \lambda_\beta\ge \lambda_\gamma$ be the
eigenvalues of $s^+_{ij}$,  
with corresponding eigenvectors $\e_i$ $(i = \al, \beta, \gamma)$. 
Due to incompressibility, we have $\lambda_\alpha + \lambda_\beta +
\lambda_\gamma = 0$. Thus  
$\lambda_\alpha$ is always non-negative whereas $\lambda_\gamma$ is
always non-positive.  
Letting $\theta_i$ be the angle between $\e_i$ and $\bv$, we may write
\begin{equation}
P  = P_\alpha + P_\beta + P_\gamma,
\end{equation}
with
\begin{equation} \label{eq:p3}
P_\alpha = - 2\lla e\lambda_\al \cos^2\theta_\al\rra, \quad 
P_\beta = - 2\lla e \lambda_\beta  \cos^2\theta_\beta  \rra, \quad
P_\gamma = - 2\lla e \lambda_\gamma \cos^2\theta_\gamma  \rra.
\end{equation}
Eq. (\ref{eq:p3}) captures the fact that the production hinges on 
the correlation among the 
eigenvalues of $s^+_{ij}$, the
alignment between $\bv$ and the eigenvectors, and the magnitude of 
$\bv$. 

\bfig
\centering
\ig[width=0.55\lnw]{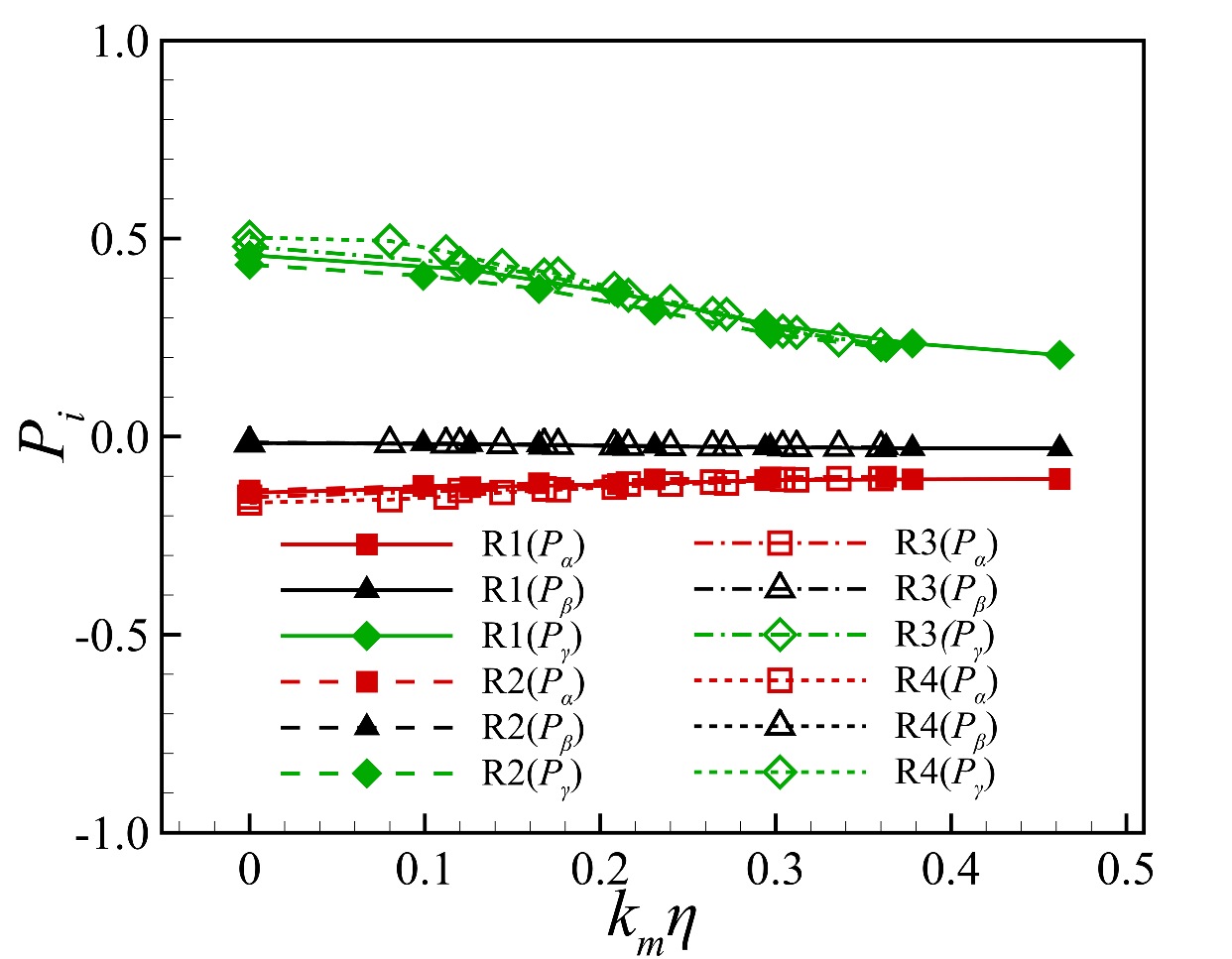}
\caption{\label{fig:P_eigensij_mean_kmeta} The $P_\alpha$,
$P_\beta$ and $P_\gamma$ components of the mean production $P$
 as functions of $k_m\eta$. }
\efig
 
\bfig
\centering
\ig[width=0.48\lnw]{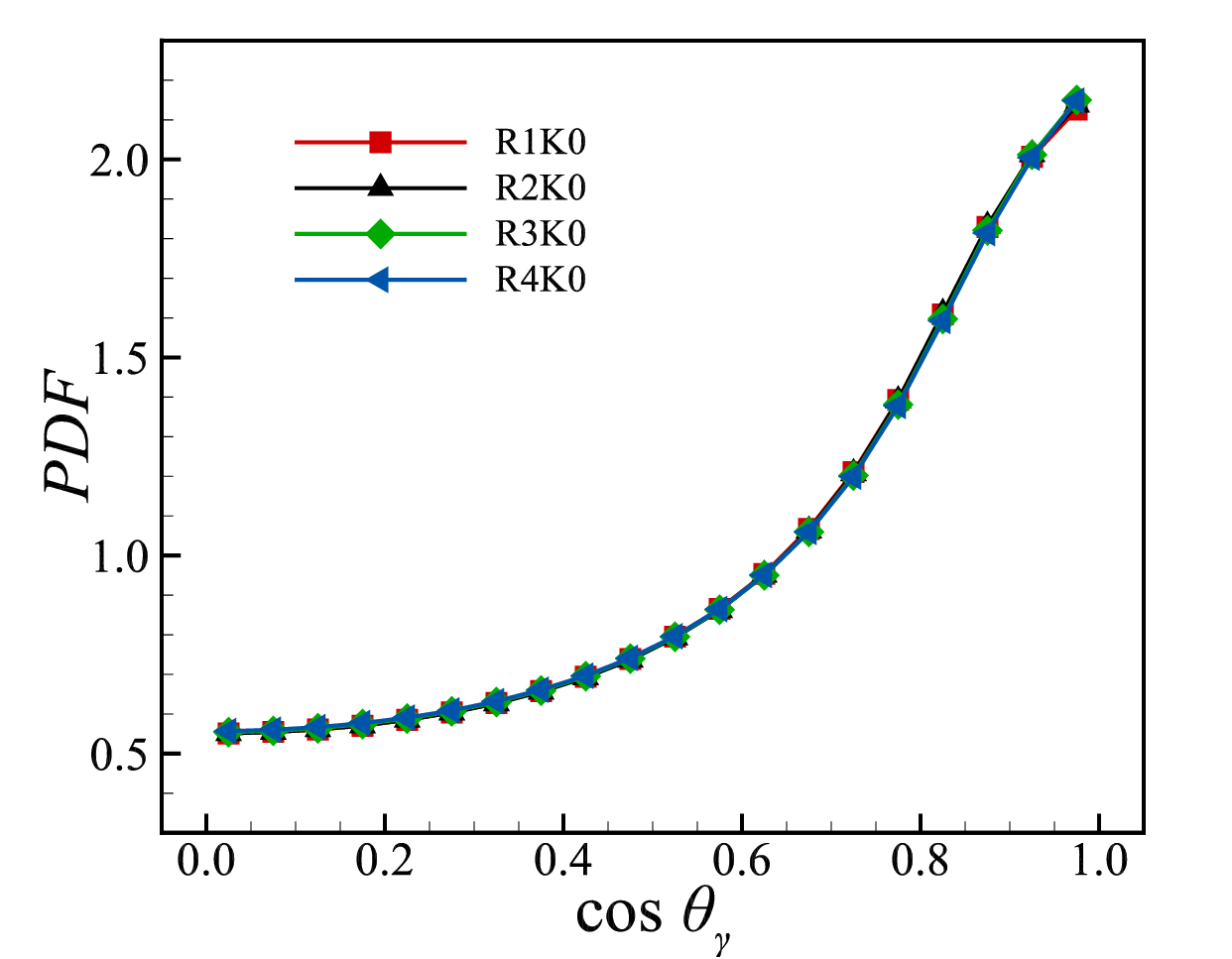} 
\ig[width=0.48\lnw]{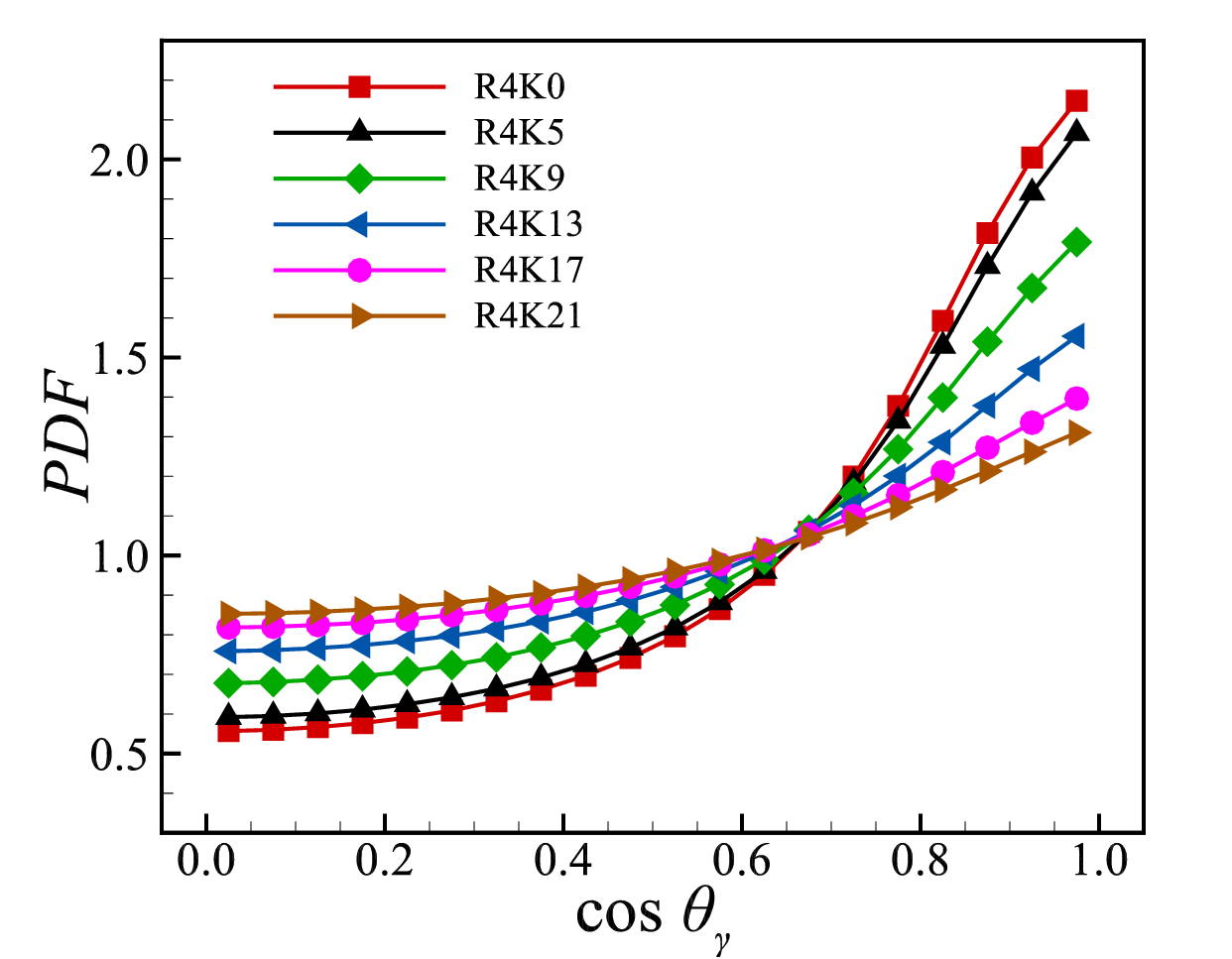}
\caption{\label{fig:alignment_v_s} The PDFs of $\cos\theta_\gamma$.
Left: cases from all groups with $k_m = 0$. Right: cases in group
R4 with different $k_m$.}
\efig 
\bfig
\centering
\ig[width=0.48\lnw]{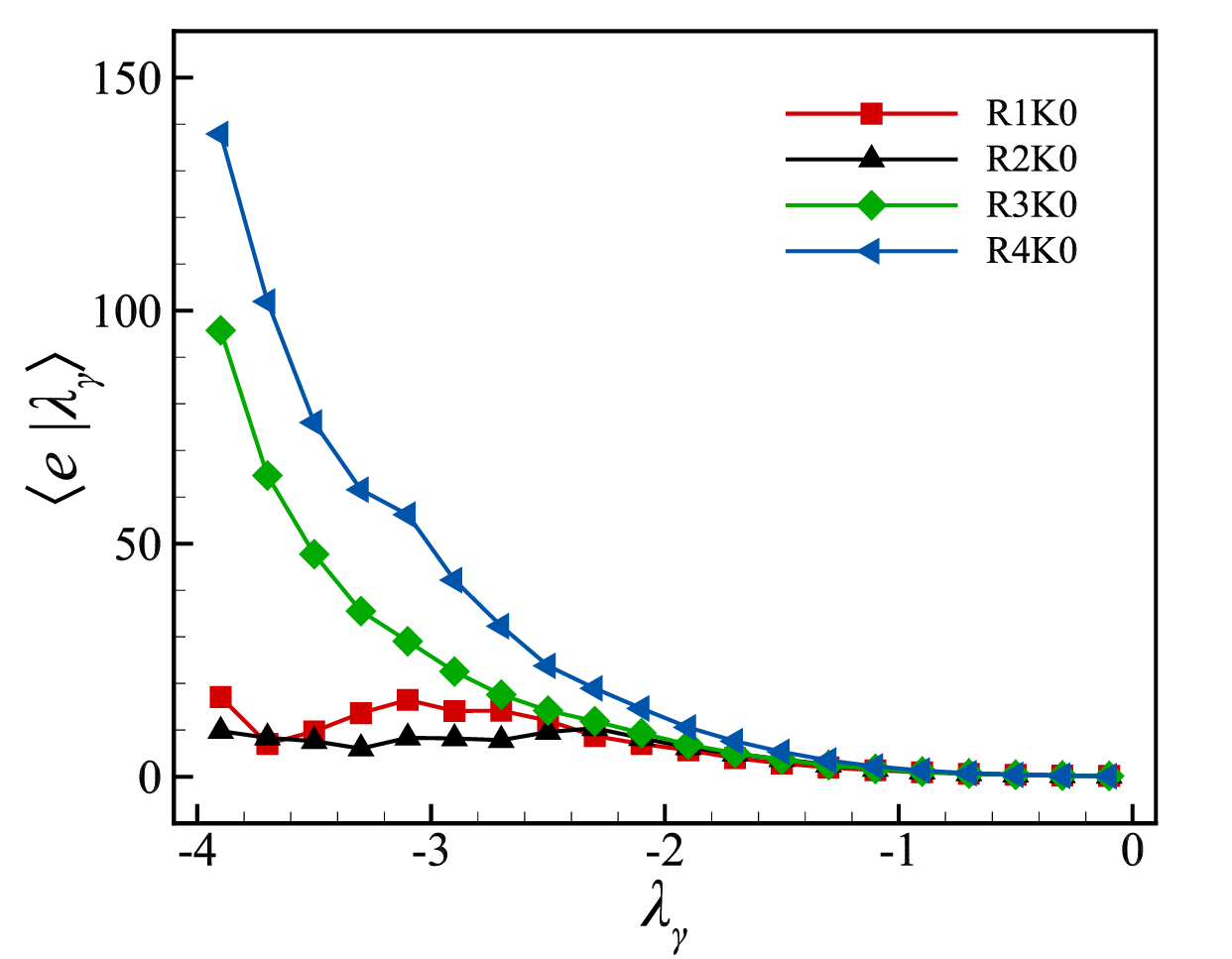} 
\ig[width=0.48\lnw]{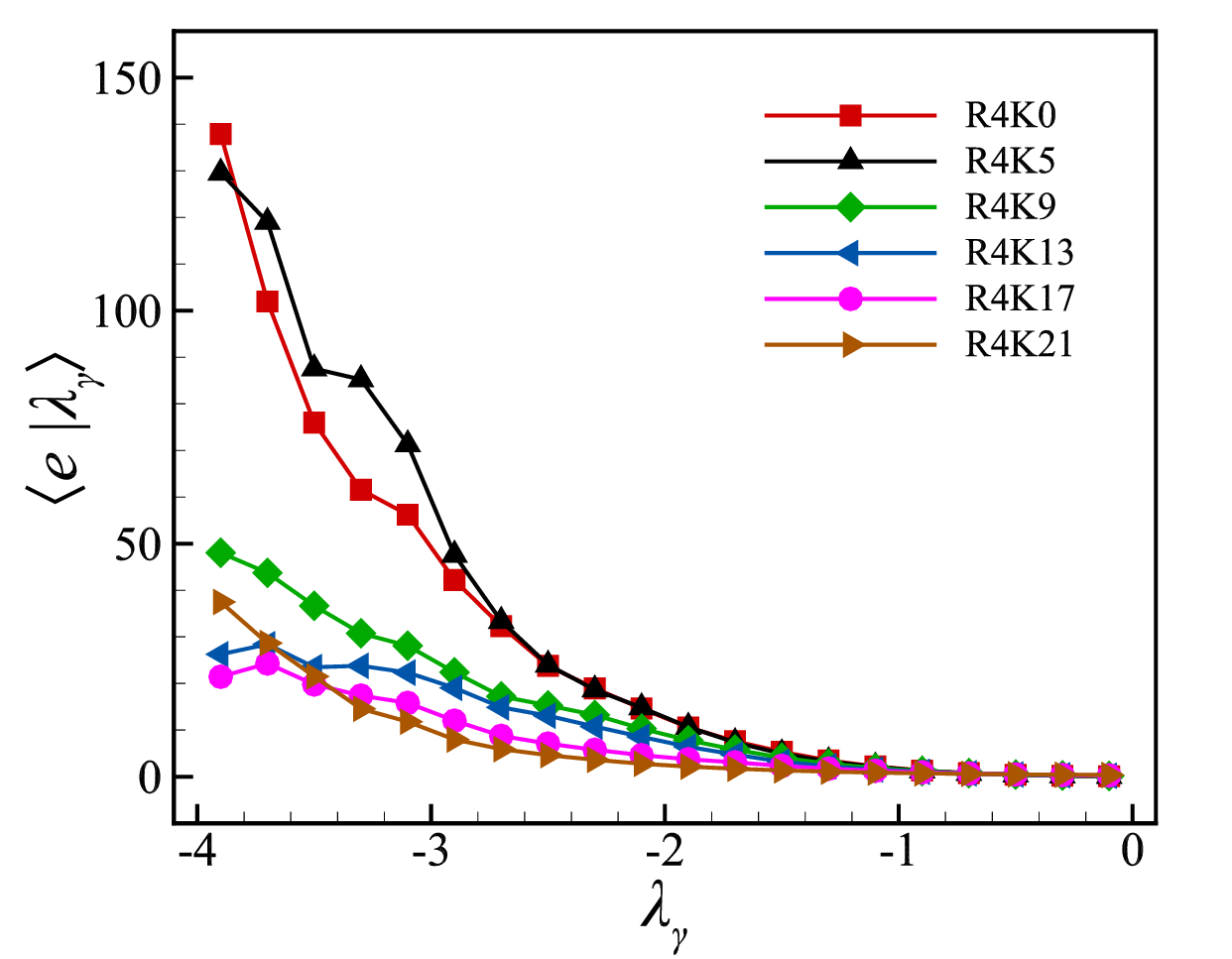}
\caption{\label{fig:conditional_average_e_gamma} The conditional average $
\lal e \vert \lambda_\gamma \ral $ as a function of $\lambda_\gamma$.
Left: cases from
all groups with $k_m = 0$. Right: cases in group R4 with different
$k_m$.}
\efig

\bfig
\centering
\ig[width=0.48\lnw]{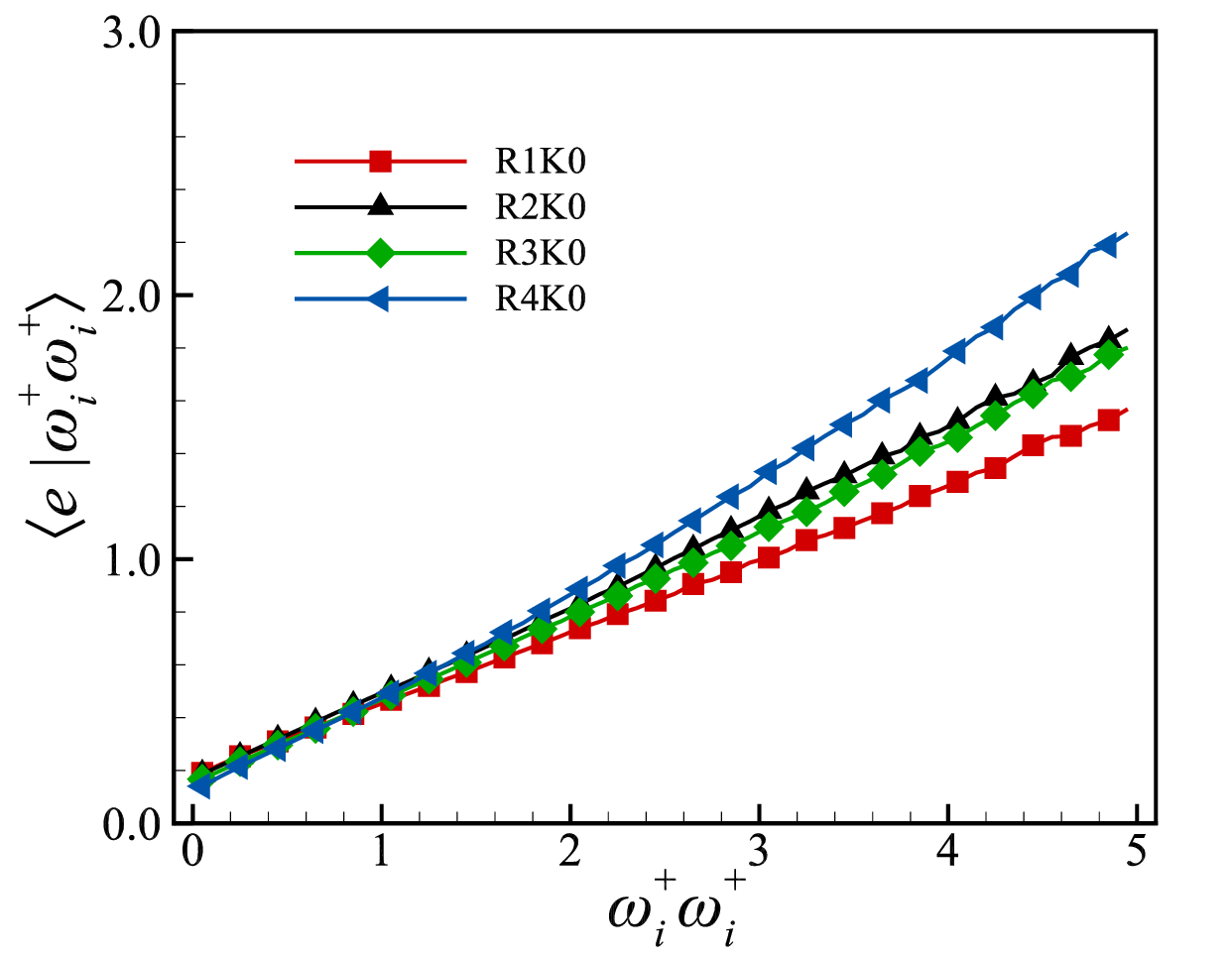} 
\ig[width=0.48\lnw]{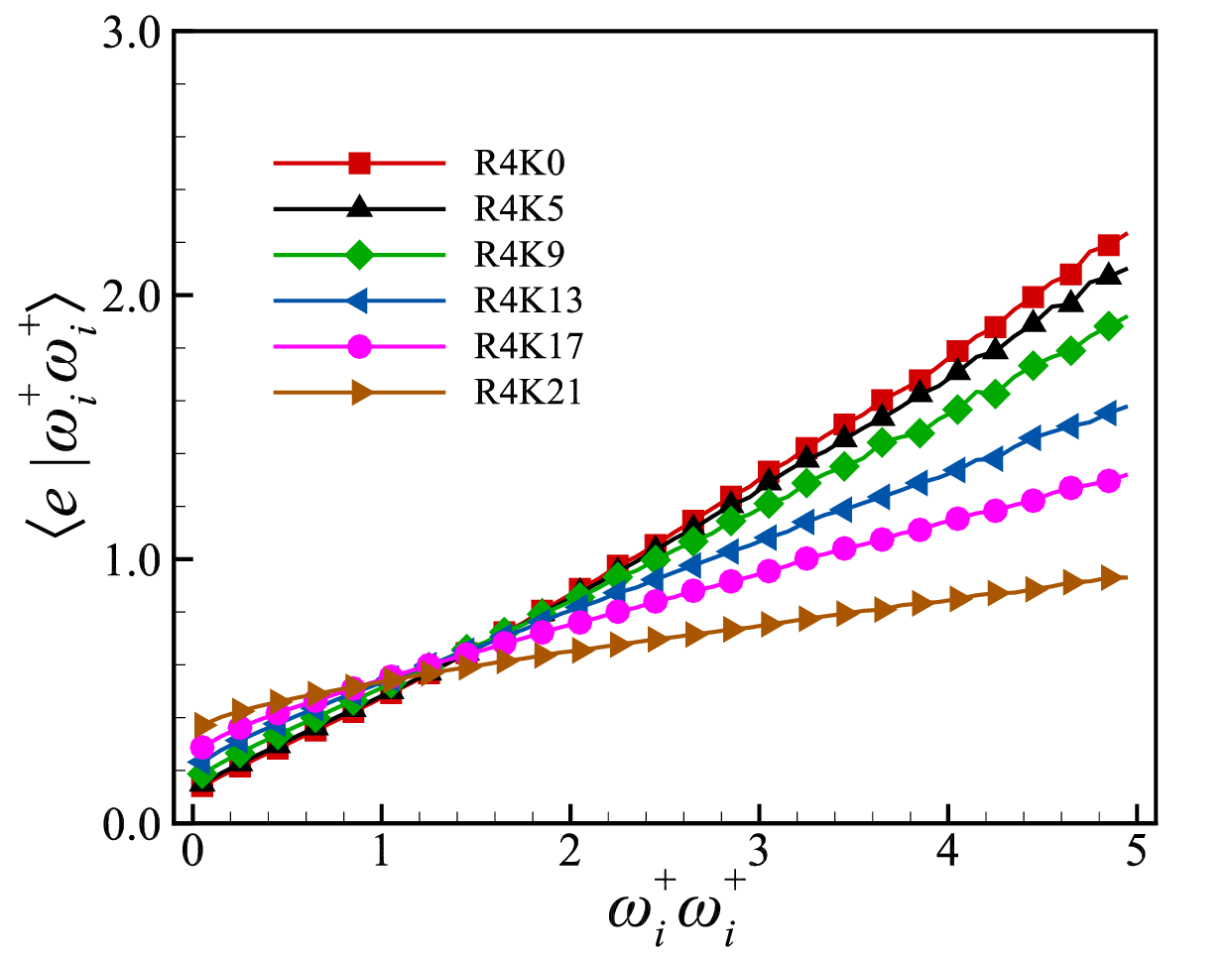}
\caption{\label{fig:conditional_average_e_o} The conditional average $
\lal e \vert \om_i^+\om_i^+ \ral $. Left: cases from all groups
with $k_m = 0$. Right: cases in group R4 with different
$k_m$.}
\efig  

The production components $P_\alpha$, $P_\beta$, and $P_\gamma$ as
functions of $k_m \eta$
are shown in Fig. \ref{fig:P_eigensij_mean_kmeta}. 
One can observe that $P_\alpha$, 
$P_\beta$, and $P_\gamma$ exhibit negligible dependence on $Re_\lambda$. 
As expected, $P_\alpha < 0$ and  $P_\gamma > 0$. We also observe
$P_\beta < 0$ with much smaller 
magnitudes, and $P_\gamma$ is the dominant one among the three terms.
Of particular note is that $P_\alpha$ and $P_\beta$ are nearly 
independent of $k_m \eta$, whereas $P_\gamma$ decreases as $k_m \eta$ increases. 
Thus, the change in $P$ with respect to $k_m \eta$ (as shown 
in Fig. \ref{fig:P_D_P-D_kmeta}) is predominantly due to the 
contribution from $P_\gamma$. As a result, we focus on $P_\gamma$ only
in what follows. 

 Fig. \ref{fig:alignment_v_s} presents the PDF of $\cos\theta_\gamma$. 
 As can be observed in the left panel of 
 Fig. \ref{fig:alignment_v_s}, results for
 different
 Reynolds numbers are essentially the same. 
 The PDFs
 increase with $\cos\theta_\gamma$, 
suggesting a strong tendency for the 
 eigenvector $\e_\gamma$ to align with the vector $\bv$.
 There are noticeable differences between the PDFs for different
 $k_m$, as shown on the right panel of  
 Fig. \ref{fig:alignment_v_s}. 
 As $k_m$ increases, the preferential alignment 
 between $\e_\gamma$ and $\bv$ is weakened, 
 manifested in the lower peaks. 
 This behaviour clearly is one of the 
 reasons why $P$ decreases with $k_m$
as shown in Fig. \ref{fig:P_D_P-D_kmeta}. 
 The PDFs at other Reynolds numbers exhibit similar 
 trends, thus not shown to avoid redundancy.
 The behaviours shown in the right panel of Fig. \ref{fig:alignment_v_s} can be
 qualitatively understood from the characteristics length scales or wavenumbers
 of $s^+_{ij}$ and $v_i$. The characteristic wavenumber of $s^+_{ij}$ can be estimated
 by the wavenumber where the dissipation spectrum of the base flow peaks, which 
 is found to be approximately $0.15\eta^{-1}$.
 The characteristic wavenumber for $v_i$ can be 
 estimated by $\max (0.2 \eta^{-1}, k_m)$, with $0.2 \eta^{-1}$ being the peak wavenumber 
 for $\lal E_v \ral$ when it is bigger than $k_m$. 
 Thus, as $k_m$ increases, 
 the mismatch between the two characteristic wavenumbers tends to  
 increases, which tends to weaken the correlation between $s^+_{ij}$ and $v_i$, 
 hence the alignment in Fig. \ref{fig:alignment_v_s} (right). 
 
Incidentally, the preferential alignment discussed above is
reminiscent of the behaviours of the
gradient of a passive scalar in
isotropic turbulence, which also tends to 
align with $\e_\gamma$ of the strain rate tensor
\citep{Ashurstetal87a}. 
However, the statistics of $\bv$ are different from those of a
passive scalar on many aspects, as we can see from the statistics of
$B_{ij}$ which will be discussed later. 

The impacts of the correlation between the perturbation and the strain
rate tensor can be explored with suitable conditional statistics. 
Note that
\begin{equation}
P_\gamma = -2 \int \lal e \cos^2 \theta_\gamma \vert \lambda_\gamma \ral \lambda_\gamma
f_\gamma(\lambda_\gamma) d \lambda_\gamma,
\end{equation}
where $f_\gamma(\lambda_\gamma)$ is the PDF of the eigenvalue
$\lambda_\gamma$. Therefore, how the correlation between the
$\lambda_\gamma$, the alignment, and $e$ contributes  to $P_\gamma$
can be understood from the conditional average $\lal e \cos^2
\theta_\gamma \vert \lambda_\gamma\ral$. Our tests show that $\lal e
\cos^2 \theta_\gamma \vert \lambda_\gamma\ral$ tends to be smaller
than $\lal e \vert \lambda_\gamma \ral$ due to the factor $\cos^2
\theta_\gamma$, but the two distributions display similar shapes. To
keep the discussion succinct, we consider only 
$\lal e \vert \lambda_\gamma \ral$, which is
shown in  Fig. \ref{fig:conditional_average_e_gamma}. 
The left panel, firstly, shows that 
$ \lal e \vert \lambda_\gamma\ral$ 
changes with the Reynolds number quite significantly, in contrast to the
alignment trend shown in Fig. \ref{fig:alignment_v_s}. The 
impact of the Reynolds numbers is especially strong for large $\vert
\lambda_\gamma \vert$, where the conditional average generally 
is larger for larger
Reynolds numbers. This behaviour is likely due to the fact that the probability 
for strong strain rate increases with the Reynolds number due to the intermittency effects. 
The right panel plots $\lal e
\vert \lambda_\gamma\ral$ 
for different $k_m$ with a fixed Reynolds number. The conditional average
increases with $\vert \lambda_\gamma\vert$ for all $k_m$, but it 
is generally smaller for larger $k_m$. 
This behaviour is another factor by which $P$ decreases
as $k_m$ increases, in addition to the weakened 
alignment shown in the right panel of Fig. \ref{fig:alignment_v_s}. The reduction in 
$\lal e \vert \lambda_\gamma\ral$ as $k_m$ increases may also be attributed to
the mismatch between the characteristics length scales of $s^+_{ij}$ and $v_i$.  
Overall, Fig. \ref{fig:conditional_average_e_gamma} shows that the perturbation tends to 
be stronger at regions with stronger strain rate (shown with larger $\vert
\lambda_\gamma\vert$), though it is reduced as $k_m$ increases.  

The descriptions of the velocity perturbation can be further elaborated 
by considering the correlation between $e$ and the 
base flow vorticity. Let $\bo = \nabla \times \bu$ be the vorticity of
the base flow, and $\bom^+ = \tau \bo$ be the non-dimensionalised
version of $\bo$. Fig. \ref{fig:conditional_average_e_o} shows the
conditional average $\lal e \vert \om_i^+\om_i^+ \ral$, which
characterises the
 correlation between the magnitude of the perturbation and the
 vorticity of the base flow. When $k_m = 0$, Fig.
 \ref{fig:conditional_average_e_o} shows that the conditional average
 increases with the magnitude of base flow vorticity, 
 and it depends only weakly on the Reynolds number. 
 When $k_m$ is increased, the dependence of $ \lal e \vert \om_i^+
 \om_i^+\ral $ on $\om_i^+ \om_i^+$ is weakened. As a result, the
 conditional average increases with $k_m$ for smaller
$\om_i^+ \om_i^+$ and decreases with $k_m$ for larger $\om_i^+
\om_i^+$.  
 Therefore, perturbations associated with 
 regions of strong vorticity in the base flow are stronger on average,
 and this trend tends to be
 weakened as the coupling wavenumber $k_m$ increases. 
 The conditional average $\lal e\vert \om_i^+\om_i^+\ral$ 
depends on the Reynolds number and $k_m$ in the same way as 
$\lal e \vert \lambda_\gamma\ral$, 
 thus its behaviours can be explained in a similar way. 
 The correlation between $e$ and 
 strong strain rate and strong vorticity  
 is also observed in channel flows to some extent \citep{Nikitin2018}.

\bfig
\centering
\ig[width=0.48\lnw]{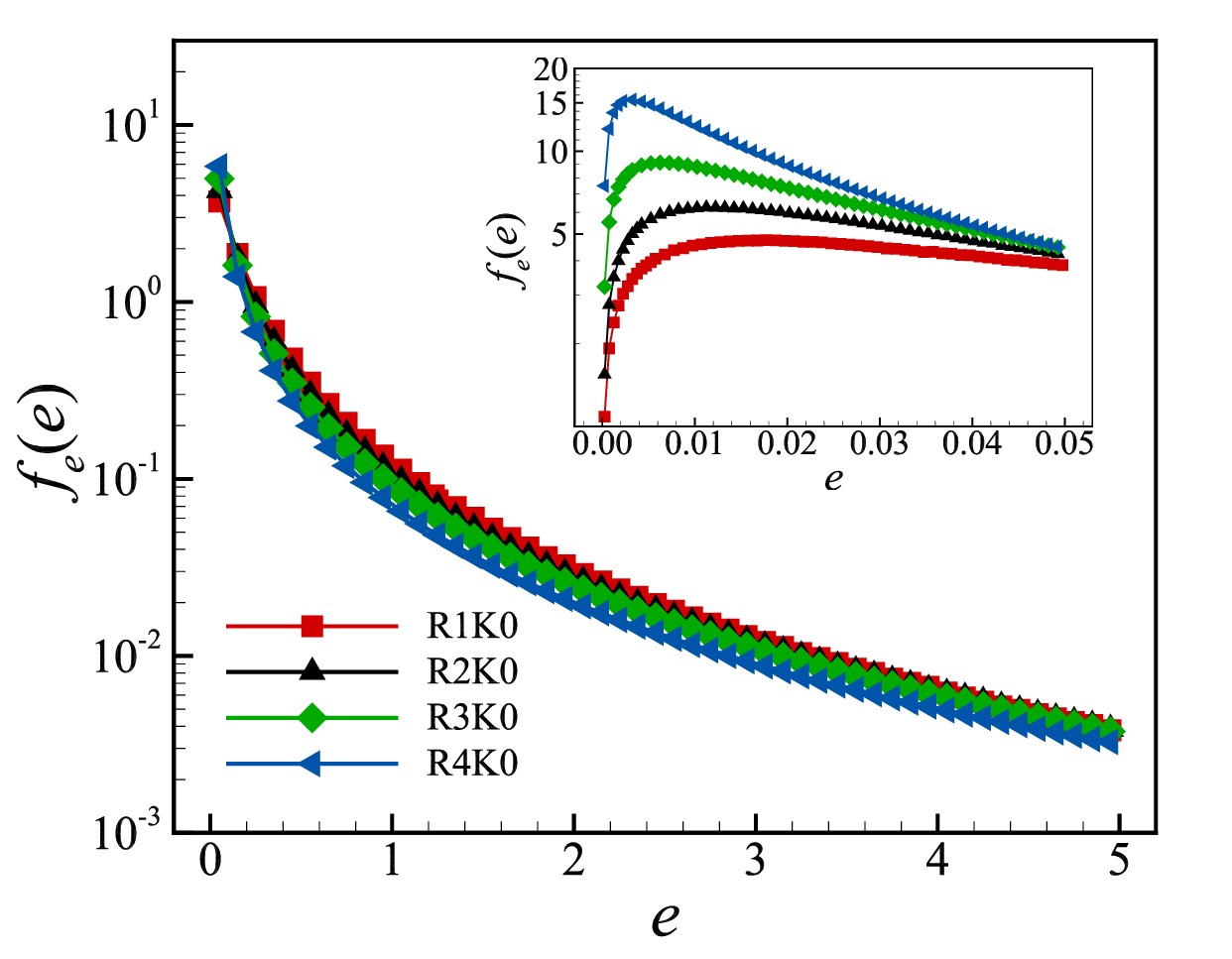}
\ig[width=0.48\lnw]{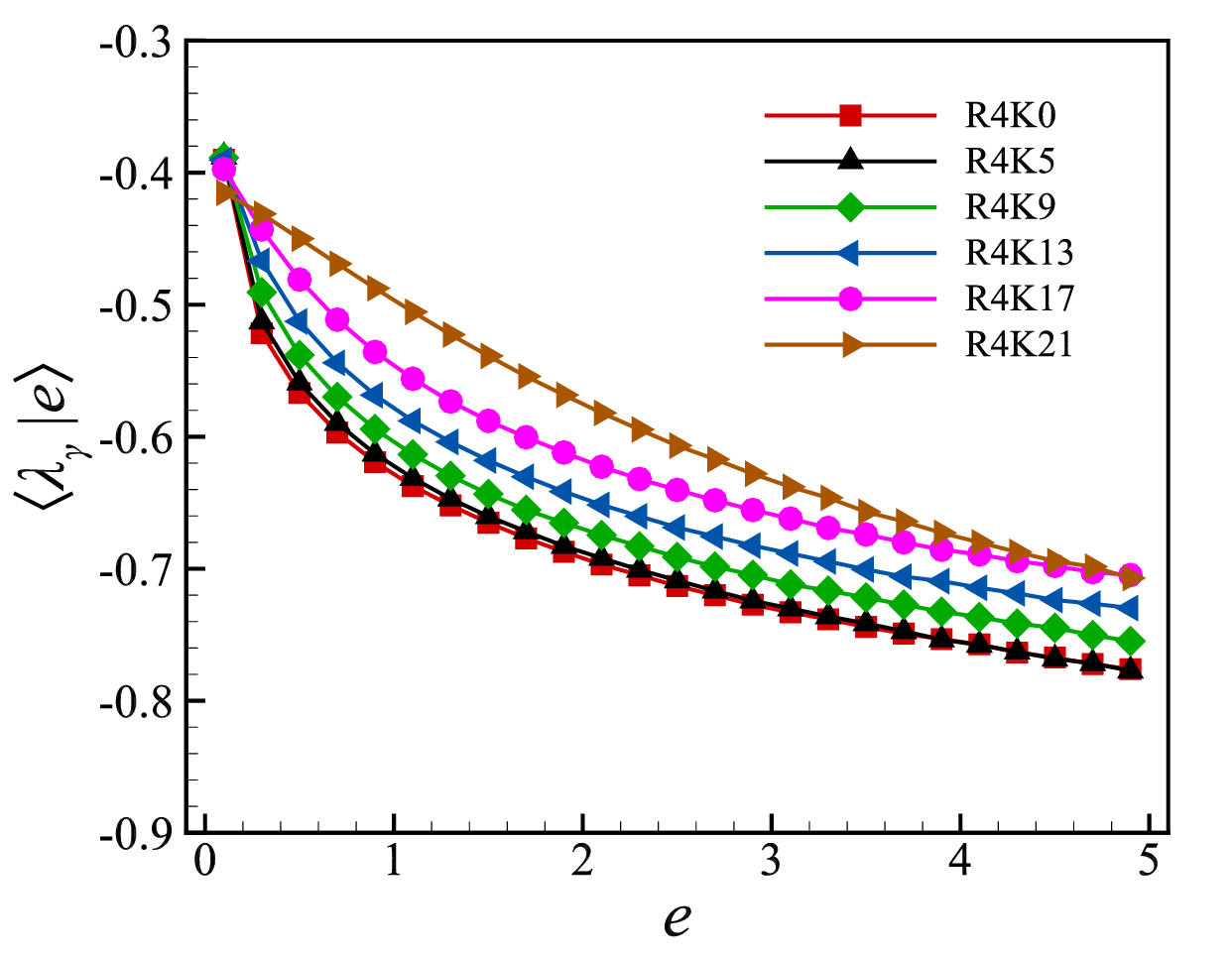} 
\caption{\label{fig:P_cnd_e} Left: PDFs $f_e(e)$ from all groups
with $k_m=0$ with the inset showing the variation of
the PDFs with the Reynolds number near
$e=0$. Right: conditional average $\lal
\lambda_\gamma\vert e \ral $ for group R4 with different
$k_m$. }
\efig 
To understand how the fluctuations in the perturbation velocity
contribute to the mean production term, we may 
write the mean production as the weighted integral of the average
conditioned on given $e$, i.e., 
\begin{equation} 
P_\gamma = -2 \int \lal \lambda_\gamma \cos^2 \theta_\gamma \vert
  e\ral e f_e (e) de, 
\end{equation} 
where $f_e(e) $ is the PDF of $e$. 
The left panel of Fig. \ref{fig:P_cnd_e} plots $f_e(e)$ for
different Reynolds numbers with $k_m=0$ only. The PDFs display very
elongated tails, showing high probabilities for large fluctuations in
$e$. The tail is very slightly fatter for higher Reynolds numbers.
The peaks of the PDFs are found at small $e$ values, and they are
slightly sharper for higher Reynolds numbers. The distributions
indicate
that the spatial distribution of the perturbation velocity is highly
intermittent, with small fluctuations covering large part of the
spatial domain and strong fluctuations observed in localised spots.  
The results for $\lal \lambda_\gamma \vert e\ral$ are shown in the
right panel of 
 Fig. \ref{fig:P_cnd_e}, The magnitude of the conditional average
 increases with $e$ and $k_m$. These
 behaviours are consistent with the results for $\lal e
 \vert \lambda_\gamma\ral$. Given the highly intermittent nature of
 the distribution of $e$, one might ask how important are the large
 fluctuations to the mean production. Though its
 figure is omitted, one can readily see that 
the product $-\lal \lambda_\gamma \cos^2\theta_\gamma \vert e \ral e
f_e(e ) \approx - \lal
 \lambda_\gamma \vert e \ral e f_e(e)$, as a function of $e$, peaks at an
 intermediate value of $e$. Therefore, the main contribution to the
 mean production does not come from the very large fluctuations.


We now explore the behaviours of the dissipation term
$\DD_e$ in the physical space. Eq. (\ref{eq:bb}) shows that $\NN_e$ is
one of the main mechanisms that determines the dissipation rate $\DD_e$,
and it will be the focus below. 
We let 
\begin{equation} \label{eq:NesNos_norm}
N_{es} = \frac{\tau\lal \NN_{es}\ral}{\lal \DD_e \ral}, \qquad
  N_{eo} = \frac{\tau \lal
\NN_{eo}\ral}{\lal \DD_e \ral},
\end{equation}
which are both dimensionless (c.f. Eq. (\ref{eq:bb})). Equivalently, we
may introduce  
a length scale $(\bar{\nu}/\lal \DD_e \ral)^{1/2}$, and
then define non-dimensional perturbation strain rate and
perturbation 
vorticity, denoted by $s_{ij}^-$ and $\om^-_{ij}$, respectively, with
\begin{equation} 
  s^-_{ij} =\frac{\bar{\nu}^{1/2}  s^v_{ij}}{\lal \DD_e \ral^{1/2}}, \qquad
  \om^-_{ij} = \frac{\bar{\nu}^{1/2} \om^v_{ij}}{\lal \DD_e \ral^{1/2}}.
\end{equation}
Recalling that $\bv$ is dimensionless, therefore the dimension of
$s^v_{ij}$ and $\om^v_i$ is that of the reciprocal of length, so that
$s^-_{ij}$ and $\om^-_i$ are dimensionless. As a consequence, we obtain
\begin{equation}
N_{es} = -4 \lal s^+_{ij} s^-_{jk} s^-_{ki} \ral, \quad N_{eo} =
 \lal s^+_{ij} \om^-_i \om^-_j\ral. 
\end{equation} 
In terms of $N_{es}$ and $N_{eo}$,
we may re-write Eq. (\ref{eq:mBB}). Assuming the correlation between
$\gamma$ and $\DD_e$ is negligible, we obtain $\lal \gamma \DD_e \ral
\approx \lal \gamma \ral \lal \DD_e \ral = \lambda \lal \DD_e \ral$.
Therefore, Eq. (\ref{eq:mBB}) becomes
\be \label{eq:mBB1}
2 \Lambda \approx N_{es} + N_{eo} + r_e 
\ee
where 
\be 
r_e \equiv - \frac{2 \bar{\nu} \tau \lal s^v_{ij}
\ptl^2_{ij} p_e \ral}{\lal \DD_e \ral} - \frac{2 \tau \lal \bar{\nu} \nabla
B_{ij} \cdot \bar{\nu} \nabla B_{ij} \ral}{ \lal \DD_e \ral} + \frac{2 \tau
\bar{\nu} \lal B_{ij} \nabla \cdot( \bv A_{ij})\ral }{ \lal \DD_e \ral}, 
\ee
is considered a `residual' term. 
Therefore, the values of $N_{eo}$ and $N_{es}$ can be compared with
$2\Lambda$ to gauge their contributions. 

The expression of $N_{eo}$ is similar to that of $P$ in form with
$v_i$ replaced by $\om^-_i$. It is
also similar in form to the vortex stretching term $s_{ij}^+ \om^+_i
\om^+_j$ for the enstrophy of the base flow. 
Using the eigen-frame defined by
the eigenvectors of $s^+_{ij}$ introduced
previously, we may write 
\begin{equation} \label{eq:Neo1}
N_{eo} = N_{eo\alpha} + N_{eo\beta} + N_{eo\gamma},
\end{equation} 
with 
\begin{equation} \label{eq:Neo2}
N_{eo\alpha} = \lal \lambda_\alpha \vert \bo^{-}\vert^2
\cos^2\theta^o_\alpha \ral, ~ N_{eo\beta} =   \lal
  \lambda_\beta \vert \bo^{-}\vert^2
\cos^2\theta^o_\beta \ral, ~ N_{eo\gamma} =   \lal
  \lambda_\gamma \vert
\bo^{-} \vert^2 \cos^2 \theta^o_\gamma \ral, 
\end{equation} 
where $\theta^o_i$ ($i=\alpha, \beta, \gamma$) denotes the angle
between $\bo^-$ and $\e_i$. Eqs. (\ref{eq:Neo1}) and
(\ref{eq:Neo2}) are similar to those for $P$. Similarly, $N_{es}$ has
an expression in terms of the 
eigenvalues and eigenvectors of $s^+_{ij}$ and
$s^-_{ij}$. Let $\e_\ell^-$ ($\ell=\alpha, \beta, \gamma$) be the
eigenvectors of $s^-_{ij}$, with corresponding eigenvalues
$\lambda^-_\ell$. We follow the same tradition where $\lambda^-_\alpha
\ge \lambda^-_\beta \ge \lambda^-_\gamma$. Letting $\theta^s_{ij}$ be
the angle between $\e^-_i$ and $\e_j$, we obtain  
\begin{align} \label{eq:Nes}
N_{es} = -4 &\left[\lal (\lambda_\alpha^-)^2 \lambda_\alpha \cos^2
  \theta^s_{\alpha \alpha}\ral + \lal (\lambda_\alpha^-)^2
  \lambda_\beta \cos^2 \theta^s_{\alpha \beta}\ral  + \lal
  (\lambda_\alpha^-)^2 \lambda_\gamma \cos^2 \theta^s_{\alpha \gamma}
  \ral\right.  \notag \\
+  & \left.\phantom{[} \lal (\lambda_\beta^-)^2 \lambda_\alpha \cos^2
  \theta^s_{\beta \alpha}\ral  + \lal (\lambda_\beta^-)^2
  \lambda_\beta \cos^2 \theta^s_{\beta \beta}\ral  + \lal
  (\lambda_\beta^-)^2 \lambda_\gamma \cos^2 \theta^s_{\beta
  \gamma}\ral  \right. \notag \\
+ &\left.\phantom{[}  \lal (\lambda_\gamma^-)^2 \lambda_\alpha \cos^2
  \theta^s_{\gamma\alpha}\ral  + \lal (\lambda_\gamma^-)^2
  \lambda_\beta \cos^2 \theta^s_{\gamma \beta}\ral  + \lal
  (\lambda_\gamma^-)^2 \lambda_\gamma \cos^2\theta^s_{\gamma
  \gamma}\ral \right].
\end{align}
The equation provides a decomposition of $N_{es}$ into contributions 
associated with different eigenvalues of the two tensors and
makes explicit how the relative orientation of the eigenvectors
affects $N_{es}$. We will use $N_{es\alpha \alpha}$, $N_{es\alpha
\beta}$, ... $N_{es\gamma \gamma}$ to denote the nine components on
the right hand side of the equation. We also let $N_{es\alpha} =
N_{es\alpha \alpha} + N_{es \alpha \beta} + N_{es\alpha \gamma}$
represent the sum of the contributions involving the eigenvalue
$\lambda^-_\alpha$,
$N_{es\beta} = N_{es\beta\alpha } + N_{es\beta\beta} +
N_{es\beta\gamma}$ represent the sum of the contributions
involving $\lambda^-_\beta$, and $N_{es\gamma} = N_{es\gamma\alpha} +
N_{es\gamma\beta} + N_{es \gamma \gamma}$ represent the 
sum of the contributions
involving $\lambda^-_\gamma$. 
Obviously, 
we have $N_{es} =
N_{es\alpha} + N_{es\beta} + N_{es\gamma}$. 

\bfig
\centering
\ig[width=0.32\lnw]{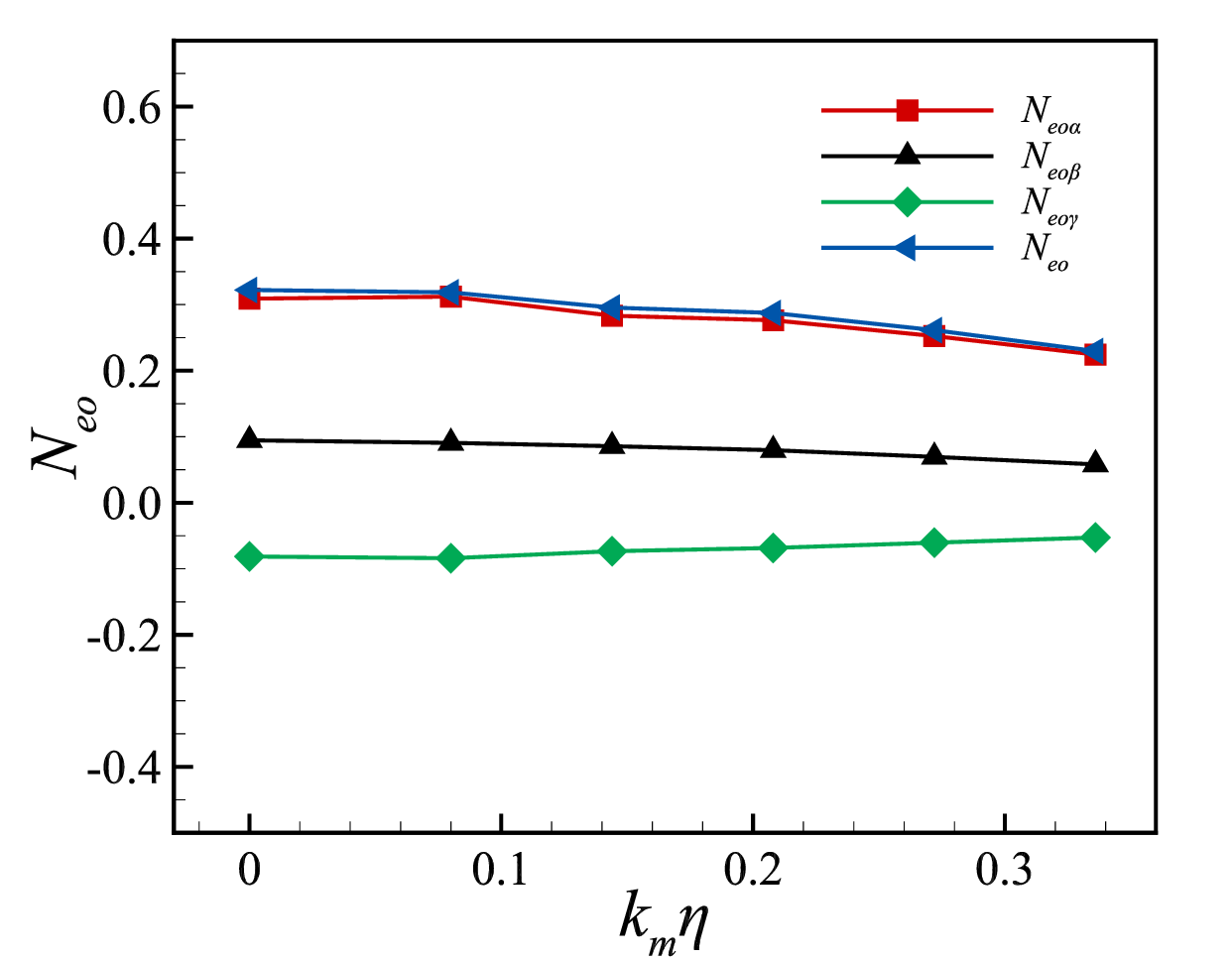} 
\ig[width=0.32\lnw]{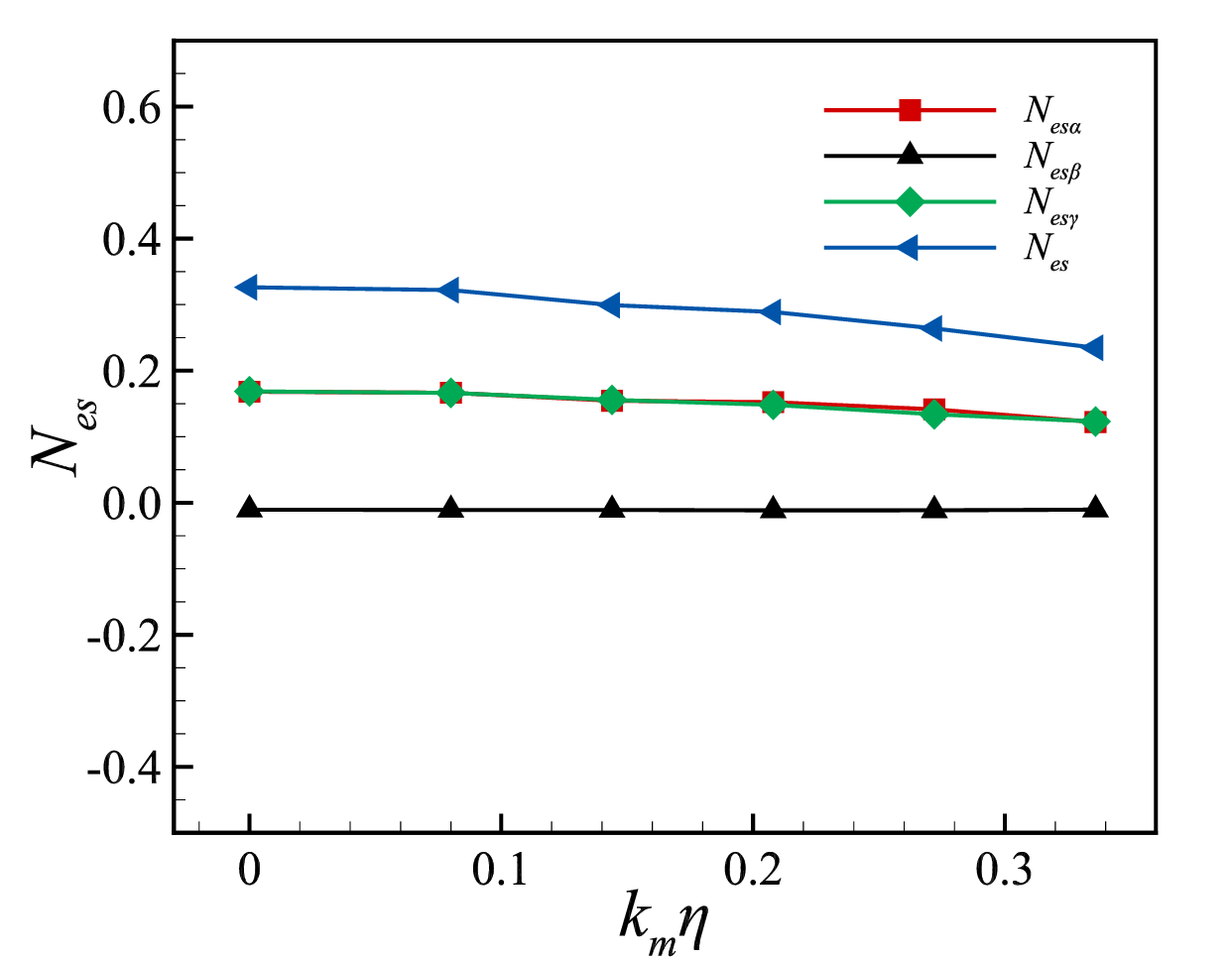} 
\ig[width=0.32\lnw]{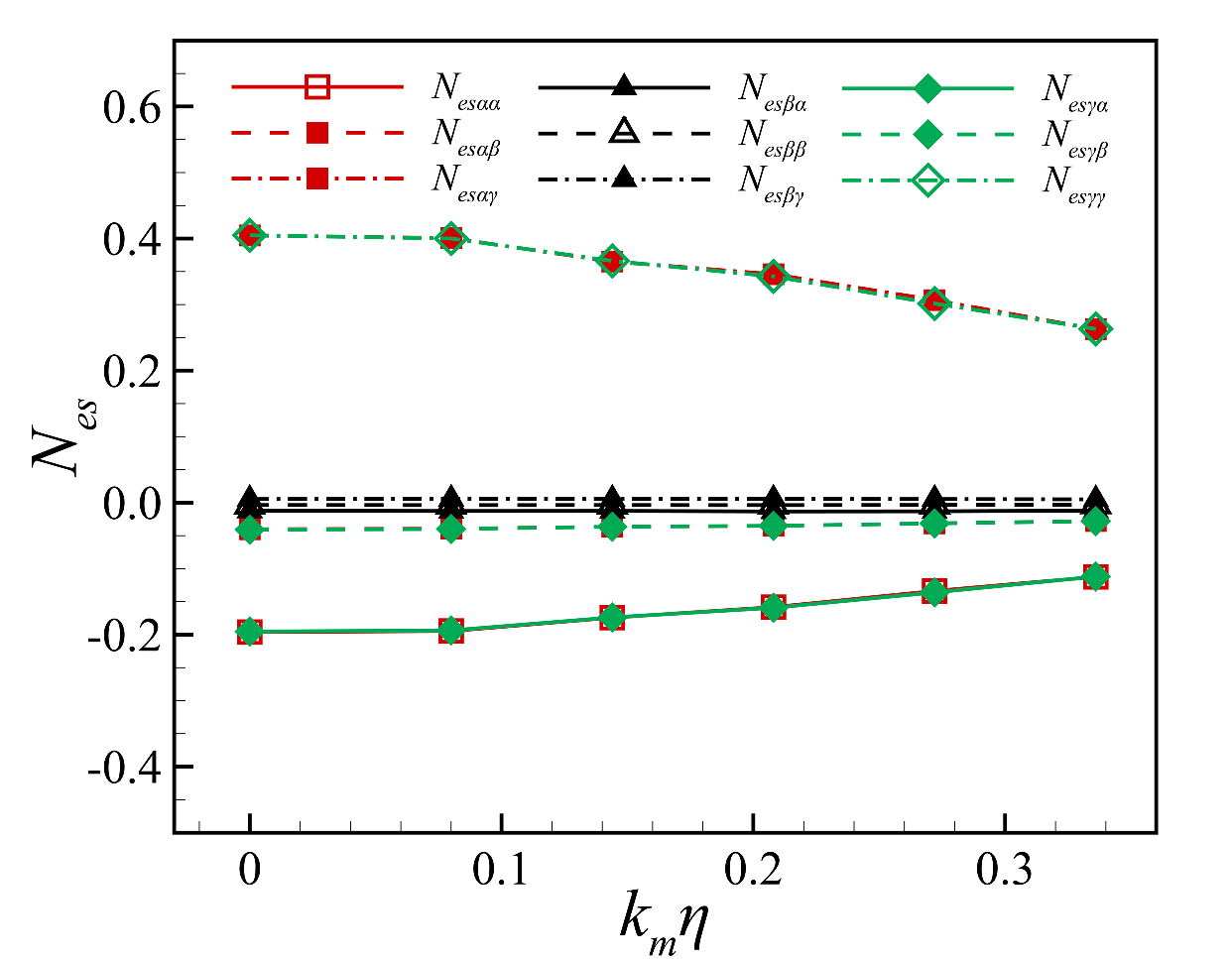} 
\caption{\label{fig:Neo} Left: $N_{eo}$ and its components
as functions of $k_m\eta$. Middle: $N_{es}$ and the 
contributions $N_{es\alpha}$, $N_{es\beta}$ and $N_{es\gamma}$. 
Right: the nine components of $N_{es}$. 
For the cases in group R4.}
\efig 

\bfig
\centering
\ig[width=0.43\lnw]{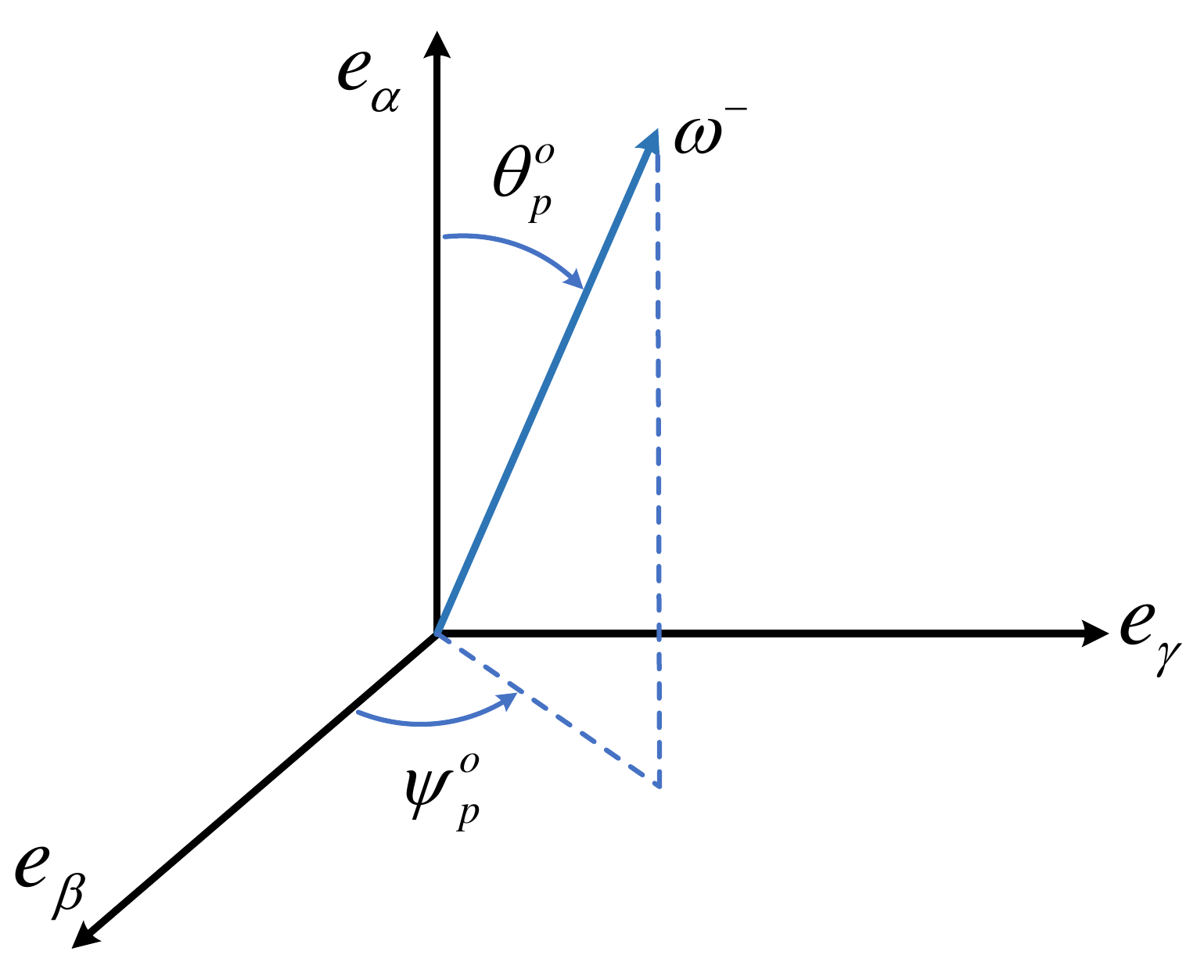}
\caption{\label{fig:eigenframe} The definitions of $\theta^o_P$
and $\psi^o_P$.}
\efig

\bfig
\centering
\ig[width=0.48\lnw]{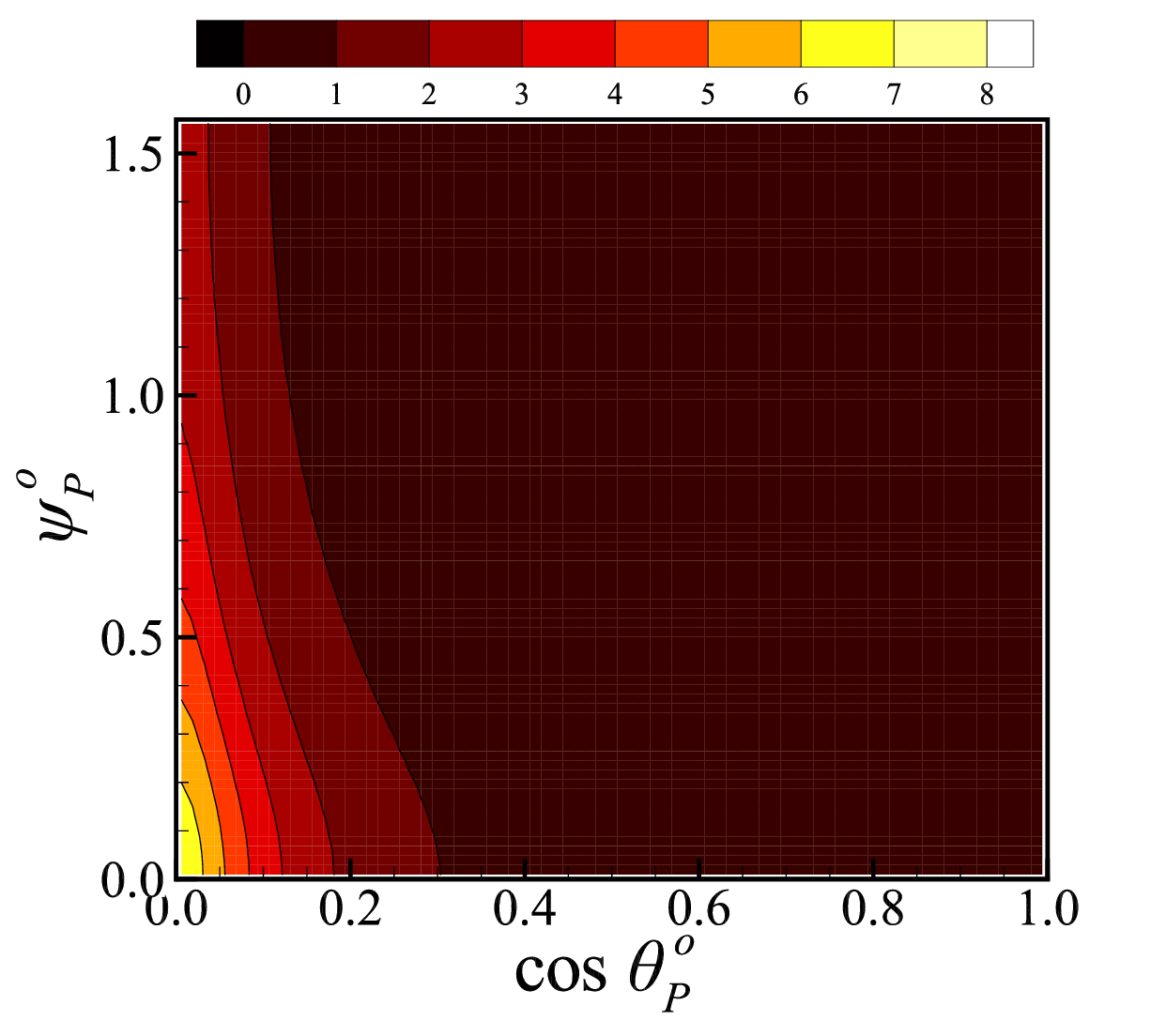} 
\caption{\label{fig:Neo_align} The alignment between $\bo^-$ and
the eigenvectors of $s^+_{ij}$, for case R$4$K$0$. }
\efig

\bfig
\centering
\ig[width=0.32\lnw]{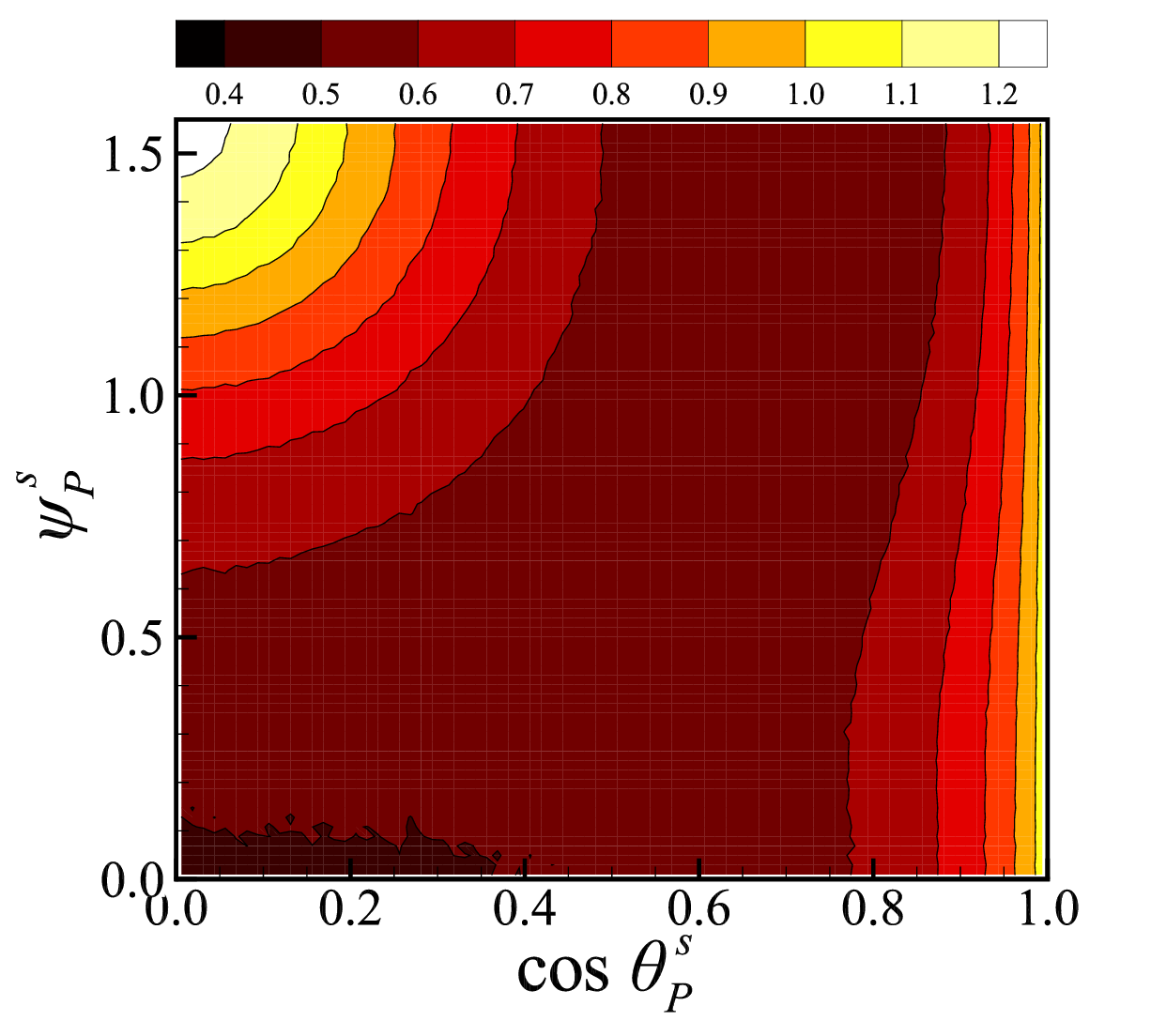} 
\ig[width=0.32\lnw]{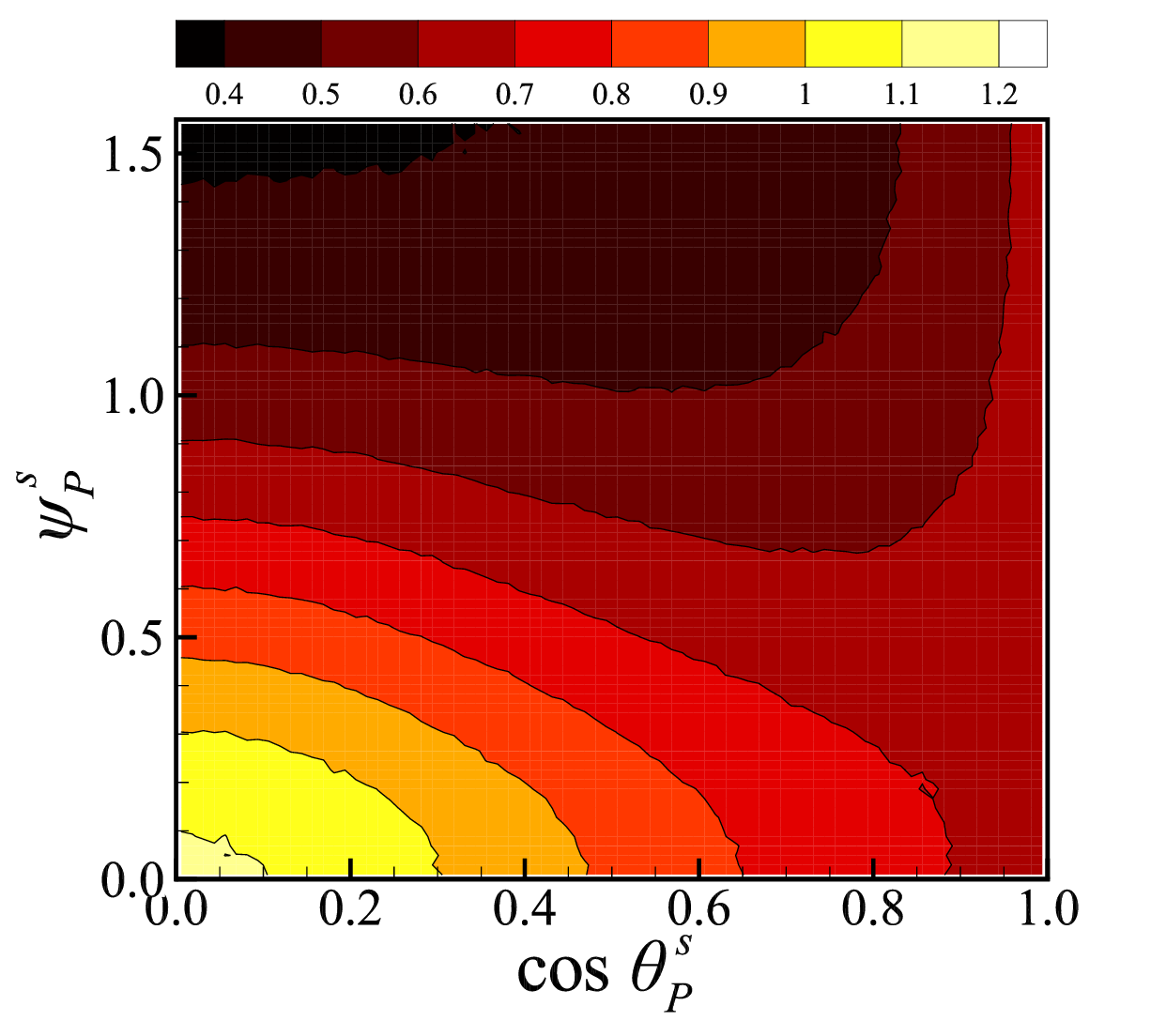} 
\ig[width=0.32\lnw]{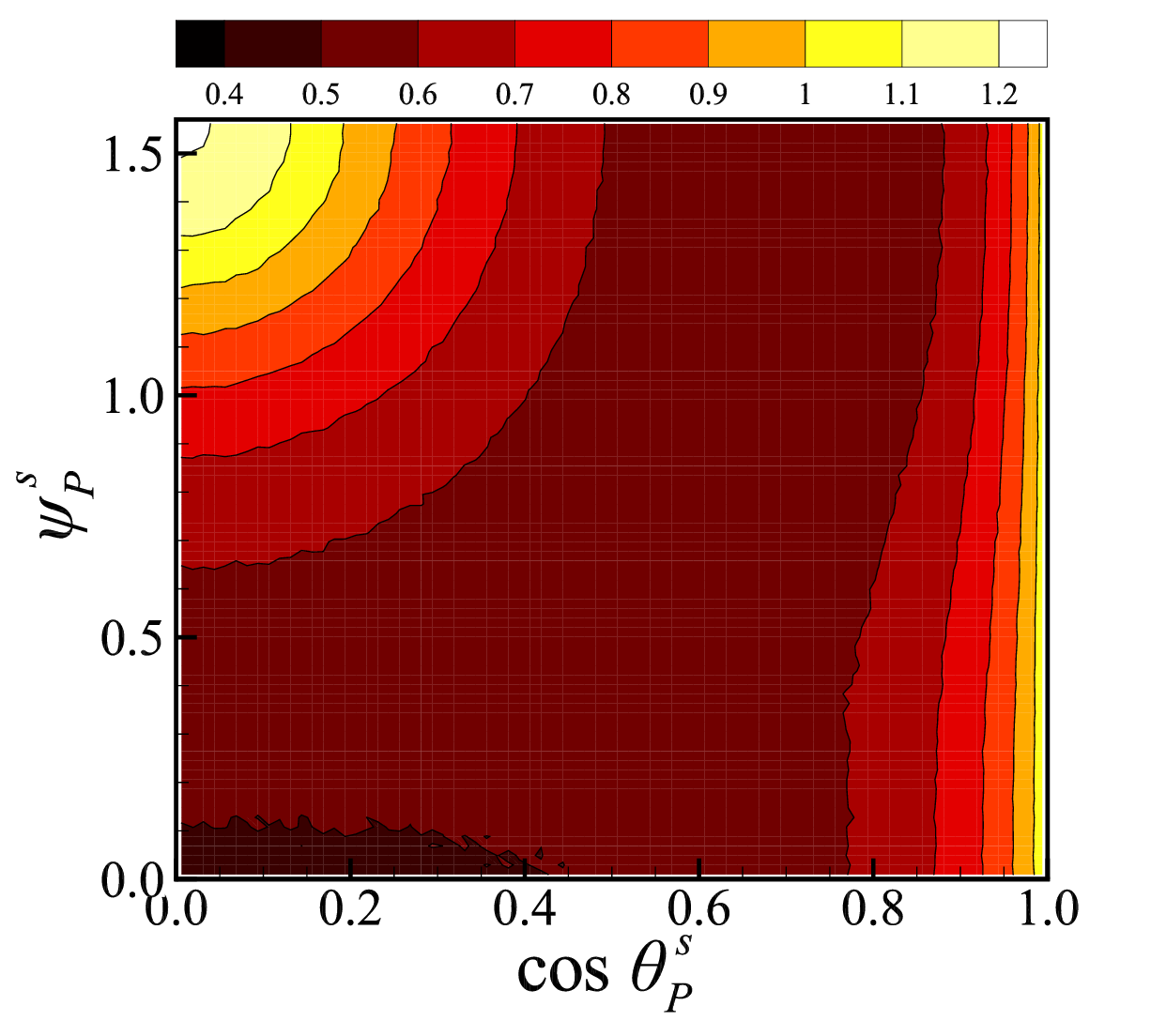}
\caption{\label{fig:Nes_align} The joint PDFs 
of $\cos\theta^s_P$ and $\psi^s_P$ for $\e^-_\al$ (left) 
$\e^-_\beta$ (middle), and $\e^-_\gamma$ (right). For case
R$4$K$0$.}
\efig 

The data for $N_{eo\alpha}$, $N_{eo \beta}$ and $N_{eo\gamma}$ are
plotted on the left panel in Fig. \ref{fig:Neo}.  The magnitudes
of these values
decrease only slightly
as $k_m\eta$ increases. Recalling the normalisation shown in 
Eq. (\ref{eq:NesNos_norm}) and the fact that $\lal \DD_e \ral$ increases
as $k_m\eta$ increases, the conclusion one may draw is thus that
the magnitudes of the \emph{non-normalised versions} of 
$N_{eo\alpha}$, $N_{eo\beta}$ and $N_{eo\gamma}$ 
all increase with $k_m
\eta$, but at rates that are slightly smaller than that of $\lal \DD_e
\ral$, so that the magnitudes of $N_{eo\alpha}$, $N_{eo\beta}$ and
$N_{eo\gamma}$ decrease slightly as $k_m\eta$ increases. 
$N_{eo\beta}$ and $N_{eo\gamma}$
have opposite signs with similar magnitudes.
Consequently, $N_{eo}$ is only slightly different from $N_{eo\alpha}$. 

$N_{es\alpha}$, $N_{es\beta}$, $N_{es\gamma}$, together with
$N_{es}$ are shown in the middle panel of Fig. \ref{fig:Neo}, 
In terms of the dependence on $k_m\eta$, these parameters 
all behaviour
similarly to $N_{eo}$, i.e., their magnitudes all decrease
with $k_m$, but only weakly. Among the three components, 
$N_{es\beta}$ is the smallest and essentially negligible.
$N_{es\alpha}$ and $N_{es\gamma}$ are both much larger and they
appear to be almost the same as each other.  
The breakdown into the nine components is given in the right panel of
Fig. \ref{fig:Neo}. 
We can see that $N_{es\beta \alpha}$,
$N_{es\beta\beta}$, $N_{es\beta\gamma}$, $N_{es\gamma\beta}$, and
$N_{es\alpha\beta}$ are all quite small. The largest contributions come
from $N_{es\alpha \gamma}$ and $N_{es\gamma\gamma}$, which are
positive by definition, and appear
to be identical. The contributions from
$N_{es\alpha\alpha}$ and $N_{es\gamma \alpha}$ are
also significant though somewhat smaller than those from $N_{es\alpha
\gamma}$ and $N_{es\gamma \gamma}$. They also appear to be identical. 

The results given in the right panel 
shows that the close agreement 
between $N_{es\alpha}$ and $N_{es\gamma}$ (middle panel) is  
a consequence of the close agreement between $N_{es\alpha \ell}$ and
$N_{es\gamma \ell}$ ($\ell = \alpha, \beta, \gamma$). 
The agreement between the latter two is, in fact, 
a mathematical consequence of the
linearity of Eq. (\ref{eq:vi}). The linearity of Eq. (\ref{eq:vi}) 
dictates
that $s^-_{ij}$ and $-s^-_{ij}$ must have same
statistics, which implies that the largest eigenvalues of
the two, $\lambda^-_\alpha$ and $-\lambda^-_\gamma$,
respectively, should have the same statistics too. As a result, 
$N_{es\alpha \ell} = N_{es\gamma \ell}$ exactly for all
$k_m$, which is reflected in the figure.

We will not discuss the residual term $r_e$ in detail to keep the
scope of this investigation manageable. Nevertheless, we may use
Eq. (\ref{eq:mBB1}) to obtain an estimate of its
impact by comparing the values of $N_{eo}$ and
$N_{es}$ with $ 2\Lambda$. Recall that, 
according to the right panel of Fig. \ref{fig:energy_decay},
$2 \Lambda \approx 0.24$ when $k_m = 0$. On the other hand,
Fig. \ref{fig:Neo} shows that 
$N_e = N_{eo} + N_{es} \approx 0.7$ for $k_m=0$.
Therefore, the residual term $r_e$ has a significant
contribution, and appears to be acting to counter the effects of 
$N_e$. Furthermore, $\Lambda$ decreases from $0.12$ to $-0.20$ as $k_m \eta$
increases according to Fig. \ref{fig:energy_decay} (for cases in
group R4). Though $N_e$ also decreases with $k_m$, the change is not large
enough to account for the change in $\Lambda$,
which shows that the dependence of $r_e$ on $k_m$ also plays a
role. 

Eqs. (\ref{eq:Neo2}) and (\ref{eq:Nes}) also suggest that 
the relative orientations
between $\bo^-$ and $s^+_{ij}$ 
or between $s^-_{ij}$ and $s^+_{ij}$ might impact the
values of $N_{eo}$ and $N_{es}$. 
We thus look into relevant results, for $N_{eo}$ to begin with.  
The alignment between $\bo^-$ and $s^+_{ij}$ can be characterised 
by the angles $\theta^o_\ell$ ($\ell =
\alpha, \beta, \gamma$) introduced previously. 
However, since this problem has not been investigated before, 
we opt for a more complete
description based on the polar angle $\theta^o_P$ and the
azimuthal angle $\psi^o_P$ that the vector $\bo^-$ make in the
eigen-frame formed by the eigenvectors of $s^+_{ij}$. 
The definitions
of the two angles are illustrated in Fig. \ref{fig:eigenframe}. Specifically,
$\theta^o_P$ is 
the angle between $\bo^-$ and the polar direction $\e_\alpha$, and
$\psi^o_P$ is the angle between $\e_\beta$ and the projection of $\bo^-$ on the
equatorial plane. The relations between $(\theta^o_P
, \psi^o_P)$ and $\theta^o_i$ can be derived readily. 
The joint PDF of $\cos\theta^o_P$ and $\psi^o_P$ is shown in 
Fig. \ref{fig:Neo_align}.  
The joint PDF has a very sharp
peak at the origin, i.e., at $\theta^o_P = 90^\circ$ and $\psi^o_P =
0$. Thus, $\bo^-$ tends to very strongly align with 
the intermediate eigenvector $\e_\beta$ of $s^+_{ij}$ (c.f. Fig.
\ref{fig:eigenframe}). This geometrical feature appears to 
have not be reported
before. 

The alignment between $s^+_{ij}$ and $s^-_{ij}$ can be
described by the polar angles that each individual eigenvector of
$s^-_{ij}$ makes in the eigen-frame of
$s^+_{ij}$. The polar angles are defined in the same way shown in
Fig. \ref{fig:eigenframe}, with $\bo^-$ replaced by one 
of the eigenvectors,
such as $\e^-_\alpha$. We use 
$\theta^s_P$ and $\psi^s_P$ to denote the angles. 
Fig. \ref{fig:Nes_align} plots the joint
PDFs of $\cos \theta^s_P$ and $\psi^s_P$ for the three eigenvectors,
$\e^-_\alpha$, $\e^-_\beta$ and $\e^-_\gamma$, in the left, middle,
and right panels, respectively. Distinct peaks can be
identified for all three distributions, 
though the peaks are not as sharp as in,
e.g., Fig. \ref{fig:Neo_align}. The left 
panel shows that 
$\e^-_\alpha$ displays a bi-modal behaviour, with the alignment
switching between $(\theta^s_P, \psi^s_P) = (90^\circ, 90^\circ)$ and
$\theta^s_P = 0^\circ$ (the value of $\psi^s_P$ is not defined
when $\theta^s_P = 0$). 
In the first configuration, $\e^-_\alpha$ aligns with $\e_\gamma$,
whereas in the second configuration, $\e^-_\alpha$ aligns with
$\e_\alpha$. 
The eigenvector $\e^-_\beta$, as shown by the middle panel, tends to
align with $\e_\beta$, since the PDF peaks at $(\theta^s_P, \psi^s_P)
= (90^\circ, 0)$. The right panel shows the joint PDF for 
$\e^-_\gamma$. Due to the
linearity of the equation for $\bv$, it should be
exactly the same as the one for $\e^-_\alpha$ shown in the left
panel. Due to statistical
fluctuations, the two joint PDFs are not exactly the same, but
they are very close, as expected. For example, they display
exactly same peak
locations. 

\section{Conclusions \label{sect:con}}	

We examine numerically the properties of the Lyapunov exponents
and conditional 
Lyapunov exponent for the Kolmogorov flow in a periodic box. The
production and dissipation of the infinitesimal velocity
perturbation (i.e., the conditional leading Lyapunov vector) are
the focus because they
determine the values of the conditional Lyapunov exponents
hence the synchronisability of the flow. 
The study mainly includes two parts, a spectral analysis and a
physical space analysis. 

In the first part, a detailed analysis
of the production spectrum and the dissipation spectrum for the
velocity perturbation is conducted. The impacts of the coupling
wavenumber are examined. We make several observations: 1) In most
cases, the
production is positive at all wavenumbers, implying
the perturbation is amplified at all scales.
2) Meanwhile, for large coupling wavenumbers, the
production spectrum may become negative for some wavenumbers,
showing the perturbation at corresponding scales are actually
weakened by the production term. 3) The conditional Lyapunov
exponents can be smaller than a lower bound proposed recently
based on a viscous estimate. 4) The production spectrum 
is attenuated by coupling and, counter-intuitively,  
this could amplify the dissipation spectrum for some wavenumbers.  

We extend previous discussions on the 
self-similar evolution of the perturbation spectrum. As a
result, 
a relation required for
self-similarity is derived between the local Lyapunov exponent and the
integral length scale of the velocity perturbation. The self-similarity of the
production spectrum is also examined; we highlight the need for simulations
with wider wavenumber range in order to observe clear
self-similarity in the production spectrum.

Regarding the peak wavenumber of the perturbation
energy spectrum, which has been related to the threshold coupling
wavenumber, an analytical relation
involving the production spectrum is given. However, to
obtain analytical solution for the peak wavenumber, a closure
model for the production spectrum is required. We discuss the
relation very briefly in a heuristic manner.

With analyses in physical space, we show that the velocity
perturbation is stronger in regions in the base flow with
strong vorticity or strong straining, but the correlation is
weakened when the coupling wavenumber is increased. We employ the
transport equation for the dissipation rate of the 
perturbation to identify two mechanisms that amplify the
dissipation: the stretching of perturbation
vorticity by the base flow strain rate, and the interaction
between the perturbation and base flow strain rates. These
observations bring to our attentions the roles of perturbation
vorticity and perturbation strain rate that appear to have been
neglected previously. The effects of the two mechanisms are then
quantified. The geometrical structures of the perturbation
vorticity and perturbation strain rate are also 
discussed. 
	
\backsection[Acknowledgement]
{The authors gratefully acknowledge the anonymous referees for their insightful
comments. The explanation for the weakened alignment observed in
Fig. \ref{fig:alignment_v_s} is 
based on the suggestions of one of the referees. 
Professor K. Yoshida is gratefully acknowledged 
for providing the data forming part of the right panel of Fig. \ref{fig:energy_decay}.  
}

\backsection[Funding]
{
Jian Li acknowledges the support of the National Natural Science
Foundation of China (No.
12102391).
} 

\backsection[Data availability statement]{The data that support the findings of this study are available 
from the corresponding author upon reasonable request.}

\backsection[Declaration of Interests]{The authors report no conflict of
interest.}

\bibliographystyle{jfm}
\bibliography{turbref}

\end{document}